\documentclass[10pt,journal]{IEEEtran}
\makeatletter
\def\ps@headings{%
\def\@oddhead{\mbox{}\scriptsize\rightmark \hfil \thepage}%
\def\@evenhead{\scriptsize\thepage \hfil \leftmark\mbox{}}%
\def\@oddfoot{}%
\def\@evenfoot{}}
\makeatother 

\usepackage{epsfig}
\usepackage{url}
\newtheorem{definition}{Definition}

\newenvironment{Proof}[1]{\emph{\textbf{Proof}}\space#1}{\hfill $\blacksquare$}
\newtheorem{Theorem}{\emph{\textbf{Theorem}}}

\newtheorem{example}{Example}
\usepackage{algorithmic}
\usepackage{algorithm}
\usepackage{color}
\usepackage{amsmath}
\usepackage{amsfonts}
\usepackage{amssymb}
\usepackage{graphicx}
\usepackage{comment}
\usepackage{subfig}
\usepackage{multirow}
\usepackage{bm}
\usepackage{epstopdf}
\usepackage[nocompress]{cite}

\newcommand{\warn}[1]{}

\def\IEEEkeywordsname{Keywords}
\def\linespaces{0.965}
\def\baselinestretch{\linespaces}
\begin{document}

\title{LoPub: High-Dimensional Crowdsourced Data Publication with Local Differential Privacy}
\author{\IEEEauthorblockN{ Xuebin Ren, Chia-Mu Yu, Weiren Yu, Shusen Yang, Xinyu Yang, Julie A. McCann, and Philip S. Yu}
\IEEEcompsocitemizethanks{
\IEEEcompsocthanksitem X. Ren, S. Yang, and X. Yang are with Xi'an Jiaotong University.
\protect\\ E-mails: \{xuebinren, shusenyang, yxyphd\}@mail.xjtu.edu.cn
\IEEEcompsocthanksitem C.-M. Yu is with National Chung Hsing University.
\protect\\ E-mail: chiamuyu@gmail.com
\IEEEcompsocthanksitem W. Yu is with Imperial College London and Aston University.
\protect\\ E-mails:weiren.yu@imperial.ac.uk, w.yu3@aston.ac.uk
\IEEEcompsocthanksitem J. McCann is with Imperial College London.
\protect\\ E-mail: j.mccann@imperial.ac.uk
\IEEEcompsocthanksitem P. Yu is with University of Illinois at Chicago.
\protect\\ E-mail: psyu@uic.edu }
}

\IEEEtitleabstractindextext{
\begin{abstract}
High-dimensional crowdsourced data collected from a large number of users produces rich knowledge for our society. However, it also brings unprecedented privacy threats to participants. Local privacy, a variant of differential privacy, is proposed as a means to eliminate the privacy concern. Unfortunately, achieving local privacy on high-dimensional crowdsourced data raises great challenges on both efficiency and effectiveness. Here, based on EM and Lasso regression, we propose efficient multi-dimensional joint distribution estimation algorithms with local privacy. Then, we develop a \underline{Lo}cally privacy-preserving high-dimensional data \underline{Pub}lication algorithm, LoPub, by taking advantage of our distribution estimation techniques. In particular, both correlations and joint distribution among multiple attributes can be identified to reduce the dimension of crowdsourced data, thus achieving both efficiency and effectiveness in locally private high-dimensional data publication. Extensive experiments on real-world datasets demonstrated that the efficiency of our multivariate distribution estimation scheme and confirm the effectiveness of our LoPub scheme in generating approximate datasets with local privacy.
\end{abstract}

}

\maketitle

\IEEEpeerreviewmaketitle

\section{Introduction}\label{sec:intro}
Nowadays, with the development of various integrated sensors and crowd sensing systems, the crowdsourced information from all aspects can be collected and analyzed among various data attributes to better produce rich knowledge about the group~\cite{7373649,7416007}, thus benefiting everyone in the crowdsourced system~\cite{Li2016Crowdsourced}. Particularly, with high-dimensional crowdsourced data (data with multiple attributes), a lot of potential information and rules behind the data can be mined or extracted to provide accurate dynamics and reliable prediction for both group and individuals~\cite{7463523}. For example, both individual electricity usage and community-wide electricity usage should be captured to achieve the real-time community-aware pricing in the smart grid, which integrates the distributed energy generation as well as consumption~\cite{liang2013udp, Shen2017Efficient}. People's historical medical records and genetic information~\cite{Akg¨¹nBayrak-770, Naveed2015Privacy} can be collected and mined to help hospital staff better diagnose and monitor patients' health status.
Various environment monitoring data collected from smart phone users can make urban planning more efficient and people's daily life more convenient~\cite{KhanXiang-435}.

However, the privacy of participants can be easily inferred or identified due to the publication of crowdsourced data~\cite{GroatEdwards-809,wang2015privacy,pham2010privacy}, especially high-dimensional data, even though some existing privacy-preserving schemes are used. The reasons for the privacy leak are two-fold:
\begin{itemize}
\item \textit{Non-local Privacy}. Most existing work for privacy protection focus on centralized datasets under the assumption that the server is trustable. However, in crowdsourced scenarios, direct data aggregation from distributed users can give adversaries (such as curious server and misbehaved insiders) the opportunity to identify individual privacy, even if end-to-end encryption is used. Despite the privacy protection against difference and inference attacks from aggregate queries, individual's data may still suffer from privacy leakage before aggregation because of no guaranteed privacy on the user side (\emph{i.e., local privacy}~\cite{NIPS2014_5392,duchi2013local,kairouz2016discrete,Chen2016Private}).

\item \textit{Curse of High-dimensionality}. In crowdsourced systems, high-dimensional data is ubiquitous. With the increase of data dimension, some existing privacy-preserving techniques like differential privacy~\cite{Dwork-405} (a \emph{de-facto} standard privacy paradigm), if straightforwardly applied to multiple attributes with high correlations, will become vulnerable, thereby increasing the success ratio of many reference attacks like cross-checking. According to the combination theorem~\cite{Mcsherry-39}, differential privacy degrades exponentially when multiple correlated queries are processed. From the aspect of utility, baseline differential privacy algorithms can hardly achieve reasonable scalability and desirable data accuracy due to the dimensional correlations~\cite{zhu2015correlated}.
\end{itemize}

In addition to the privacy vulnerability, a large scale of various data records collected from distributed users can imply the inefficiency of data processing. Especially in IoT applications, the ubiquitous but resource-constrained sensors and infrastructures require extremely high efficiency and low overhead. For example, privacy-preserving real-time pricing mechanism requires not only effective privacy guarantee for individuals' electricity usages but also fast response to the dynamic changes of the demands and supplies in the smart grid~\cite{liang2013udp}. Thus, it is important to provide an efficient privacy-preserving method to publish the crowdsourced high-dimensional data.

In addressing the above issues, various existing schemes demonstrated their effectiveness in terms of different perspectives. One is to provide local privacy for distributed users~\cite{Erlingsson-2014}, while some recent work \cite{sun2016pristream}\cite{fanti2015building} have been proposed to provide a local privacy guarantee for individuals in data aggregations. However, these schemes are either inefficient or not applicable for high-dimensional data~\cite{zhang2014privbayes} because of the high computation complexity and great utility loss. The other is to privately release high dimensional data~\cite{day2015differentially}\cite{zhang2014privbayes}\cite{chen2015differentially}. For example, Chen~\emph{et al.}~\cite{chen2015differentially} proposed to achieve differential privacy on compacted attribute clusters after dimension reduction according to the attribute correlations. However, these schemes mainly deal with the centralized dataset without local privacy guarantee for distributed data contributors. To overcome the compatibility problem between those schemes, we propose a novel scheme to publish high-dimensional crowdsourced data while guaranteeing local privacy.

\subsection{Contribution}\label{sec:Contribution}
Our major contributions are summarized as follows.
\begin{itemize}
  \item In this paper, we propose a locally privacy-preserving scheme for crowdsensing systems to collect and build high dimensional data from the distributed users. Particularly, we propose an efficient algorithm for multivariate joint distribution estimation and a combined algorithm to improve the existing estimation algorithms in order to achieve both high efficiency and accuracy.
  \item Based on the distribution and correlation information, the dimensionality and sparsity in the original dataset can be reduced by splitting the dataset into many compacted clusters.
  \item After learning the marginal distribution in these compacted clusters, the server can draw a new dataset from these compacted clusters, thus achieving an approximation of the whole original crowd data while guaranteeing local privacy for individuals.
  \item We implement and evaluate our schemes on different real-world datasets, experimental results show that our scheme is both efficient and effective in marginal estimation as well as high-dimensional data releasing.
\end{itemize}

\subsection{Organization}\label{sec:Organization}
The paper is organized as follows: In Section \ref{sec:model}, we introduce the system model in our paper. In Section \ref{sec:pre}, we present preliminaries of our approach. Then, Section \ref{sec:LoPub} presents our proposed schemes in detail. In Section \ref{sec:evaluation}, we present the experiment results to validate the efficiency and effectiveness of our schemes. Last, we concludes this paper in Section \ref{sec:conclusion}.

\section{Related Work}\label{sec:related}
\subsection{High-dimensional data}\label{sec:hdd}
To solve the privacy issue in data releasing, differential privacy~\cite{Dwork-405} was proposed to lay a mathematical privacy foundation by adding proper randomness. Examples of the use of differential privacy include the privacy-preserving data aggregation operations, where differential privacy of individuals can be guaranteed by injecting carefully-calibrated Laplacian noise~\cite{LiyueLi-278,kellaris2013practical,zhang2014privbayes,chen2015differentially,li2015differentially,7839941, YWR+17}. The techniques for \emph{non-interactive differential privacy}~\cite{Dwork-50,DworkMcsherry-19} suffer from "curse of dimensionality"~\cite{zhang2014privbayes,chen2015differentially}. Particularly, the combination theorems~\cite{Mcsherry-39} have pointed out that differential privacy degrades when multiple related queries are processed~\cite{sarathy2011evaluating}. To deal with the correlations in high-dimensional data, different schemes have been proposed~\cite{dependencemake,rescuedp2016,zhang2014privbayes,chen2015differentially,kellaris2013practical,day2015differentially,zhu2015correlated}.

For example, Liu \emph{et al.}~\cite{dependencemake} proposed the notion of dependent differential privacy (DDP) and dependent perturbation mechanism (DPM) to both measure and preserve the privacy with consideration of the dependence and correlations. Another solution to mitigate correlation problem is to group the correlated records into clusters and then achieve privacy on each low-dimensional cluster. Based on the data structure of dimensions, these solutions can be categorized into two classes: homogeneous and heterogeneous.
\begin{enumerate}
  \item \textbf{Homogeneous data} means the data structure of all dimensions is the same. For example, in the massive text document, each dimension represents the frequency histogram of a text on the distinct words. Since the data structure is the same, basic distance metrics can be simply used to cluster the similar dimensions. For example, Kellaris \emph{et al.}~\cite{kellaris2013practical} proposed to group the data attributes with similar sensitivities together to reduce the overall sensitivity. Based on the similar idea, Wang \emph{et al.}~\cite{rescuedp2016} proposed to group the data streams with the similar trend to achieve efficient privacy on data streams. However, both value and trend similarities cannot truly reflect the correlations in heterogeneous data.

  \item \textbf{Heterogeneous data} refers to the high-dimensional data with different dimensions, i.e., different attribute domains. For example, personal profiles may include various attributes like gender, age, education status. Intuitive distance metrics cannot measure the correlations among heterogeneous. Therefore, Zhang \emph{et al.}~\cite{zhang2014privbayes} proposed to calculate the mutual information of attribute-parent pairs and then model the correlations via a Bayesian network. Li \emph{et al.}~\cite{li2014differentially} proposed to use copula functions to model joint distribution for high-dimensional data. However, Copula functions cannot handle attributes with small domains, which limits its application. Chen \emph{et al.}~\cite{chen2015differentially} proposed to directly calculate the mutual information between pairwise dimensions and build a dependency graph and junction tree to model the correlations.
\end{enumerate}

\begin{figure*}[t]
\centering\epsfig{file=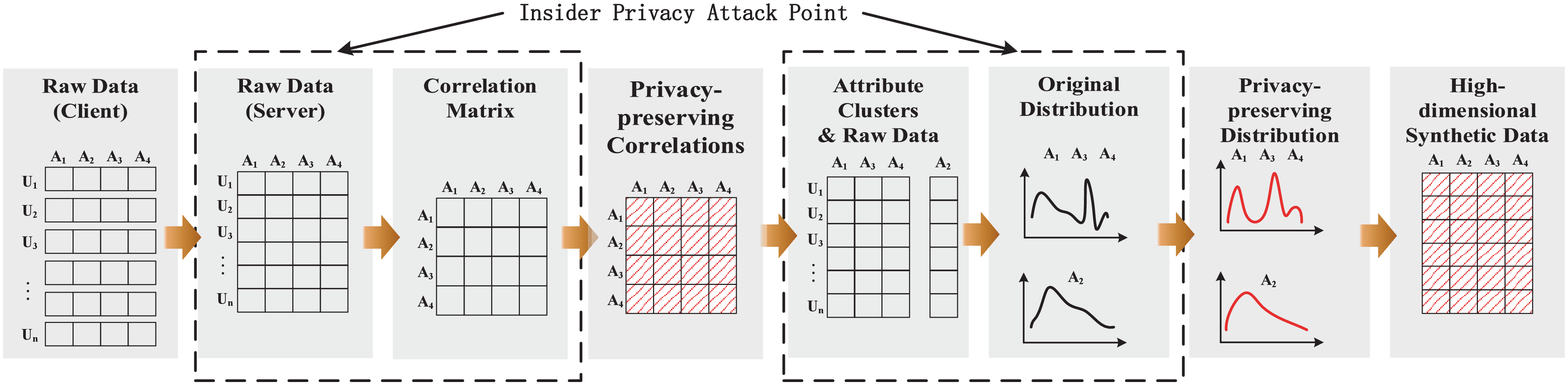, width=\textwidth}
\caption{Main procedures of high-dimensional data publishing without local privacy \label{MP1}}
\end{figure*}

However, in existing schemes~\cite{chen2015differentially,zhang2014privbayes} as depicted by Figure~\ref{MP1}, the original dataset are plainly accessed twice to learn the correlations among attributes and generate the distributions for clusters. Although calibrated noises are added separately (the $4$th and $7$th step) on the aggregation results, individual's privacy can only be protected against the third parties from the aggregation but still be threatened by the server, who collects and aggregates the data. In addition, the two accesses are computed separately without a consistent privacy guarantee. That is, two differential privacy budgets were allocated separately but it is not clear about how to allocate the privacy budget to achieve both sufficient privacy guarantee and utility maximization. Besides, although the entire high dimensional data can be reduced into several low-dimensional clusters. The sparsity caused by combinations in each cluster still exists and may lead to lower utility. In contrast to the totally centralized setting in~\cite{chen2015differentially}, Su \emph{et al.}~\cite{su2016differentially} proposed a distributed multi-party setting to publish new dataset from multiple data curators. However, their multi-party computation can only protect the privacy between data servers. Instead, individual's local privacy in a data server cannot be guaranteed.

Also, for high-dimensional data, to show the crowd statistics and draw the correlations between attributes (variables), both privacy-preserving histogram (univariate distribution)~\cite{acs2012differentially} and contingency table\footnote{A contingency table is a visualised table widely used in statistic areas that displays multivariate joint distribution of attributes.} (multivariate joint distribution)~\cite{qardaji2014priview} are widely investigated. However, these work can not provide local privacy guarantee, thus being unable to apply to crowdsourced systems.

\subsection{Local privacy}\label{sec:lp}
However, the schemes mentioned above mainly deal with the centralized dataset. There could be scenarios, where distributed users contribute to the aggregate statistics. Despite the privacy protection against difference and inference attacks from aggregate queries, individual's data may also suffer from privacy leakage before aggregation. Hence, the notion of local privacy has been proposed to provide local privacy guarantee for distributed users. In addition, local privacy from the end user can ensure the consistency of the privacy guarantee when there are multiple access to users' data. Instead, non-local privacy schemes like~\cite{zhang2014privbayes} and~\cite{chen2015differentially} have to split and assign proper privacy budgets to different steps while accessing to the original data.

Local privacy schemes mainly focus on achieving differentially private noise via combining the data from distributed source. Many different distributed differential privacy techniques were proposed \cite{eigner2014differentially}. However, the local noise may be too small to cover the original data. Recently, local privacy aims to learn particular aggregation features from distributed users with some public knowledge \cite{GroatEdwards-809,Erlingsson-2014,fanti2015building,sun2016pristream}. Sun \emph{et al.}~\cite{sun2016pristream} proposed a percentile aggregation scheme for monitoring distributed stream. In this scheme, the possible range information is public known. For example, Groat\emph{et al.}~\cite{GroatEdwards-809} proposed the technique negative surveys, which is based on randomized response techniques~\cite{warner1965randomized,7967624}, to identify the true distributions from noisy participants' data. Similarly, Erlingsson \emph{et al.}~\cite{Erlingsson-2014} proposed RAPPOR to estimate the frequencies of different strings in a candidate set. Their subsequent research~\cite{fanti2015building} proposed to learn the correlations between dimensions via EM based learning algorithm. However, when the dimension is high, sparsity of data will lead to great utility loss, and moreover, the EM algorithm will have a exponentially higher complexity.


Different from these work, we propose a novel mechanism to publish high-dimensional crowdsourced data with local privacy for individuals. We compare our work with three similar existing work in the Table~\ref{table:comparison}. In terms of privacy model, our method is local privacy while JTree~\cite{chen2015differentially} is not. Our method supports high dimensional data better than RAPPOR~\cite{Erlingsson-2014} and EM~\cite{fanti2015building}. Also, our method has relatively lower communication cost, time and storage complexity\footnote{Detailed analysis of time and storage complexity can be referred to Section~\ref{sec:MDME}}.

\begin{table}[htb]\centering\caption{Comparison of LoPub with existing methods}\label{table:comparison}
\tiny
\begin{tabular}{|c|c|c|c|c|}
  \hline
  Comparison & LoPub (Our method) & RAPPOR~\cite{Erlingsson-2014} & EM~\cite{fanti2015building} & JTree~\cite{chen2015differentially} \\
  \hline
  Local privacy & Y & Y & Y & N \\
  \hline
  High Dimension & Y & N & N & Y \\
  \hline
  Communication & $O(\sum\nolimits_{j} |\Omega_j|)$ & $O(\prod\nolimits_j |\Omega_j|)$  & $O(\sum\nolimits_{j}|\Omega_j|)$ & - \\
  \hline
  Time Complexity & Low & Low & Large & -\\
  \hline
  Space Complexity & Low &Large &Large & - \\
  \hline
\end{tabular}
\end{table}

\section{System Model}\label{sec:model}

\begin{figure}[t]
\centering\epsfig{file=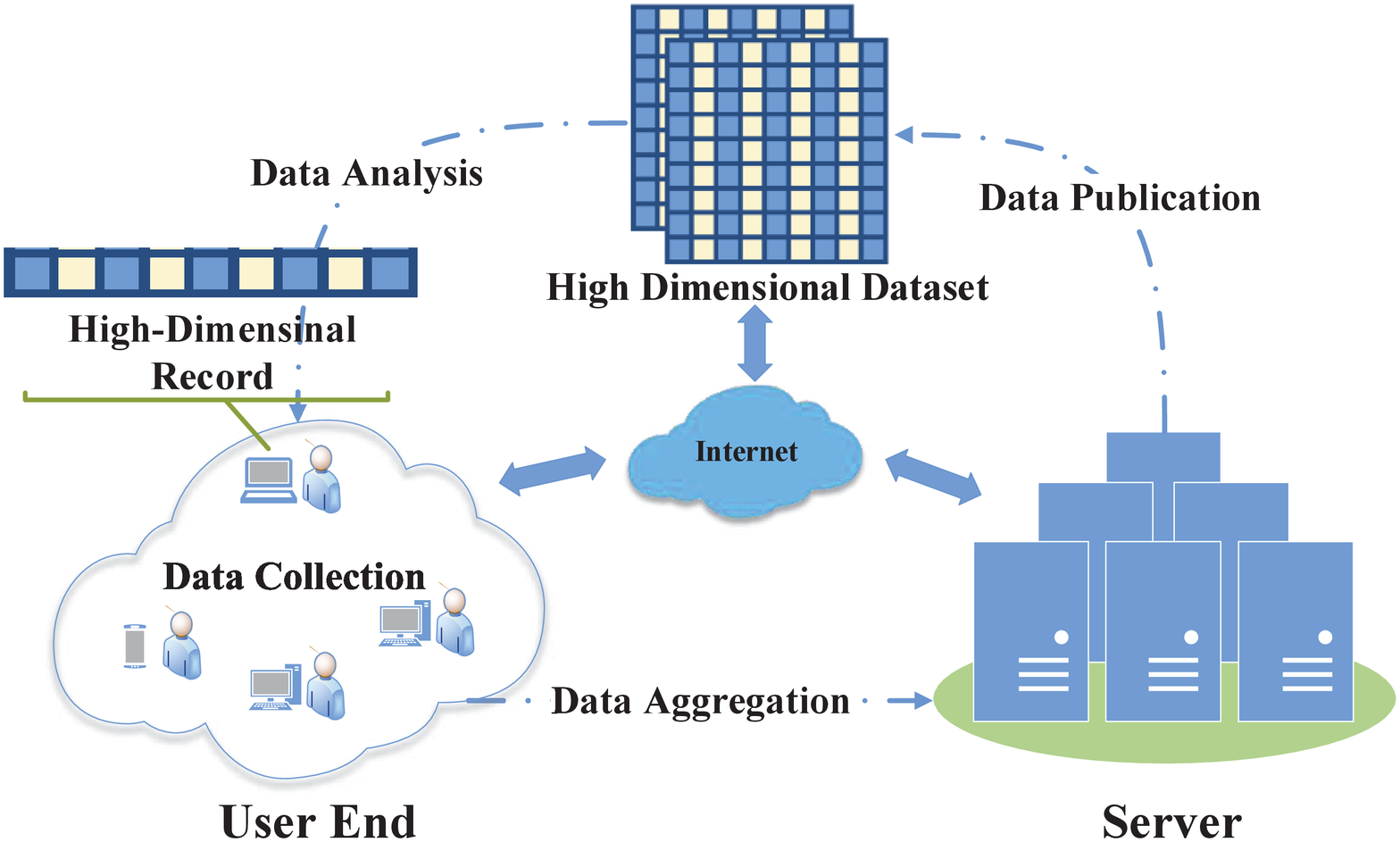, width=0.5\textwidth}
\caption{An architecture of distributed high dimensional private data collecting and publishing \label{scenario}}
\end{figure}

The system model is depicted in Figure~\ref{scenario}, where a lot of users and a central server are interconnected, constituting a crowdsourcing system. The users first collect high-dimensional data records from multiple attributes at the same time, and then send these data to the central server. The server gathers all the data and estimates high-dimensional crowdsourced data distribution with local privacy, aiming to release a privacy-preserving dataset to third-parties for conducting data analysis. In this paper, we mainly focus on data privacy, thus the detailed network model is not emphasised.

\textbf{Problem Statement}. Given a collection of data records with $d$ attributes from different users, our goal is to help the central server publish a synthetic dataset that has the approximate joint distribution of $d$ attributes with local privacy. Formally, let $N$ be the total number of users (i.e., data records\footnote{For brevity, we assume that each user sends only one data record to the central server.}) and sufficiently large. Let $X=\{X^1,X^2,\dots,X^N\}$ be the crowdsourced dataset, where $X^i$ denotes the data record from the $i$th user. We assume that there are $d$ attributes $A=\{A_1,A_2,\dots,A_d\}$ in $X$. Then each data record $X^i$ can be represented as $X^i=\{x^i_1,x^i_2,\dots,x^i_d\}$, where $x^i_j$ denotes the $j$th element of the $i$th user record. For each attribute $A_j~(j=1,2,\dots,d)$, we denote $\Omega_j=\{\omega_j^1,\omega_j^2,\dots,\omega_j^{|\Omega_j|}\}$ as the domain of $A_j$, where $\omega_j^i$ is the $i$th attribute value of $\Omega_j$ and $|\Omega_j|$ is the cardinality of $\Omega_j$.

With the above notations, our problem can be formulated as follows: Given a dataset $X$, we aim to release a locally privacy-preserving dataset $X^\star$ with the same attributes $A$ and number $N$ of users in $X$ such that
\begin{align}\label{problem}
P_{X^\star}(A_{1} \dots A_{d})&\approx P_{X}(A_{1} \dots A_{d})
\end{align} and
\begin{align}
&P_{X}(A_{1} \dots A_{d}) \triangleq P_{X}(x^i_{1}=\omega_{1},\dots,x^i_{d}=\omega_{d})\\
&\text{for}~i=1,\dots,N, \text{and} ~~\omega_{1},\dots,\omega_{d}\in \Omega_{d} \nonumber
\end{align}
where $P_{X}(x^i_{1}=\omega_{1},\dots,x^i_{d}=\omega_{d})$ is defined as the $d$-dimensional (multivariate) joint distribution on $X$.

To focus our research on data privacy, we assume that the central server and users are all \textit{honest-but-curious} in the sense that they will honestly follow the protocols in the system without maliciously manipulating their received data. However, they may be curious about others' data privacy and even collide to infer others' data. In addition, the central server and users share the same public information, such as the privacy-preserving protocols (including the hash functions used) and the domain $\Omega_j$ for each attribute $A_j$ ($j=1,2,\dots,d$).

\section{Preliminaries}\label{sec:pre}

\subsection{Differential Privacy}
Differential privacy is the de-facto standard for privacy guarantee on . It limits the adversaries' ability of inferring the participation or absence of any user in a data set via adding carefully calibrated noise to the query results on the data set. The definition of $\epsilon$-differential privacy can be found in \cite{Dwork-405}. $\epsilon$ is the privacy budget (or privacy parameter) to specify the level of privacy protection and smaller $\epsilon$ means better privacy. According to the combination theorem~\cite{sarathy2011evaluating}, extra privacy budget will be required when multiple differential privacy mechanisms are applied on related queries. Once $\epsilon$ is run out, no more differential privacy can be guaranteed.

If a mechanism $M$ is $\epsilon$-differential privacy on dataset $D$ and a new dataset $D'$ sampled from $D$ with uniformly sampling rate $\beta$. Then, querying $M$ on $D'$ can guarantee a $\ln(1+\beta(e^\epsilon-1))-$differential privacy for dataset $D$~\cite{li2012sampling}.

\subsection{Local Differential Privacy}
Generally, differential privacy focuses on centralized database and implicitly assumes data aggregation is trustworthy. Aiming to eliminate this assumption, local privacy was proposed for crowdsourced systems to provide a stringent privacy guarantee that data contributors trust no one~\cite{NIPS2014_5392,duchi2013local}.  A formal definition of local differential privacy is given below.
\begin{definition}
For any user $i$, a mechanism $\mathcal{M}$ satisfies $\epsilon$-local differential privacy (or simply local privacy) if for any two data records $X^i, Y^i \in \Omega_1\times\dots\times\Omega_d$, and for any possible privacy-preserving outputs $\tilde{X}^i \in Range(\mathcal{M})$,
\begin{align}
P\left(\mathcal{M}(X^i)= \tilde{X}^i\right)\leq e^{\epsilon} \cdot P\left(\mathcal{M}(Y^i)= \tilde{X}^i\right),
\end{align}
where the probability is taken over $\mathcal{M}'s$ randomness and $\epsilon$ has similar impact on privacy as in the ordinary differential privacy.
\end{definition}

Local privacy has many applications. A typical example is the randomized response technique~\cite{warner1965randomized}, which is widely used in the survey of people's ``yes or no'' opinions about a private issue. Participants of the survey are required to give their true answers with a certain probability or random answers with the remaining probability. Due to the randomness, the surveyor cannot determine the true answers of participants individually (i.e., local privacy is guaranteed) while can predict the true proportions of alternative answers.

\subsection{RAPPOR based Local Privacy}\label{sec:rappor}

Recently, RAPPOR (Randomized Aggregatable Privacy-Preserving Ordinal Response) has been proposed for statistics aggregation~\cite{Erlingsson-2014}. RAPPOR is only applicable to one or two dimensional crowdsourced data for estimating data distribution with local privacy. The basic idea of RAPPOR is the extension of the randomized response technique. In the randomized response technique, the candidate set of the domain $\Omega$ is only a binary input $\{0,1\}$ denoting ``yes or no'', whereas RAPPOR generalizes the candidate set of $\Omega$ to multiple inputs such that a sufficiently long $\{0,1\}$ bit string can be applied to the domain $\Omega$.

On users, RAPPOR consists of two phases:
\begin{enumerate}
  \item \textbf{Feature Assignment.} In this phase, the dataset $X$ has only one attribute with the domain $\Omega$. For each candidate value $\omega \in \Omega$, several hash functions are used to transform $\omega$ into a $m$-bit $\{0,1\}$ string $S$ (a.k.a. a Bloom filter). Once $m$ and the number of hash functions are well chosen, this transformation can maximize the uniqueness of a bit string $S$ to represent any attribute value $\omega \in \Omega$.As suggested by~\cite{starobinski2003efficient}, \textbf{$m$ is proportional to the domain size $|\Omega|$, i.e., $m \propto |\Omega|$.}
  \item \textbf{Feature Cloaking.} After the Bloom filter $S$ is obtained, each bit in $S$ will be randomized to $0$ or $1$ with a certain probability, or remains unchanged with the remaining probability. This randomness is important as it endows the privacy on the original data. The more randomness, the better privacy.
\end{enumerate}

On the central server, RAPPOR first gathers randomized Bloom filters from different users, and then estimates the univariate distribution as follows:
\begin{enumerate}
  \item \textbf{Aggregation.} Once gathered by the central server, all the bit strings will be summed up bitwise. Then the true count of each bit can be estimated based on the randomness of the bit strings.
  \item \textbf{Feature Rebuilding.} The hash functions used on the user side are replayed on the server side to reconstruct the Bloom filters for each $\omega \in \Omega$.
  \item \textbf{Distribution Estimation.} Taking the Bloom filters as the feature variables, the server can estimate the univariate distribution of the single attribute via linear regression.
\end{enumerate}
It is important to note that RAPPOR is only efficient in low dimensional data because when the dimension $k$ is high, the length $m_{RAPPOR}$ of Bloom filters over the multi-attribute domain $\Omega_1 \times \Omega_2 \times \dots \times \Omega_k$ will become
\begin{align}\label{eq:exponent}
m_{RAPPOR} \propto |\Omega_1 \times \Omega_2 \times \dots \times \Omega_k|=\prod\limits_{j=1}^k |\Omega_j|,
\end{align}
which \textbf{requires exponential storage space in terms of $k$}.

To address this problem, Fanti {\em et al.} \cite{fanti2015building} proposed an EM (Expectation Maximization) based association learning scheme, which extends the $1$-dimensional RAPPOR to estimate the $2$-dimensional joint distribution. First, a bivariate joint distribution is initialized uniformly. Then, for each record, the conditional probability distribution of the true $2$-dimensional Bloom filters given the observed noisy record is calculated according to the Bayes' theorem. Finally, the bivariate joint distribution is updated as the expectation of the conditional probability distributions over all records. By iterating the above steps several rounds, an estimation of the bivariate joint distribution can be obtained. \textbf{However, repeating scanning all the collected RAPPOR strings in each round of EM algorithm incurs considerable computational complexity}.

Some notations used in this paper are listed in Table~\ref{Notation}.

\begin{table}
\centering \caption{Notation}\label{Notation}\vspace{-2mm}
\fontsize{8pt}{\baselineskip}\selectfont
\begin{tabular}{lp{0.35\textwidth}} \hline
$N$& number of users (data records) in the system\\
$X$& entire crowdsourced dataset on the server side\\
$X^i$& data record from the $i$th user\\
$x^i_j$& $j$th element of $X^i$\\
$d$& number of attributes in $X$\\
$\mathbb{R}$& set of all attribute clusters\\
$A_j:$& $j$th attribute of $X$\\
$\Omega_j$& domain of $A_j$\\
$\omega_j$& candidate attribute value in $\Omega_j$\\
$\mathcal{H}_j(x)$& hash functions for $A_j$ that map $x$ into a Bloom filter \\
$s^i_j$& Bloom filter of $x^i_j$ ($S^i_j=\mathcal{H}_j(x^i_j)$)\\
$s^i_j[b]$& $b$th bit of $s^i_j$\\
$\hat{s}^i_j$& randomized Bloom filter of $s^i_j$\\
$\hat{s}^i_j[b]$& $b$th bit of $\hat{s}^i_j$\\
$m_j$&  length of $s^i_j$ \\
$f$& probability of randomly flipping a bit of a Bloom filter\\

\hline
\end{tabular}
\vspace{-3mm}
\end{table}

\section{LoPub: High-dimensional Data Publication with Local Privacy}\label{sec:LoPub}
We propose LoPub, a novel solution to achieve high-dimensional crowdsourced data publication with local privacy. In this section, we first introduce the basic idea behind LoPub and then elaborate the algorithmic procedures in more details.

\subsection{Basic idea}
Privacy-preserving high-dimensional crowdsourced data publication aims at releasing an approximate dataset with similar statistical information (i.e., statistical distribution defined in Equation~(\ref{problem})) while guaranteeing the local privacy. This problem can be considered in four aspects:

First, to achieve local privacy, some local transformation should be designed on the user side to cloak individuals' original data record. Then, the central server needs to obtain the statistical information, a.k.a, the distribution of original data. There are two plausible solutions. One is to obtain the $1$-dimensional distribution on each attribute independently. Unfortunately, the lack of consideration of correlations between dimensions will lose the utility of original dataset. Another is to consider all attributes as one and compute the $d$-dimensional joint distribution. However, due to combinations, the possible domain will increase exponentially with the number of dimensions, thus leading to both low scalability and the signal-noise-ratio problems~\cite{zhang2014privbayes}. Particularly, with fixed local privacy guarantee, the statistical accuracy of distribution estimation will degrade significantly with the increase of the possible domain and dimensionality. Therefore, next crucial problem is to find a solution for reducing the dimensionality while keeping the necessary correlations. Finally, with the statistical distribution information on low-dimensional data, how to synthesize a new dataset is the remaining problem.

\begin{figure}[t]
\centering\epsfig{file=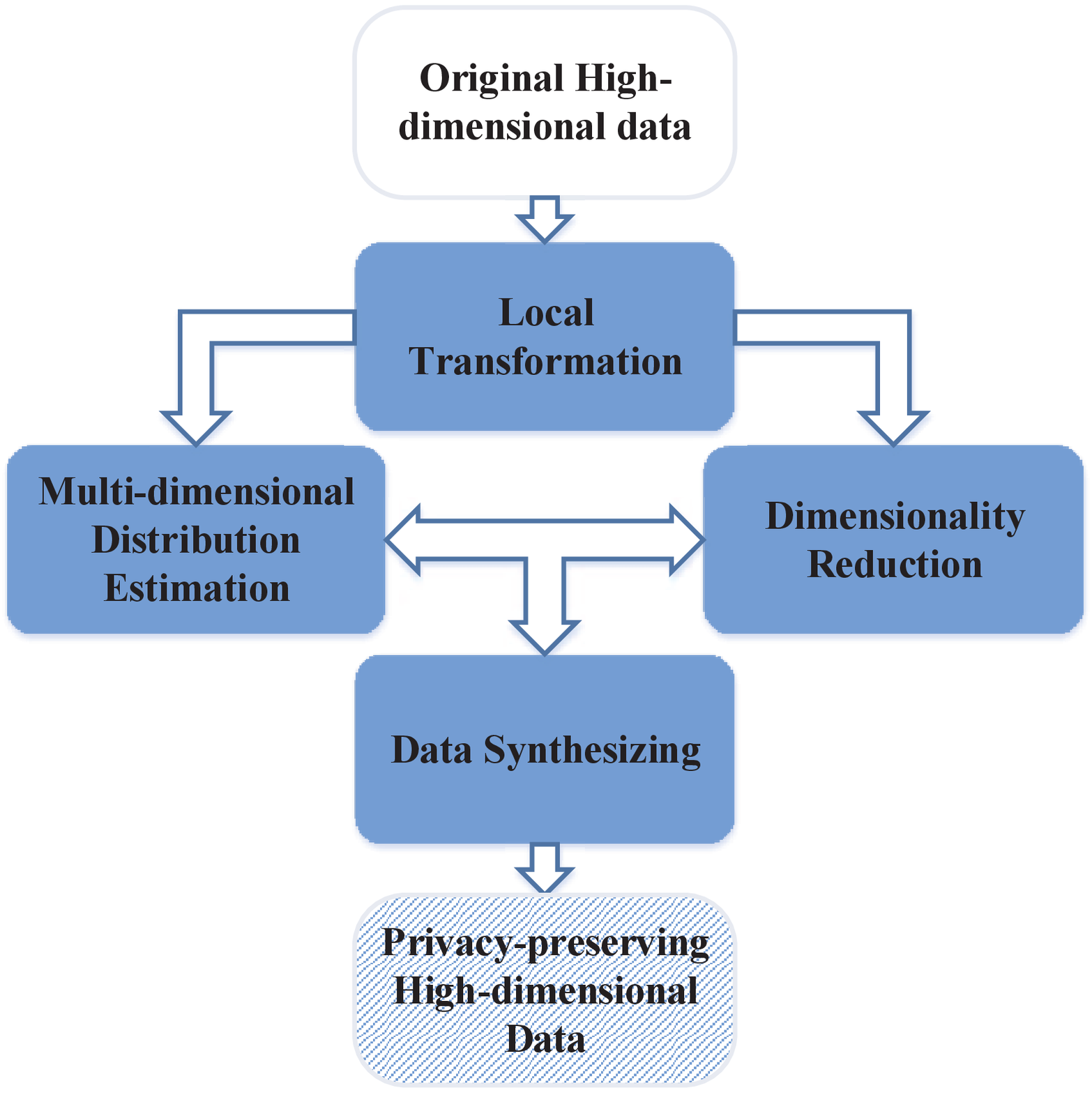, width=0.3\textwidth}
\caption{An overview of LoPub \label{lopubroad}}
\vspace{-0.3cm}
\end{figure}

To this end, we present LoPub, a locally privacy-preserving data publication scheme for high-dimensional crowdsourced data. Figure~\ref{lopubroad} shows the overview of LoPub, which mainly consists of four mechanisms: local privacy protection, multi-dimensional distribution estimation, dimensionality reduction, and data synthesizing.

\begin{enumerate}
  \item \textbf{Local Privacy Protection.} We first propose the local transformation process that adopts randomized response technique to cloak the original multi-dimensional data records on distributed users to provide local privacy for all individuals in the crowdsourced systems. Particularly, we locally transform each attribute value to a random bit string. Then, the locally privacy-preserved data is sent to and aggregated at the central server.
  \item \textbf{Multi-dimensional Distribution Estimation.} We then propose multi-dimensional joint distribution estimation schemes to obtain both the joint and marginal probability distribution on multi-dimensional data. Inspired by~\cite{fanti2015building}, we present an EM-based approach for high-dimensional estimation. Moreover, we present Lasso-based approach for fast estimation at the cost of slight accuracy degradation. Finally, we propose a hybrid approach striking the balance between the accuracy and efficiency.
  \item \textbf{Dimensionality Reduction.} Based on the multi-dimensional distribution information, we then propose to reduce the dimensionality by identifying mutual-correlated attributes among all dimensions and split the high-dimensional attributes into several compact low-dimensional attribute clusters. In this paper, considering the heterogeneous attributes, we adopt mutual information and undirected dependency graph to measure and model the correlations of attributes, respectively. In addition, we also propose a heuristic pruning scheme to further boost the process of correlation identification.
  \item \textbf{Synthesizing New Dataset.} Finally, we propose to sample each low-dimensional dataset according to the estimated joint or conditional distribution on each attribute cluster, thus synthesizing a new privacy-preserving dataset.
\end{enumerate}

\subsection{Local Transformation}\label{sec:transform}


\subsubsection{Design Rationale}\label{sec:basic}
A common framework of locally private distribution estimation is that each individual user applies a local transformation on the data for privacy protection and then sends the transformed data to the server. The server estimates the joint distribution according to the transformed data. Local transformation in our design includes two key steps: one is mapping into Bloom filters and the other is adding randomness. Particularly, Bloom filter over $\Omega$ with multiple hash functions can hash all the variables in the domain into a pre-defined space. Thus, the unique bit strings are the representative features of the original report. Then, after privacy protection by randomized responses, a large number of samples with various levels of noise are generated by individual users. After aggregation, the central server obtain a large sample space with random noises. As a result, one may estimate the distribution from the noised sample space by taking advantage of machine learning techniques such as EM algorithm and regression analysis.

Under the above framework, a key observation can be made: if features are mutual-independent, one can easily conclude that the combinations of features from different candidate sets are also mutual-independent. Therefore, when Bloom filters of each attribute are mutual-independent (i.e., no collisions for all bits), then the Cartesian product of Bloom filters of different attributes are mutual-independent. In this sense, with mutual-independent features of Bloom filters, existing machine learning techniques like EM and Lasso regression are effective for the multivariate distribution estimation. Some notations used in this paper are listed in Table~\ref{Notation}.

\subsubsection{Algorithmic Procedures of Local Transformation}
Before describing the distribution estimation, we present that details about the local transformation. In essence, local transformation consists of three steps:
\begin{enumerate}
  \item
  On each $i$th user, suppose we have an original data record $X^i=\{x_1^i,x_2^i,\dots,x_d^i\}$ with $d$ attributes. For each attribute $A_j \ (j=1,\dots,d)$ , we employ hash functions ${\cal H}_j(\cdot)$ to map $x_j^i$ to a length-$m_j$ bit string $s_j^i$ (called \emph{a Bloom filter}). That is,
  \[
   s_j^i = {\cal H}_j (x_j^i) \quad (\forall j=1,\dots,d)
  \]
  \item
  Each bit $s_j^{i}[b]~(b=1,2,\dots,m_j)$ in $s_j^i$ is randomly flipped into $0$ or $1$ according to the following rule:
\begin{align}\label{rr}
\hat{s}_j^{i}[b]=
\begin{cases}
  s_j^{i}[b], ~~& \text{with probability of}~ 1-f\\
  1, ~~& \text{with probability of}~ f/2\\
  0, ~~& \text{with probability of}~ f/2\\
\end{cases}
\end{align}
where $f\in (0,1)$ is a user-controlled flipping probability that quantifies the level of randomness for local privacy.
  \item
  Having flipped each bit string $s_j^{i}$ bit by bit into randomized Bloom filter $\hat{s}_j^{i} \ (j=1,\dots,d)$,
  we concatenates $\hat{s}_1^{i},\dots, \hat{s}_d^{i}$ to obtain a stochastic $(d \cdot m_j)$-bit vector:
\[ \big[\hat{s}_1^{i}[1],\dots,\hat{s}_1^{i}[m_1]~\vline~ \hat{s}_2^{i}[1],\dots,\hat{s}_2^{i}[m_2]~\vline~\dots~\vline~ \hat{s}_d^{i}[1],\dots,\hat{s}_d^{i} [m_d]\big] \]
and send it to the server with guaranteed local privacy.
\end{enumerate}

\begin{example}
\begin{table}[h!]
\footnotesize
\centering\caption{An example of census data $X$ of college alumni}
\begin{tabular}{c|c|c|c|c|}
  \cline{2-5}
   & Age & Gender & Education & Income Level \\
  \cline{2-5}
  $u_1$ & 29 & M & college & working  \\
  $u_2$ & 35 & F & master & low-middle \\
  $u_{3}$ & 45 & F & college & working \\
  $\dots$ & \dots & \dots & \dots & \dots \\
  $u_N$ & 49 & M & phd & up-middle \\
  \cline{2-5}
\end{tabular}\label{origin}
\end{table}
\begin{table}[h!]
\footnotesize
\centering\caption{Attribute domain information of dataset $X$}
\begin{tabular}{|c|c|c|c|}
  \hline
  j & $A_j$ & $\Omega_j$ & $|\Omega_j|$ \\
  \hline
  1 & Age & $[25,60]$ &35\\
  2 & Gender & $\{\text{M},\text{F}\}$ &2\\
  3 & Education & $\{\text{college}, \text{master}, \text{phd}\}$ &3\\
  4 & Income & $\{\text{working}, \text{low-middle},\text{up-middle},\text{affluent}\}$ &4\\
  \hline
\end{tabular}\label{domain}
\end{table}
\begin{table}[h!]
\footnotesize
\centering\caption{Bloom strings of attribute domain ${\cal H}_j (\Omega_j)$}
\begin{tabular}{|c|c|c|c|}
  \hline
  j & $A_j$ & ${\cal H}_j(\Omega_j)$ & $|{\cal H}_j(\Omega_j)|$ \\
  \hline
  1 & Age & $\{100100111,\dots,01010110\}$ &35\\
  2 & Gender & $\{01,10\}$ &2\\
  3 & Education & $\{ 0101, 0110, 1100\}$ &3\\
  4 & Income & $\{0110, 0011,1001,1100\}$ &4\\
  \hline
\end{tabular}\label{domain}
\end{table}

\begin{table}[h!]
\footnotesize
\centering\caption{Privacy-preserving bit strings of $X$}
\begin{tabular}{c|c|c|c|c|}
  \cline{2-5}
   & Age & Gender & Education & Income Level \\
  \cline{2-5}
  $u_1$ & 10\underline{0}101\underline{1}1 & \underline{10} & 01\underline{1}1 & 01\underline{0}0  \\
  $u_2$ & 01\underline{1}1\underline{0}001 & \underline{0}0 & \underline{1}110 & 0\underline{1}11 \\
  $u_{3}$ & \underline{0}1\underline{0}1\underline{1}100 & 10 & 010\underline{0} & 0\underline{0}10 \\
  $\dots$ & \dots & \dots & \dots & \dots \\
  $u_N$ & 11010\underline{1}\underline{0}0 & 0\underline{0} & \underline{0}1\underline{1}0 & 1\underline{1}\underline{1}1 \\
  \cline{2-5}
\end{tabular}\label{transform}
\end{table}
Tabel~\ref{origin} shows a simplified example of original census dataset $X$ with 4 attributes \{``Age'',``Gender'', ``Education'', "Income Level"\}, where each record is contributed by user $u_i$, $1\leq i\leq N$. Consider data record $X^2=\{\emph{\text{35}, \text{F}, \text{Master}, \text{Low-middle}} \}$  on $2$nd user.
To guarantee local privacy, we first use hash functions ${\cal H}_j (\cdot)$ to map $j$th element of $X^2$ into a bit string. Next, having been randomly flipped bit by bit, these bit strings become
\[\begin{array}{lllll}
 {\cal H}_1 (\textrm{35}) &=& \{ 0101 1001\} & \Rightarrow \ \ & \{ 10\underline{0}101\underline{1}1\} \\
 {\cal H}_2 (\textrm{F}) &=& \{ 10\} & \Rightarrow \ \ & \{ \underline{0}0\} \\
 {\cal H}_3 (\textrm{master}) &=& \{ 0110\} & \Rightarrow \ \ & \{ \underline{1}1 10\} \\
 {\cal H}_4 (\textrm{low-middle}) &=& \{ 0011\} & \Rightarrow \ \ & \{ 0\underline{1}11\} \\
\end{array}\]
\end{example}
where underlined bits have been changed according to Equation (\ref{rr}). Finally, the concatenation of these randomized bit strings for each attribute yields a privacy-preserving bit string for the entire record $X^2$
\[
[ 10\underline{0}101\underline{1}1 \underline{1}011 \ \vline \ \underline{0}0 \ \vline \ \underline{1}1 10\ \vline \ 0\underline{1}11],
\]
which will be sent to the central server under local privacy. Similarly, other users should transform their data in a same way.  Table~\ref{transform} shows the example of transformed privacy-preserving report strings received by the server.  $\blacksquare$

\textbf{Parameters Setup:} According to the characteristic of Bloom filter~\cite{starobinski2003efficient}, given the false positive probability $p$ and the number $|\Omega_i|$ of elements to be inserted, the optimal length $m_j$ of Bloom filter can be calculated as
\begin{align}\label{eq:m}
m_j=\frac{ \ln (1/p)}{(\ln 2)^2} |\Omega_j|.
\end{align}
Furthermore, the optimal number $h_j$ of hash functions is
\begin{align}\label{eq:h}
h_j=\frac{m_j}{|\Omega_j|}\ln 2=\frac{\ln (1/p)}{(\ln 2)}.
\end{align}
So, the optimal $h=\frac{\ln (1/p)}{(\ln 2)}$ for all dimensions.

\textbf{Privacy Analysis:} Because local transformation is performed by the individual user, no one can obtain the original record $X^i$, local privacy can be easily achieved and we only have to analyze the privacy guarantee on the user side. According to the conclusion in~\cite{Erlingsson-2014}, differential privacy obtained for each attribute on the user side is $2h \ln \left( (2-f)/f \right)$, where $h$ is the number of hash functions in the Bloom filter and $f$ is the probability that a bit vector was flipped.

Since both hash operations and randomized response on all attributes are independent, then as pointed by the composition theorem~\cite{Mcsherry-39}, the overall differential privacy achieved on the user side should be \begin{align}
\epsilon=2d h \ln \left( (2-f)/f \right),\label{eq:dp}
\end{align} where $d$ is the number of dimensions.

Overall, since the same transformation is done by all users independently, this $\epsilon$-local privacy guarantee is equivalent for all distributed users. In the rest of our paper, we will focus on how to achieve a better utility-privacy tradeoff from users' privacy-preserving high-dimensional data with this privacy guarantee $\epsilon$.

\textbf{Communication Overhead:}
\begin{Theorem}
The minimal communication cost $C$ after the local transformation is
\begin{align}\label{comcost:our}
C=\sum_{j=1}^d m_j=\frac{\ln(1/p)}{(\ln 2)^2} \sum_{j=1}^d|\Omega_j|.
\end{align}
\end{Theorem}

\begin{Proof}
If we assume that the domain of each attribute is publicly known by both users and the server, then the communication cost of non-private collection is basically $\sum_{j=1}^d \ln |\Omega_j|$, which is related to the domain size. Nevertrheless, in our method with local privacy, the communication cost is $\sum_{j=1}^d m_j$, which is related to the length of Bloom filters because only randomly flipped bit strings (not original data record) are sent.
\end{Proof}


For comparison, under the same condition, when RAPPOR~\cite{Erlingsson-2014} is directly applied to the $k$-dimensional data, all $\Omega_1\times \dots\times \Omega_k$ candidate value will be regarded as $1$-dimensional data, then the cost is
\begin{align}
C_{RAPPOR}=\frac{\ln (1/p)}{(\ln 2)^2} \prod_{j=1}^k |\Omega_j|,
\end{align} where $\prod_{j=1}^k |\Omega_j|$ is due to the size of the candidate set $\Omega_1\times \dots\times \Omega_k$. Difference between Equation~\ref{comcost:our} and \ref{comcost:rappor} is because our LoPub, compared with straightforward RAPPOR, considers the mutual independency between multiple attributes. It should be noted that the Bloom filter length $m_j$ as well as communication cost $C_{LoPub}$ (or $C_{RAPPOR}$) is independent from the privacy level achieved.

\subsection{Multivariate Distribution Estimation with Local Privacy} \label{sec:MDME}

\subsubsection{EM-based Distribution Estimation}\label{sec:EM}
After receiving randomized bit strings, the central server can aggregate them and estimate their joint distribution. However, the existing EM-based estimation~\cite{fanti2015building} for joint distribution estimation is restricted to $2$ dimensions, which is impractical to many real-world datasets with high dimensions.
Here, we propose an alternative EM-based estimation that can applies to $k$-dimensional dataset ($2\leq k\leq d$) with provable complexity analysis.

Before illustrating our algorithm, we first introduce the following notations.
Without loss of generality, we consider $k$ specified attributes as $A_{1}$, $A_{2}$, $\dots$, $A_{k}$ and their index collection $\mathcal{C}=\{1,2,...,k\}$. For simplicity, the event $A_j=\omega_j$ or $x_j=\omega_j$ 
is abbreviated as $\omega_j$. For example, the prior probability $P(x_1=\omega_1, x_2=\omega_2, \dots , x_k=\omega_k)$ can be simplified into $P(\omega_1 \omega_2 \dots \omega_k)$ or $P(\omega_{\mathcal{C}})$.

Algorithm~\ref{alg:EM} depicts our EM-based approach for estimating $k$-dimensional joint distribution. More specifically, it consists of the following five main steps.

\begin{algorithm}[htb]\footnotesize
\caption{EM-based $k$-dimensional Joint Distribution}
\label{alg:EM}
\begin{algorithmic}[1]
\REQUIRE
\begin{tabular}[t]{p{0mm}l}
 $\mathcal{C}$&: attribute indexes cluster, i.e., $\mathcal{C}=\{1,2,...,k\}$\\
 $A_{j}$&: $k$-dimensional attributes $(1 \leq j \leq k)$,\\
 $\Omega_j$&: domain of $A_j$ $(1 \leq j \leq k)$,\\
 $\hat{s}_j^i$&: observed Bloom filters $(1 \leq i \leq N)$ $(1 \leq j \leq k)$,\\
 $f$&: flipping probability,\\
 $\delta$&: convergence accuracy.
\end{tabular}
\ENSURE $P(A_{\mathcal{C}})$: joint distribution of $k$ attributes specified by $\mathcal{C}$.

\STATE initialize $P_0(\omega_{\mathcal{C}})=1/(\prod \limits_{j \in \mathcal{C}} |\Omega_{j}|)$.
\FOR {each $i=1,\dots,N$}
\FOR {each $j \in \mathcal{C}$}
\STATE compute $P(\hat{s}^i_j|\omega_j)=\prod \nolimits_{b=1}^{m_j} (\frac{f}{2})^{\hat{s}_j^i[b]} (1-\frac{f}{2})^{1-\hat{s}_j^i[b]}$.
\ENDFOR
\STATE compute $P(\hat{s}^i_{\mathcal{C}}|\omega_{\mathcal{C}})=\prod \limits_{j \in \mathcal{C}} P(\hat{s}^i_j|\omega_j)$.
\ENDFOR
\STATE initialize $t=0$ ~~~~~~~~~~~~~~~~~~~~~~~~/* number of iterations */
\REPEAT
\FOR {each $i=1,\dots,N$}
\FOR {each $(\omega_{\mathcal{C}})\in \Omega_1\times\Omega_2\times \dots \times \Omega_k$}
\STATE compute $P_t(\omega_{\mathcal{C}}|\hat{s}^i_{\mathcal{C}})=\frac{P_{t}(\omega_{\mathcal{C}}) \cdot P(\hat{s}_{\mathcal{C}}^i |\omega_{\mathcal{C}})}{\sum \limits_{\omega_{\mathcal{C}}} P_t(\omega_{\mathcal{C}}) P(\hat{s}^i_{\mathcal{C}}|\omega_{\mathcal{C}})}$ \\
\ENDFOR
\ENDFOR
\STATE set $P_{t+1}(\omega_{\mathcal{C}})=\frac{1}{N}\sum_{i=1}^N P_t(\omega_{\mathcal{C}}|\hat{s}^i_{\mathcal{C}})$
\STATE update $t=t+1$
\UNTIL {$\max \limits_{\omega_{\mathcal{C}}} P_{t}(\omega_{\mathcal{C}})-\max\limits_{\omega_{\mathcal{C}}} P_{t-1}(\omega_{\mathcal{C}})\leq \delta$.}
\RETURN $P(A_{\mathcal{C}})=P_{t}(\omega_{\mathcal{C}})$
\end{algorithmic}
\end{algorithm}

\begin{enumerate}
  \item Before executing EM procedures, we set an uniform distribution $P(\omega_1 \omega_2 \dots \omega_k)=1/(\prod \limits_{j=1}^k |\Omega_{j}|)$ as the initial prior probability.

\begin{example} For simplicity, we consider the joint distribution on attributes $A_2 A_3$ (a.k.a. ``Gender'' and ``Education''). First, we initialize the probability on each combination as $P(A_2 A_3)=1/(|\Omega_2|\times |\Omega_3|)=1/6$.
\end{example}

\item According to Equation~(\ref{rr}), each bit $s_j^i[b]$ will be changed with probability $\frac{f}{2}$ and remains unchanged with probability $1-\frac{f}{2}$. By comparing the bits $\mathcal{H}_j (\omega_j)$ with the randomized bits, the conditional probability $P(\hat{s}^i_j|\omega_j)$ can be computed (see line 4 of Algorithm~\ref{alg:EM}).

\begin{example} We calculate the conditional probability $P(\hat{s}^i_j|\omega_j)$ of each attribute $A_j~(j=2,3)$. For example, on attribute $A_3$ (or $\text{Education}$), in the case of privacy parameter $f=1/2$, we have
\begin{align}
&P(\hat{s}^1_3 | A_3=\text{college}) \\ \nonumber
=&P(\hat{s}^1_3=0111 | {\cal H}_3=0101) \\ \nonumber
=&P(\hat{s}^1_3[1]=0| {\cal H}_3[1]=0)\times  \\ \nonumber
&P(\hat{s}^1_3[2]=1| {\cal H}_3[2]=1) \times  \\ \nonumber
&P(\hat{s}^1_3[3]=1| {\cal H}_3[3]=0) \times  \\ \nonumber
&P(\hat{s}^1_3[4]=1| {\cal H}_3[4]=1) \\ \nonumber
=&(1-\frac{f}{2})(1-\frac{f}{2})(\frac{f}{2})(1-\frac{f}{2})=\frac{27}{256}
\end{align}
Similarly, the conditional probabilities on other combinations can also be obtained.
\end{example}

\item Due to the independence between attributes (and their Bloom filters), the joint conditional probability can be easily calculated by combining each individual attribute, so $P(\hat{s}^i_{1} \hat{s}^i_{2} \dots \hat{s}^i_{k}|\omega_{1} \omega_{2} \dots \omega_{k})=\prod \nolimits_{j=1}^k P(\hat{s}^i_j|\omega_j)$.

\begin{example} The joint conditional probability $P(\hat{s}^1_{2} \hat{s}^1_{3} |A_{2} A_{3})$ of attributes $A_2 A_3$ can be enumerated. For example, the probability that $u_1$ is ``a female (Gender=F) alumni with phd degree and low-middle income'' can be computed as
\begin{align}
&P(\hat{s}^1_{2}\hat{s}^1_{3} | A_{2}=\text{F}, A_{3}=\text{phd}) \\ \nonumber
=&P(\hat{s}^1_{2}| A_{2}=\text{F})\times P(\hat{s}^1_{3} |A_{3}=\text{phd})
\end{align}
Similarly, the conditional probability $P(\hat{s}^1_{2} \hat{s}^1_{3} |A_{2} A_{3})$ on other $5$ candidate combinations of $A_2 A_3$ can be obtained.
\end{example}

\item Given all the conditional distributions of one particular combination of bit strings, their corresponding posterior probability can be computed by the Bayes' Theorem,
\begin{align}
&P_t(\omega_{1}\omega_2\dots \omega_{k}|\hat{s}^i_{1} \hat{s}^i_{2} \dots \hat{s}^i_{k}) \\ \nonumber
=&\frac{P_{t}(\omega_1\omega_2 \dots \omega_k) \cdot P(\hat{s}_{1}^i \hat{s}_{2}^i \dots \hat{s}_{k}^i|\omega_{1} \omega_{2} \dots \omega_{k})}{\sum \limits_{\omega_1} \sum \limits_{\omega_2}\dots \sum \limits_{\omega_k} P_t(\omega_1\omega_2 \dots \omega_k) P(\hat{s}^i_{1} \hat{s}^i_{2} \dots \hat{s}^i_{k}|\omega_{1}, \omega_{2} \dots \omega_{k})}.
\end{align}
Where $P_{t}(\omega_1\omega_2 \dots \omega_k)$ is the $k-$dimensional joint probability at the $t$th iteration.

\begin{example}
Given the privacy-preserving bit string $\hat{s}^1_{2}\hat{s}^1_{3}$, the posterior probability $P(A_2 A_3 |\hat{s}^1_{2}\hat{s}^1_{3})$ at the first iteration can be computed such as
\begin{align}
&P(A_2=\text{F}, A_3=\text{phd} |\hat{s}^1_{2}\hat{s}^1_{3})\\ \nonumber
=&\frac{P_{1}(A_2=\text{F}, A_3=\text{phd}) \cdot P(\hat{s}^1_{2}\hat{s}^1_{3} |A_2=\text{F}, A_3=\text{phd})}{\sum \limits_{\omega_2 \in \Omega_2} \sum \limits_{\omega_3 \in \Omega_3}P_1(\omega_2 \omega_3) \cdot P(\hat{s}^1_{2}\hat{s}^1_{3}|\omega_2 \omega_3)}
\end{align}
Similarly, after observing the privacy-preserving bit string $10~ 0111~ 0100$ ($\hat{S}^1$), all posterior probabilities of different combinations of $A_2 A_3$ can be obtained.
\end{example}

\item After identifying posterior probability for each user, we calculate the mean of the posterior probability from a large number of users to update the prior probability. The prior probability is used in another iteration to compute the posterior probability in the next iteration. The above EM-like procedures are executed iteratively until convergence, i.e., the maximum difference between two estimations is smaller than the specified threshold $\max P_{t}(\omega_{1} \omega_{2}\dots \omega_{k})-\max P_{t-1}(\omega_{1} \omega_{2}\dots \omega_{k})\geq \delta$. 

\begin{example}
For each observed privacy-preserving bit string $\hat{S}^i~(i=1,2,\dots,N)$, the posterior probability $P(A_2=\text{F}, A_3=\text{phd} |\hat{S}^i)$ for each combination can be obtained by the above procedures,
Then, the prior probability $P(A_2=\text{F}, A_3=\text{phd})$ is updated by the mean value of $N$ (here we take the $4$ records $u_1,u_2,u_3, \text{and}~ u_N$ in the table) posterior probabilities as
\begin{align}
&P_1(A_2=\text{F}, A_3=\text{phd}) \\ \nonumber
=&\frac{1}{N}\sum_{i=1}^N P(A_2=\text{F}, A_3=\text{phd}, |\hat{S}^i)
\end{align}
\end{example}
So, instead of initial probability $P_0(A_2=\text{F},A_3=\text{phd})=1/6=0.1667$, the updated prior probabilities $P_1(A_2=\text{F}, A_3=\text{phd})$ (should be about $0.1564$ in our example) will be used in the next iteration. Similarly, the initial probabilities of other $5$ combinations will be updated for the next iteration.
\end{enumerate}

The above algorithm can converge to a good estimation when the initial value is well chosen. EM-based $k$-dimensional joint distribution estimation will also fail when converging to local optimum. Especially when $k$ increases, there will be many local optimum to prevent good convergence because sample space of all combinations in $\Omega_{j_1}\times \Omega_{j_2}\times \dots\times \Omega_{j_k}$ explodes exponentially.

\textbf{Complexity:}
Before the analysis of complexity, we should note that number of user records $N$ needs to be sufficiently large according to the analysis in \cite{Erlingsson-2014}, i.e., $N\gg v^k$, where $v$ denotes the average size of $|\Omega_j|$, otherwise it is difficult to estimate reliably from a small sample space with low signal-noise-ratio.

\begin{Theorem}\label{theorem:emtime}
Suppose that the average length of $m_j$ is $m$ and the average $|\Omega_j|$ is $v$. Then, the time complexity of Algorithm~\ref{alg:EM} is 
\begin{align}\label{eq:EMtimecomp}
O\big(Nkm v^{k}+tN v^{2k}\big).
\end{align}
\end{Theorem}

\begin{Proof}
EM-based estimation will scan all $N$ users' bit strings with the length of $km$ one by one to compute the conditional probability for $v^k$ different combinations, the time complexity basically can be estimated as $O(N (km) (v^k))$.
Also, in the $t$th iteration, computing the posterior probability of each combination when observing each bit string will incur the time complexity of $O(tN (v^k)^2)$.
As a consequence, the overall time complexity is $O\big(tN v^{2k}+Nkm v^{k}\big)$.
\end{Proof}

\begin{Theorem}\label{theorem:emspace}
The space complexity of Algorithm~\ref{alg:EM} is 
\begin{align}\label{spacecomp:EM}
O\big(N k m+2N v^k\big).
\end{align}
\end{Theorem}

\begin{Proof}
In Algorithm~\ref{alg:EM}, the necessary storage includes $N$ users' bit strings with the length of $k m$, so it is $O(N k m)$. 
The prior probabilities on $k$ dimensions is $O(v^k)$. 
The conditional probabilities and posterior probabilities on $v^k$ candidates for all bit strings is $O(2Nv^k)$. 
So, the overall complexity is $O\big(N k m+2N v^k+ v^k \big)=O\big(N k m+2N v^k\big)$ since $N$ is the dominant variable.
\end{Proof}

According to Theorem~\ref{theorem:emtime}, the space overhead could be daunting when either $N$ or $k$ is large. This makes the performance of EM-based $k$-dimensional distribution estimation degrade dramatically and not applicable to high dimensional data.

\subsection{Lasso-based Multivariate Distribution Estimation}\label{sec:Lasso}
To improve the efficiency of the $k$-dimensional joint distribution estimation, we present a Lasso regression-based algorithm here. As mentioned in Section~\ref{sec:basic}, the bit strings are the representative features of the original report. After randomized responses and flipping, a large number of samples with various levels of noise will be generated by individual users. So, one may consider that the central server receives a large number of samples from specific distribution, however, with random noise. In this sense, one may estimate the distribution from the noised sample space by taking advantage of linear regression $\vec{\mathbf{y}}=\mathbf{M} \beta$, where $\mathbf{M}$ is predictor variables and $\vec{\mathbf{y}}$ is response variable, and $\beta$ is the regression coefficient vector. The determinism of Bloom filter can guarantee that the features (predictor variables $\mathbf{M}$) re-extracted at the server side are the same as the user side. Moreover, response variable $\vec{\mathbf{y}}$ can be estimated from the randomized bit strings according to the statistic characters of known $f$. Therefore, the only problem is to find a good solution to the linear regression $\vec{\mathbf{y}}=\mathbf{M} \beta$. Obviously, $k$-dimensional data may incur a output domain $\Omega_1 \times ... \times \Omega_k$ with the size of $|\Omega_1|\times...\times|\Omega_k|$, which increases exponentially with $k$. With fixed $N$ entries in the dataset $X$, the frequencies of many combination $\omega_1 \omega_2 ...\omega_k \in \Omega_1 \times ... \times \Omega_k$ are rather small or even zero. So, $\mathbf{M}$ is actually sparse and only part of the sparse but effective predictor variables need to be chosen. Otherwise, the general linear regression techniques will lead to overfitting problem. Fortunately, Lasso regression~\cite{tibshirani1996regression} is effective to solve the sparse linear regression by choosing predictor variables.

\begin{algorithm}[htb]\footnotesize
\caption{Lasso-based $k$-dimensional Joint Distribution}
\label{alg:lasso}
\begin{algorithmic}[1]
\REQUIRE ~~
\begin{tabular}[t]{p{0mm}l}
 $\mathcal{C}$&: attribute indexes cluster i.e., $\{1,2,...,k\}$,\\
 $A_{j}$&: $k$-dimensional attributes $(1 \leq j \leq k)$,\\
 $\Omega_j$&: domain of $A_j$ $(1 \leq j \leq k)$,\\
 $\hat{s}_j^i$&: observed Bloom filters $(1 \leq i \leq N)$ $(1 \leq j \leq k)$,\\
 $f$&: flipping probability.
\end{tabular}
\ENSURE ~~

$P(A_{\mathcal{C}})$: joint distribution of $k$ attributes specified by $\mathcal{C}$ .
\FOR {each $j \in \mathcal{C}$}
\FOR {each $b=1,2,\dots,m_j$}
\STATE compute $\hat{y}_j [b]= \sum \nolimits_{i=1}^N \hat{s}_j^i [b]$
\STATE compute $y_{j} [b]=(\hat{y}_j [b]-fN/2)/(1-f)$
\ENDFOR
\STATE set $\mathcal{H}_j(\Omega_j)=\{\mathcal{H}_j(\omega) ~\vline ~\forall \omega\in \Omega_j \}$
\ENDFOR
\STATE set $\vec{\mathbf{y}}=\big[y_{1} [1],\dots,y_{1} [m_1]~\vline~ y_2 [1],\dots,y_2 [m_2]~\vline~\dots~\vline~ y_k [1],\dots,y_k [m_k]\big]$
\STATE set $\mathbf{M}=\big[ \mathcal{H}_1(\Omega_1)\times \mathcal{H}_2(\Omega_2)\times \dots \times \mathcal{H}_k(\Omega_k)\big]$
\STATE compute $\vec{\bm{\beta}}=\textsf{Lasso\_regression}(\mathbf{M},\vec{\mathbf{y}})$
\RETURN $P(A_{\mathcal{C}})=\vec{\bm{\beta}}/N$
\end{algorithmic}
\end{algorithm}

Our Lasso-based estimation is described in Algorithm \ref{alg:lasso} and consists of the following four major steps.

\begin{figure*}[t]
\centering\epsfig{file=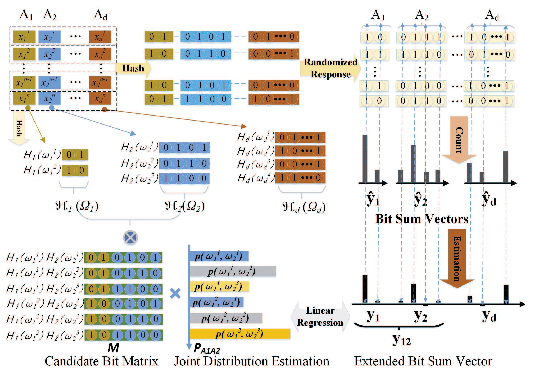, width=0.8\textwidth}
\caption{Illustration of Lasso-based Multivariate Joint Distribution Estimation \label{lasso}}
\end{figure*}

\begin{enumerate}
  \item After receiving all randomized Bloom filters from $N$ nodes, for each bit $b$ in each attribute $j$, the central server counts the number of $1's$ as $\hat{y}_j [b]= \sum \nolimits_{i=1}^N \hat{s}_j^i [b]$.

\begin{example}
In Table~\ref{transform}, each bit is counted to obtain the sum. With current $4$ records, the count vector is $(2,0 \ \vline \ 1,4,3,1 )$.
\end{example}

\item The true count sum of each bit $y_{j} [b]$ can be estimated as $y_{j} [b]=(\hat{y}_j [b]-fN/2)/(1-f)$ according to the randomized response applied to the true count. These count sums of all bits form a vector $\vec{\mathbf{y}}$ with the length of $\sum\nolimits_{j=1}^k m_j$.

\begin{example}
With the count vector $(2,0 \ \vline \ 1,4,3,1 )$, the true counts on each bit can be estimated as $(2-0.5*4/2)/(0.5)=2$, $(0-0.5*4/2)/0.5=-2$, $(1-0.5*4/2)/0.5=0$, $(4-0.5*4/2)/0.5=6$, and $(3-0.5*4/2)/0.5=4$. Therefore, the true count vector is $(2,-2 \ \vline \ 0,6,4,0)$.
\end{example}

\item To construct the features of the overall candidate set of attribute $\omega_1\dots\omega_k$, the Bloom filters on each dimension $\Omega_j$ is re-implemented by the server with the same hash functions $\mathcal{H}_j ()$ on the user end. Suppose all distinct Bloom filters on $\Omega_j$ are $\mathcal{H}_j(\Omega_j)=\{\mathcal{H}_j(\omega) ~\vline ~\forall \omega\in \Omega_j \}$, where they are orthogonal with each other. The candidate set of Bloom filters is then $\mathbf{M}=\big[ \mathcal{H}_1(\Omega_1)\times \mathcal{H}_2(\Omega_2)\times \dots \times \mathcal{H}_k(\Omega_k)\big]$ and the members in $\mathbf{M}$ are still mutual orthogonal. 

\begin{example}
By Cartesian product of multiple Bloom filters of different individual attributes, the Bloom filters of all candidate attribute combinations can be reconstructed. For example, the case that ``a male alumni with master degree and high-middle income'' can be represented by the concatenated Bloom filter ${\cal{H}}_2 (\text{M}){\cal{H}}_3 (\text{master})=010110 $. Similarly, all $|\Omega_2|\times |\Omega_3|=2*3=6$ candidate combinations can reconstruct their corresponding Bloom filters as $010101$, $010110$, $011100$, $100101$, $100110$, $101100$.
Therefore, the candidate matrix $\mathbf{M}$ can be represented as
\begin{align}
\mathbf{M}=\left[\begin{matrix}
   0&1&0&1&0&1 \\0&1&0&1&1&0\\0&1&1&1&0&0\\1&0&0&1&0&1\\1&0&0&1&1&0\\1&0&1&1&0&0
  \end{matrix}\right]
\end{align}
Similar demonstration is also shown in Figure~\ref{lasso}.
\end{example}

\item Fit a Lasso regression model to the counter vector $\vec{\mathbf{y}}$ and the candidate matrix $\mathbf{M}$, and then choose the non-zero coefficients as the corresponding frequencies of each candidate string. By reshaping the coefficient vector into a $k$-dimensional matrix by natural order and dividing with $N$, we can get the $k$-dimensional joint distribution estimation $P(A_{1} A_{2}\dots A_{k})$. For example, in Figure.~\ref{lasso}, we fit a linear regression to $\mathbf{y}_{12}$ and the candidate matrix $\mathbf{M}$ to estimate the joint distribution $P_{A_1 A_2}$.

\begin{example}
Suppose 
$\vec{\mathbf{\beta}}=[p_{11},p_{12},p_{13},p_{21},p_{22},p_{23}]$ represents the frequency of $6$ combinations of $A_2$ and $A_3$. Then, without noises caused by random flipping, there should be the only $\vec{\mathbf{\beta}}$ satisfying the linear equations
\begin{align}
\vec{\mathbf{\beta}} \cdot \mathbf{M}=\vec{\mathbf{y}},
\end{align}
However, since $\vec{\mathbf{y}}$ is an estimated vector with noises, $\vec{\mathbf{\beta}}$ cannot be directly solved. Therefore, linear regression techniques can be used to capture the best $\vec{\mathbf{\beta}}$, which includes the frequency entries for possible combinations of $A_2$ and $A_3$. So $\vec{\mathbf{\beta}}/N$ is the ratio of frequencies and the estimated probability of $P(A_2 A_3)$.
\end{example}
\end{enumerate}

Generally, the regression operation, the core of the estimation, will lose accuracy only when there are many collisions between Bloom filter strings. However, as mentioned in Section~\ref{sec:basic}, if there is no collision in the bit strings of each single dimension, then there is no collision in conjuncted bit strings of different dimensions. In fact, the probability of collision in conjuncted bit strings will not increase with dimensions. For example, suppose the collision rate of Bloom filter in one dimension is $p$, then the collision rate will decrease to $p^k$ when we connect bit strings of $k$ dimensions together. Therefore, we only need to choose proper $m$ and $h$ according to Equation~(\ref{eq:m}) and (\ref{eq:h}) to lower the collision probability for each dimension and then we are guaranteed to have a proper estimation for multiple dimensions.

\textbf{Complexity:}
Compared with Algorithm \ref{alg:EM}, our Lasso-based estimation can effectively reduce the time and space complexity.

\begin{Theorem}\label{theorem:lassotime}
The time complexity of Algorithm~\ref{alg:lasso} is
\begin{align}\label{eq:lassotimecomp}
O\big(v^{3k}+kmv^{2k}+Nkm \big).
\end{align}
\end{Theorem}

\begin{Proof}
Algorithm \ref{alg:lasso} involves two parts: to compute the bit counter vector, $N$ bit strings with each length of $km$ will be summed up and this operation at most incurs the complexity of $O(Nkm)$; and Lasso regression with $v^k$ candidates (total domain size) and $km$ samples (the length of the bit counter vector is $km$) has the complexity of $O \big( (v^k)^3+(v^k)^2 (km)\big)$.
\end{Proof}

Based on the general assumption that $N$ dominates Equation~(\ref{eq:lassotimecomp}), then we can see the complexity in Equation~(\ref{eq:lassotimecomp}) is much less than Equation~(\ref{eq:EMtimecomp}) in Theorem \ref{theorem:emtime}.

\begin{Theorem}\label{theorem:lassospace}
The space complexity of Algorithm~\ref{alg:lasso} is
\begin{align}\label{spacecomp:Lasso}
O \big( Nk m+v^k km \big).
\end{align}
\end{Theorem}

\begin{Proof}
In Algorithm~\ref{alg:lasso}, the storage overhead consists of three parts: users' bit strings $O(Nkm)$, 
a count vector with size $O(km)$, 
and the candidate bit matrix $\mathbf{M}$ with size $O(km v^k)$. 
Therefore, the overall space complexity of our proposed Lasso based estimation algorithm is $O \big( Nk m+km+v^k km \big)=O\big( Nk m+v^k km \big)$, 
which is also smaller than Equation~(\ref{spacecomp:EM}) as $N$ is dominant.
\end{Proof}

The empirical results are shown in Section~\ref{sec:evaluation}. The efficiency comes from the fact that the $N$ bit strings of length $m$ will be scanned to count sum only once and then one-time Lasso regression is fitted to estimate the distribution. In addition, Lasso regression could extract the important (i.e., frequent) features with high probability, which fits well with the sparsity of high-dimensional data.


\subsubsection{Hybrid Algorithm}
Recall that, with sufficient samples, EM-based estimation could have good convergence but also high complexity. Instead, Lasso-based estimation can be very efficient with some estimation deviation compared with EM-based algorithm. The high complexity of EM algorithm stems from two parts: Firstly, it iteratively scans user's reports and builds a prior likely distribution table, which has the size of $N \cdot \prod |\Omega_j|$. And for each record of table, the computation has to compare $\sum m_j$ bits. However, when the dimension is high, the combination of $\Omega_j$ will be very sparse and has lots of zero items.
Secondly, Without prior knowledge, the initial value of the random assignment (i.e., uniform distribution) will lead to too many iterations for final convergence to occur.

To achieve a balance between the EM-based estimation and Lasso-based estimation, we also propose a hybrid algorithm Lasso+EM in Algorithm~\ref{alg:ELM} that first eliminates the redundant candidates and estimates the initial value with Lasso based algorithm~\ref{alg:lasso} and then refines the convergence using EM-based algorithm~\ref{alg:EM}. The hybrid algorithm has two advantages:
\begin{enumerate}
  \item The sparse candidates will be selected out by the Lasso based estimation algorithm, as shown in Steps 1,2,7,8 of Algorithm~\ref{alg:ELM}. So the EM algorithm can just compute the conditional probability on these sparse candidates instead of all candidates, which can greatly reduce both time and space complexity.
  \item Lasso-based algorithm can give a good initial estimation of the joint distribution. Compared with using initial values with random assignments, using the initial value estimated with the Lasso-based algorithm can further boost the convergence of the EM algorithm, which is sensitive to the initial value especially when the candidate space is sparse.
\end{enumerate}

\begin{Theorem}
The time complexity of Algorithm~\ref{alg:ELM} is
\begin{align}\label{eq:hybridcomp}
O\big((v^{3k}+kmv^{2k}+Nkm)+(tN(v')^{2}+Nkm(v')) \big),
\end{align}
where $v'$ is the average size of sparse items in $\Omega_1 \times ... \times \Omega_k$, and $v'<v^k$.
\end{Theorem}

\begin{Proof}
See Theorem \ref{theorem:emtime} and Theorem \ref{theorem:lassotime}, the only difference is that after the Lasso based estimation, only sparse items in $\Omega_1 \times ... \times \Omega_k$ are selected.
\end{Proof}

\begin{Theorem}
The space complexity of Algorithm~\ref{alg:ELM} is
\begin{align}\label{eq:hybridspace}
O \big( Nk m+v^k km +2Nv' \big).
\end{align}
\end{Theorem}

\begin{Proof}
See Theorem \ref{theorem:emspace} and Theorem \ref{theorem:lassospace}.
\end{Proof}

\begin{algorithm}[htb]\footnotesize
\caption{Lasso+EM $k$-dimensional Joint Distribution (Lasso+EM\_JD)}
\label{alg:ELM}
\begin{algorithmic}[1]
\REQUIRE
\begin{tabular}[t]{p{0mm}l}
 $A_{j}$&: $k$-dimensional attributes $(1 \leq j \leq k)$,\\
 $\Omega_j$&: domain of $A_j$ $(1 \leq j \leq k)$,\\
 $\hat{s}_j^i$&: observed Bloom filters $(1 \leq i \leq N)$ $(1 \leq j \leq k)$,\\
 $f$&: flipping probability.
\end{tabular}
\ENSURE $P(A_{1} A_{2}\dots A_{k})$: $k$-dimensional joint distribution.

\STATE compute $P_0(\omega_{1} \omega_{2}\dots\omega_{k})=\textsf{Lasso\_JD}(A_j,\Omega_j,\{\hat{s}_j^i\}_{i=1}^N,f)$
\STATE set $\mathcal{C}'=\{x|x \in \mathcal{C}, P_0(x)=0 \}$.

\FOR {each $i=1,...,N$}
\FOR {each $j=1,...,k$}
\STATE compute $P(\hat{s}^i_j|\omega_j)=\prod \nolimits_{b=1}^{m_j} (\frac{f}{2})^{\hat{s}_j^i[b]} (1-\frac{f}{2})^{1-\hat{s}_j^i[b]}$.
\ENDFOR
\IF { $\omega_{1} \omega_{2} \dots \omega_{k} \in \mathcal{C}'$ }
\STATE $P(\hat{s}^i_{1} \hat{s}^i_{2} \dots \hat{s}^i_{k}|\omega_{1} \omega_{2} \dots \omega_{k})=0$
\ELSE
\STATE compute $ P(\hat{s}^i_{1} \hat{s}^i_{2} \dots \hat{s}^i_{k}|\omega_{1} \omega_{2} \dots \omega_{k})=\prod \nolimits_{j=1}^k P(\hat{s}^i_j|\omega_j)$.
\ENDIF
\ENDFOR
\STATE initialize $t=0$ ~~~~~~~~~~~~~~~~~~~~~~~~/* number of iterations */
\REPEAT
\STATE ... ...
\STATE /* (similar to Algorithm~\ref{alg:EM}) */
\STATE ... ...
\UNTIL {$P_{t}(\omega_{1} \omega_{2}\dots \omega_{k})$ converges.}
\RETURN $P(A_{1} A_{2}\dots A_{k})=P_{t}(\omega_{1} \omega_{2}\dots \omega_{k})$
\end{algorithmic}
\end{algorithm}
\vspace{-0.1cm}

\subsection{Dimension Reduction with Local Privacy}

\subsubsection{Dimension Reduction via $2$-dimensional Joint Distribution Estimation} \label{dimension_reduction}

The key to reducing dimensionality in high-dimensional dataset is to find the compact clusters, within which all attributes are tightly correlated to or dependent on each other. Inspired by \cite{zhang2014privbayes,chen2015differentially} but without extra privacy budget on dimension reduction, our dimension reduction based on locally once-for-all privacy-preserved data records consists of the following three steps:

\begin{algorithm}[htb]\footnotesize
\caption{Dimension reduction with local privacy}
\label{alg:dr}
\begin{algorithmic}[1]
\REQUIRE ~~
\begin{tabular}[t]{p{0mm}l}
 $A_{j}$&: $k$-dimensional attributes $(1 \leq j \leq k)$,\\
 $\Omega_j$&: domain of $A_j$ $(1 \leq j \leq k)$,\\
 $\hat{s}_j^i$&: observed Bloom filters $(1 \leq i \leq N)$ $(1 \leq j \leq k)$,\\
 $f$&: flipping probability,\\
 $\phi$&: dependency degree\\
\end{tabular}
\ENSURE ~~
$\mathcal{C}_1,\mathcal{C}_2,...,\mathcal{C}_l$: attribute indexes clusters
\STATE initialize $\mathbf{G}_{d \times d}=\mathbf{0}$.
\FOR {each $j=1,2,\dots ,d$}
\STATE estimate $P(A_j)$ by Lasso based Algorithm~\ref{alg:lasso}
\ENDFOR
\FOR {each attribute $m = 1,2,\dots ,d$}
\FOR {each attribute $n = m+1,m+2,\dots ,d$}
\STATE estimate $P(A_m A_n)$ by Lasso based Algorithm~\ref{alg:lasso}
\STATE compute $I_{m,n}=\sum_{i\in \Omega_m}\sum_{i\in \Omega_n}p_{ij}\ln{\frac{p_{ij}}{p_{i\cdot}p_{\cdot j}}}$
\STATE compute $\tau_{m,n}=\min (|\Omega_m|-1,|\Omega_n|-1)*\phi ^2/2$
\IF {$I(m,n)\geq \tau_{m n}$}
\STATE set $G_{m,n}=G_{n,m}=1,~~G \in \mathbf{G}_{d \times d}$
\ENDIF
\ENDFOR
\ENDFOR
\STATE build dependency graph with $\mathbf{G}_{d \times d}$
\STATE triangulate the dependency graph into a junction tree
\STATE split the junction tree into several cliques $\mathcal{C}_1,\mathcal{C}_2,...,\mathcal{C}_l$ with elimination algorithm.
\RETURN $\mathbb{C}=\{\mathcal{C}_1,\mathcal{C}_2,...,\mathcal{C}_l \}$
\end{algorithmic}
\end{algorithm}

\begin{enumerate}
  \item \textbf{Pairwise Correlation Computation.} We use mutual information to measure pairwise correlations between attributes. The mutual information is calculated as
\begin{equation}\label{eq:MI}
I_{m,n}=\sum_{i\in \Omega_m}\sum_{j\in \Omega_n}p_{ij}\ln{\frac{p_{ij}}{p_{i\cdot}p_{\cdot j}}}
\end{equation}
where, $\Omega_m$ and $\Omega_n$ are the domains of attributes $A_m$ and $A_n$, respectively. $p_{i\cdot}$ and $p_{\cdot j}$ represent the probability that $A_m$ is the $i$th value in $\Omega_m$ and the probability that $A_n$ is the $j$th value in $\Omega_n$, respectively. Then, $p_{ij}$ is their joint probability. Particulary, both $p_{i\cdot}$ and $p_{\cdot j}$ can be learned from the direct RAPPOR scheme with Lasso regression~\cite{Erlingsson-2014}. Their joint distribution $p_{ij}$ then can be efficiently obtained with our proposed multi-dimensional marginal estimation algorithm in Section \ref{sec:Lasso}.

\begin{example}
  A $4\times 4$ correlation matrix $\mathbf{I}$ can be built for the example dataset of Table~\ref{origin} to record the mutual information measure.
  \begin{align}
  \begin{matrix}
  \mathbf{I}=\left[\begin{matrix}
   1.00&0.05&0.03&0.16 \\0.05&1.00&0.02&0.04\\0.03&0.02&1.00&0.15\\0.16&0.04&0.15&1.00
  \end{matrix}\right]
  \end{matrix}
  \end{align}
  where $\mathbf{I}_{m,n}$ is the mutual information between $A_m$ and $A_n$, e.g., $\mathbf{I}_{2,3}=0.02$ means the mutual information between $A_2$ and $A_3$ is $0.02$.
  \end{example}


  It should be noted that the correlations need to be learnt between all attributes pairs in heterogeneous multi-attribute data. That is to say the basic complexity is $O(d^2)$, where $d$ is the number. And in each learning process, the complexity is decided by the distribution estimation algorithm. So, the overall complexity is quite high if the distribution estimation has large complexity. Therefore, the joint distribution estimation algorithm must be light and efficient. In addition, to further overcome the high complexity of pairwise correlation learning when $d$ is large, we also proposed a heuristic pruning scheme, which can be referred to Section~\ref{sec:pruning}

  \item \textbf{Dependency Graph Construction.} Based on mutual information, the dependency graph between attributes can be constructed as follows. First, an adjacent matrix $\mathbf{G}_{d \times d}$ is initialized with all $0$. Then, all the attribute pairs $(A_m,A_n)$ are chosen to compare their mutual information with an threshold $\tau_{m,n}$, which is defined as
      \begin{align}
      \tau_{m,n}=\min (|\Omega_m|-1,|\Omega_n|-1)*\phi ^2/2
      \end{align}
      and $\phi~(0 \leq \phi \leq 1)$ is a flexible parameter determining the desired correlation level. $G_{m,n}$ and $G_{n,m}$ are both set to be $1$ if and only if $I_{m,n}>\tau_{m,n}$.

  \begin{example}
  By comparing the correlation matrix $I$ with the dependency threshold $\tau_{m,n}$, a dependency graph represented by the adjacent matrix $\mathbf{G}$ can be built.
    \begin{align}
  \begin{matrix}
  \mathbf{G}_{4\times 4}=\left[\begin{matrix}
   1&0&0&1 \\0&1&0&0\\0&0&1&1\\1&0&1&1
  \end{matrix}\right]
  \end{matrix}\label{dependencymatrix}
  \end{align}
  \end{example}

\item \textbf{Compact Clusters Building.} By triangulation, the dependency graph $\mathbf{G}_{d \times d}$ can be transformed to a junction tree, in which each node represents an attribute $A_j$. Then, based on the junction tree algorithm, several clusters $\mathcal{C}_1, \mathcal{C}_2,\dots , \mathcal{C}_l$ can be obtained as the compact clusters of attributes, in which attributes are mutually correlated. Hence, the whole attributes set can be divided into several compact attribute clusters and the number of dimensions can be effectively reduced.

\begin{example}
  By triangulation, the dependency graph can be transformed to a junction tree, in which nodes are split into different clusters. Figure~\ref{jtree} demonstrates an example of the junction tree built from the dependency graph. The attribute set $\{A_1,A_2,A_3,A_4,A_5,A_6\}$ is split into $4$ clusters $A_1 A_2$, $A_2 A_3 A_5$, $A_2 A_4 A_5$ and $A_6$, where $A_2$ and $A_2 A_5$ is the separators and clusters $A_1 A_2$, $A_2 A_3 A_5$, and $A_2 A_4 A_5$ are also called cliques.
\begin{figure}[t]
\centering\epsfig{file=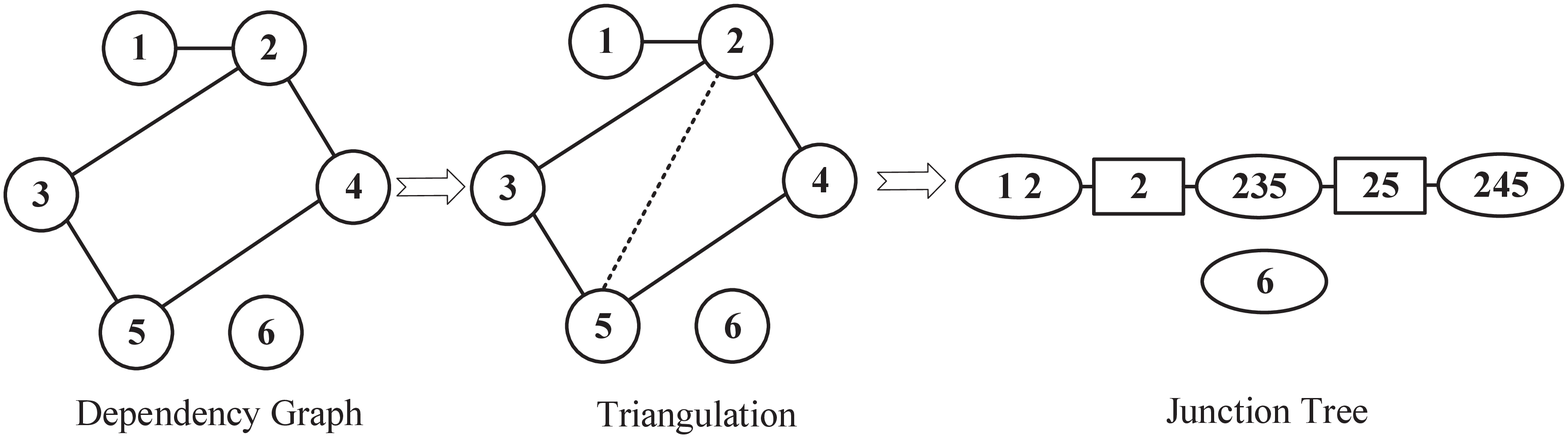, width=0.5\textwidth}
\caption{Build junction tree from dependency graph \label{jtree}}
\end{figure}
\end{example}
\end{enumerate}

\textbf{Complexity:}

\begin{Theorem}
The time complexity of Algorithm~\ref{alg:dr} is
\begin{align}
O(d^2(v^{6}+2mv^{4}+2Nm+tN(v')^{2}+2Nm(v')) ).
\end{align}
\end{Theorem}

\begin{Proof}
The core of the dimension reduction process is the $d \choose 2$ times of $2$-dimensional joint distribution estimation. The complexity of each $2$-dimensional joint distribution estimation can be derived from Equation~(\ref{eq:hybridcomp}) when adopting the hybrid algorithm (Algorithm~\ref{alg:ELM}). The complexity of building junction tree on $d\times d$ dependency graph is negligible when compared with the joint distribution estimation.
\end{Proof}

\begin{Theorem}
The space complexity of Algorithm~\ref{alg:dr} is
\begin{align}
O( 2Nm+2v^2m+2Nv').
\end{align}
\end{Theorem}

\begin{Proof}
When we compute the mutual correlations between any pairs, a $2$-dimensional joint distribution estimation algorithm will be triggered with the space complexity of $O(2Nm+2mv^2+2Nv')$, since $k=2$ is substituted into Equation~(\ref{eq:hybridspace}). This maximum complexity dominates Algorithm~\ref{alg:dr}. The space complexity of building junction tree on $d\times d$ dependency graph is negligible when compared with the joint distribution estimation.
\end{Proof}
\subsubsection{Entropy based Pruning Scheme}\label{sec:pruning}
In existing work~\cite{kellaris2013practical,rescuedp2016} on homogeneous data, the correlations can be simply captured by distance metrics. 
However, in our work, mutual information is used to measure general correlations since heterogenous attributes (a.k.a., attributes with different domains) are also considered.

As shown in Equation~(\ref{eq:MI}), to calculate the mutual information of variables $X$ and $Y$, the joint probability on the joint combination is inevitable, thus making the pairwise computation of dependency necessary. Although mutual information is already simple than Kendall rank coefficients in the similar work~\cite{li2014differentially}, here, we still propose a pruning-based heuristic to boost this pairwise correlation learning process. 

Intuitively, there are different situations in Algorithm~\ref{alg:dr}: 1. When $\phi=0$ or $\phi=1$, all attributes will be considered mutually correlated or independent. Thus, there is no need to compute pairwise correlation. 2. With the increase of $\phi$ ($0< \phi < 1$), less dependencies will be included in the adjacent matrix $\mathbf{G}_{d \times d}$ of dependency graph, which will become sparser. This also means that we may selectively neglect some pairs. 
Inspired by the relationship between mutual information and information entropy\footnote{The relationship between mutual information and information entropy can be represented as $I(X;Y)=H(X)+H(Y)-H(X,Y)$, where $H(X)$ and $H(X,Y)$ denote the information entropy of variable $X$ and their joint entropy of $X$ and $Y$, respectively.}, we first heuristically filter out some portion of attributes $A_x$ with least relative information entropy $RH(A_x)=H(A_x)/|\Omega_x|$, and then verify the mutual information among the remaining attributes, thus reducing the pairwise computations.

Furthermore, the adjacent matrix $\mathbf{G}_{d \times d}$ of dependency graph varies in different datasets. For example, the adjacent matrix $\mathbf{G}_{d \times d}$ is rarely sparse in binary datasets but very sparse in non-binary datasets. Based on this observation, we can further simplify the calculation by finding the independency in binary datasets or finding the dependency in non-binary datasets. For example, we first set all entries of $\mathbf{G}_{d \times d}$ for a binary datasets as $1$'s and start from the attributes with least relative information entropy $RH(A_x)=H(A_x)/|\Omega_x|$ to find the uncorrelated attributes. While for non-binary datasets, we first set $\mathbf{G}_{d \times d}$ as $0$'s and then start from the attributes with largest average entropy to find the correlated attributes.

\begin{algorithm}[htb]\footnotesize
\caption{Entropy based Pruning Scheme}
\label{alg:pruning}
\begin{algorithmic}[1]
\REQUIRE ~~
\begin{tabular}[t]{p{0mm}l}
 $A_{j}$&: $k$-dimensional attributes $(1 \leq j \leq k)$,\\
 $\Omega_j$&: domain of $A_j$ $(1 \leq j \leq k)$,\\
 $\hat{s}_j^i$&: observed Bloom filters $(1 \leq i \leq N)$ $(1 \leq j \leq k)$,\\
 $f$&: flipping probability,\\
 $\phi$&: dependency degree\\
\end{tabular}

\ENSURE ~~
$\mathbf{G}_{d \times d}$: adjacent matrix $\mathbf{G}_{d \times d}$ of dependency graph of attributes $A_j~(j=1,2,...,d)$

\STATE{initialize $\mathbf{G}_{d \times d}=0$}
\FOR {each $j=1,2,\dots,k$}
\STATE compute $P(A_j)=\textsf{JD}(A_j,\Omega_j,\{\hat{s}_j^i\}_{i=1}^N,f)$
\STATE compute $RH(A_j)=-\frac{1}{|\Omega_j|}\sum\limits_{p \in P(A_j)} p \log p$
\ENDFOR
\STATE sort $list_A=\{A_1, A_2,...,A_j\}$ according to entropy $H(A_j)$
\STATE pick up the previous $\lfloor length(list_A)*(1-\phi)\rfloor$ items from $list_A$ as a new list $list_{A'}$
\STATE ...
\STATE compute pairwise mutual information among $list_{A'}$ and set dependency graph $\mathbf{G}_{d \times d}$ as in Algorithm~\ref{alg:dr}.

\RETURN $\mathbf{G}_{d \times d}$
\end{algorithmic}
\end{algorithm}

\subsection{Synthesizing New Dataset}
For brevity, we first define $A_C=\{A_j|j \in C\}$ and $\hat{X}_C=\{x_j|j \in C\}$. Then the process of synthesizing the new dataset via sampling is shown in the following Algorithm~\ref{alg:NDS}.

\begin{algorithm}[htb]\footnotesize
\caption{New Dataset Synthesizing}
\label{alg:NDS}
\begin{algorithmic}[1]
\REQUIRE ~~
\begin{tabular}[t]{p{0mm}l}
 $\mathbb{C}$&: a collection of attribute index clusters $\mathcal{C}_1,...\mathcal{C}_l$,\\
 $A_{j}$&: $k$-dimensional attributes $(1 \leq j \leq k)$,\\
 $\Omega_j$&: domain of $A_j$ $(1 \leq j \leq k)$,\\
 $\hat{s}_j^i$&: observed Bloom filters $(1 \leq i \leq N)$ $(1 \leq j \leq k)$,\\
 $f$&: flipping probability,\\
\end{tabular}

\ENSURE ~~
$\hat{X}$: Synthetic Dataset of $X$
\STATE initialize $R=\varnothing$
\REPEAT
\STATE randomly choose an attribute index cluster $C \in \mathbb{C}$
\STATE estimate joint distribution $P(A_{C})$ by $\textsf{JD}$
\STATE sample $\hat{X}_{C}$ according to $P(A_C)$
\STATE $\mathbb{C}=\mathbb{C}-C$, $R=R\cup C$, $\mathbb{D}=\{D\in \mathbb{C}|D \cap R \neq \varnothing \}$
\FOR {each $D \in \mathbb{D}$}
\STATE estimate joint distribution $P(A_D)$ by $\textsf{JD}$
\STATE obtain conditional distribution $P(A_{D-R}|A_{D\cap R})$ from $P(A_D)$
\STATE sample $\hat{X}_{D-R}$ according to $P(A_{D-R}|A_{D\cap R})$ and $\hat{X}_{D \cap R}$
\STATE $\mathbb{C}=\mathbb{C}-D$, $R=R\cup D$, $\mathbb{D}=\{D\in \mathbb{C}|D \cap R \neq \varnothing \}$
\ENDFOR
\UNTIL {$\mathbb{C} = \varnothing$}
\RETURN $\hat{X} $
\end{algorithmic}
\end{algorithm}

We first initialize a set $R$ to keep the sampled attribute indexes. Then, we randomly choose an attribute index cluster $C$ to estimate the joint distribution and sample new data $\hat{X}$ in the attributes $A_{j}, \forall j \in C$. Next, we remove $C$ from the cluster collection $\mathbb{C}$ into $R$, and find the connected component $\mathbb{D}$ of $C$. In the connected component, each cluster $D$ is traversed and sampled as follows. first estimate the joint distribution on the attributes $A_D$ by our proposed distribution estimations and obtain the conditional distribution $P(A_{D-R}|A_{D\cap R})$. Then, sample $\hat{X}_{D-R}$ according to this conditional distribution and the sampled data $\hat{X}_{D \cap R}$. After the traverse of $\mathbb{D}$, the attributes in the first connected components are sampled. Then randomly choose cluster in the remaining $\mathbb{C}$ to sample the attributes in the second connected components, until all clusters are sampled. Finally, a new synthetic dataset $\hat{X}$ is generated according to the estimated correlations and distributions in origin dataset $X$.

\begin{Theorem}
The time complexity of Algorithm~\ref{alg:NDS} is
\begin{align}
O( l(v^{3k}+kmv^{2k}+Nkm+tN(v')^{2}+Nkm(v'))  ),
\end{align}
where $l$ is the number of clusters after dimension reduction and $k$ here refers to average number of dimensions in these clusters.
\end{Theorem}

\begin{Proof}
The core of the dataset synthesizing is actually multiple ($l$ times) $k$-dimensional joint distribution estimation.
\end{Proof}

\begin{Theorem}
The space complexity of Algorithm~\ref{alg:NDS} is
\begin{align}
O( Nkm+v^k km+2Nv'+Nd).
\end{align}
\end{Theorem}

\begin{Proof}
Every time, a $k$-dimensional joint distribution estimation algorithm (with space complexity of $O(Nk m+v^k km +2Nv')$) is processed to draw a new dataset. A new dataset with the size $O(Nd)$ is maintained while synthesizing.
\end{Proof}

\begin{figure*}[t]
\centering\epsfig{file=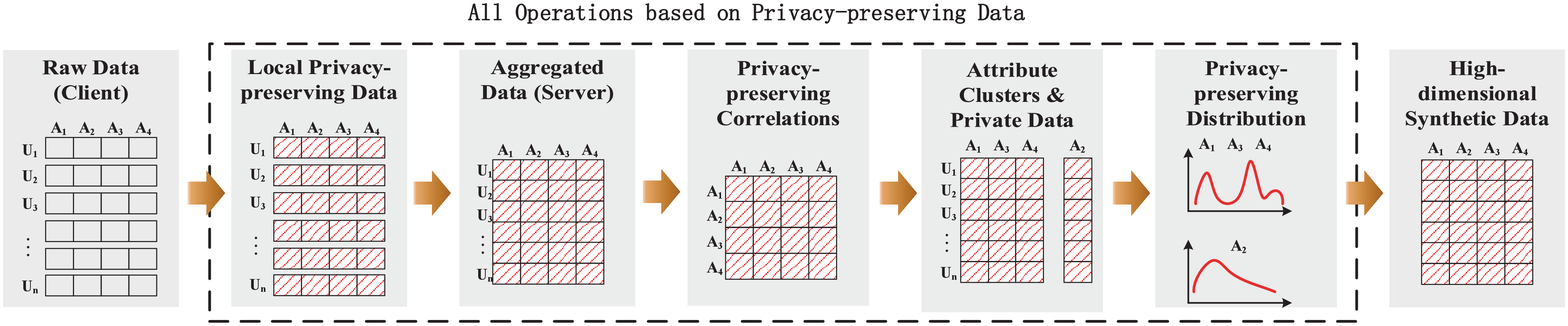, width=\textwidth}
\caption{Main procedures of high-dimensional data publishing with local privacy \label{MP2}}
\end{figure*}

The overall process of LoPub can also be summarized as in Figure~\ref{MP2}. Clearly, all the processed are conducted on the locally privacy-preserved data. Therefore, local privacy is guaranteed on all the crowdsourced users.

\section{Evaluation}\label{sec:evaluation}
In this section, we conducted extensive experiments on real datasets to demonstrate the efficiency of our algorithms in terms of computation time and accuracy.

We used three real-world datasets: \textsf{Retail} \cite{retail}, \textsf{Adult} \cite{adult}, and \textsf{TPC-E} \cite{tpc}.
\textsf{Retail} is part of a retail market basket dataset. Each record contains distinct items purchased in a shopping visit. \textsf{Adult} is extracted from the 1994 US Census. This dataset contains personal information, such as gender, salary, and education level. \textsf{TPC-E} contains trade records of ``Trade type'', ``Security'', ``Security status'' tables in the TPC-E benchmark.
It should be noted that some continuous domain were binned in the pre-process for simplicity.
\begin{center}
\scriptsize
\begin{tabular}{|c|c|r|c|c|}
  \hline
 \textbf{Datasets} & \textbf{Type} & \textbf{\#. Records} ($N$) & \textbf{\#. Attributes} ($d$) & \textbf{Domain Size}\\
  \hline
  \textsf{Retail} & Binary & 27,522 & 16 & $2^{16}$ \\
  \textsf{Adult} & Integer & 45,222 & 15 & $2^{52}$ \\
  \textsf{TPC-E} & Mixed & 40,000 & 24 & $2^{77}$ \\
  \hline
\end{tabular}
\end{center}


All the experiments were run on a machine with Intel Core i5-5200U CPU 2.20GHz and 8GB RAM, using Windows 7. We simulated the crowdsourced environment as follows. First, users read each data record individually and locally transform it into privacy-preserving bit strings. Then, the crowdsourced bit strings are gathered by the central server for synthesizing and publishing the high-dimensional dataset.

LoPub can be realized by combining distribution estimations and data synthesizing techniques. Thus, we implemented different LoPub realizations using Python 2.7 with the following three strategies.
\begin{enumerate}
  \item \textsf{EM\_JD}, the generalized EM-based multivariate joint distribution estimation algorithm.
  \item \textsf{Lasso\_JD}, our proposed Lasso-based multivariate joint distribution estimation algorithm.
  \item \textsf{Lasso+EM\_JD}, our proposed hybrid estimation algorithm that uses the \textsf{Lasso\_JD} to filter out some candidates to reduce the complexity and replace the initial value to boost the convergence of \textsf{EM\_JD}.
\end{enumerate}
It is worth mentioning that we compared only the above algorithms since our algorithm adopts a novel local privacy paradigm on high-dimensional data. Other competitors are either for non-local privacy or on low-dimension data.

For fair comparison, we randomly chose 100 combinations of $k$ attributes from $d$ dimensional data. For simplicity, we sampled\footnote{It should be noted that, with sampled data, the differential privacy level can be further enhanced~\cite{li2012sampling}. But sampling used here is for simplicity.} $50\%$ data from dataset \textsf{Retail} and $10\%$ data from datasets \textsf{Adult} and \textsf{TPC-E}, respectively. The efficiency of our algorithms is measured by \emph{computation time} and \emph{accuracy}. The computation time includes CPU time and IO cost. Each set of experiments is run 100 times, and the average running time is reported. To measure accuracy, we used the distance metrics AVD (average variant distance) on the three datasets, as suggested in~\cite{chen2015differentially}, to quantify the closeness between the estimated joint distribution $P(\omega)$ and the origin joint distribution $Q(\omega)$.
The AVD error is defined as
\begin{align}
Dist_{\text{AVD}} (P,Q)=\frac{1}{2} \sum \limits_{\omega\in \Omega} |P(\omega)-Q(\omega)|.
\end{align}

The default parameters are described as follows. In the binary dataset \textsf{Retail}, the maximum number of bits and the number of hash functions used in the bloom filter are $m=32$ and $h=4$, respectively. In the non-binary datasets \textsf{Adult} and \textsf{TPC-E}, the maximum number of bits and the number of hash functions used in bloom filter are $m=128$ and $h=4$, respectively. The convergence gap is set as $0.001$ for fast convergence.

\subsection{Multivariate Distribution Estimation}
Here, we show the performance of our proposed distribution estimations in terms of both efficiency and effectiveness. The efficiency is measured by computation time, and the effectiveness is measured by estimation accuracy.

\subsubsection{Computation Time}
We first evaluate the computation time of \textsf{EM\_JD}, \textsf{Lasso\_JD}, and \textsf{Lasso+EM\_JD} for the $k$-dimensional joint distribution estimation on three real datasets.

\begin{figure*}[htbp]
\begin{minipage}[t]{4.2cm}
\centering \epsfig{file=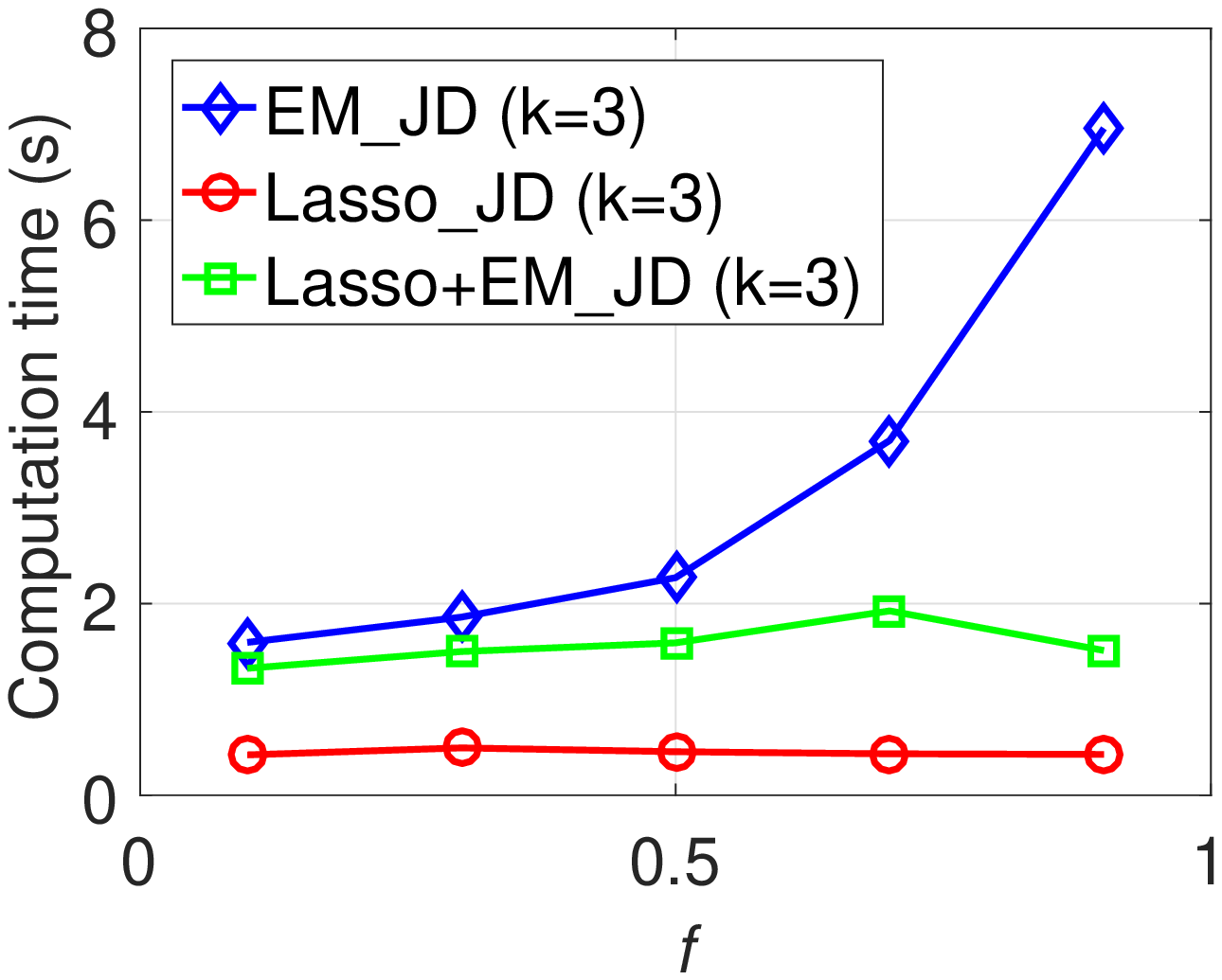, width=123pt}
\caption{Estimation Time (\textsf{Retail}, $k=3$)\label{timer3}}
\end{minipage}
\ \
\begin{minipage}[t]{4.2cm}
\centering \epsfig{file=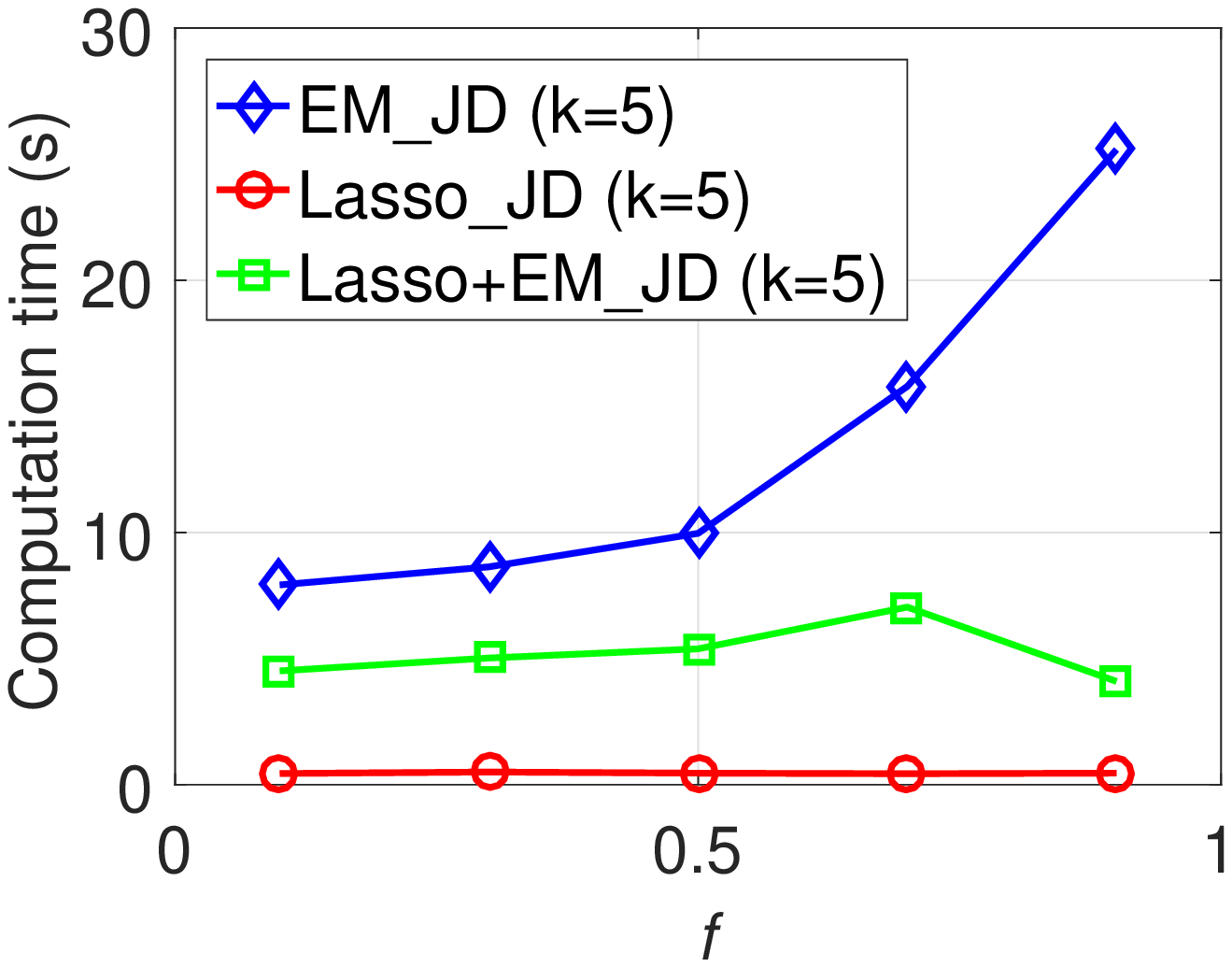, width=125pt}
\centering\caption{Estimation Time (\textsf{Retail}, $k=5$)\label{timer5}}
\end{minipage}
\ \
\begin{minipage}[t]{4.2cm}
\centering \epsfig{file=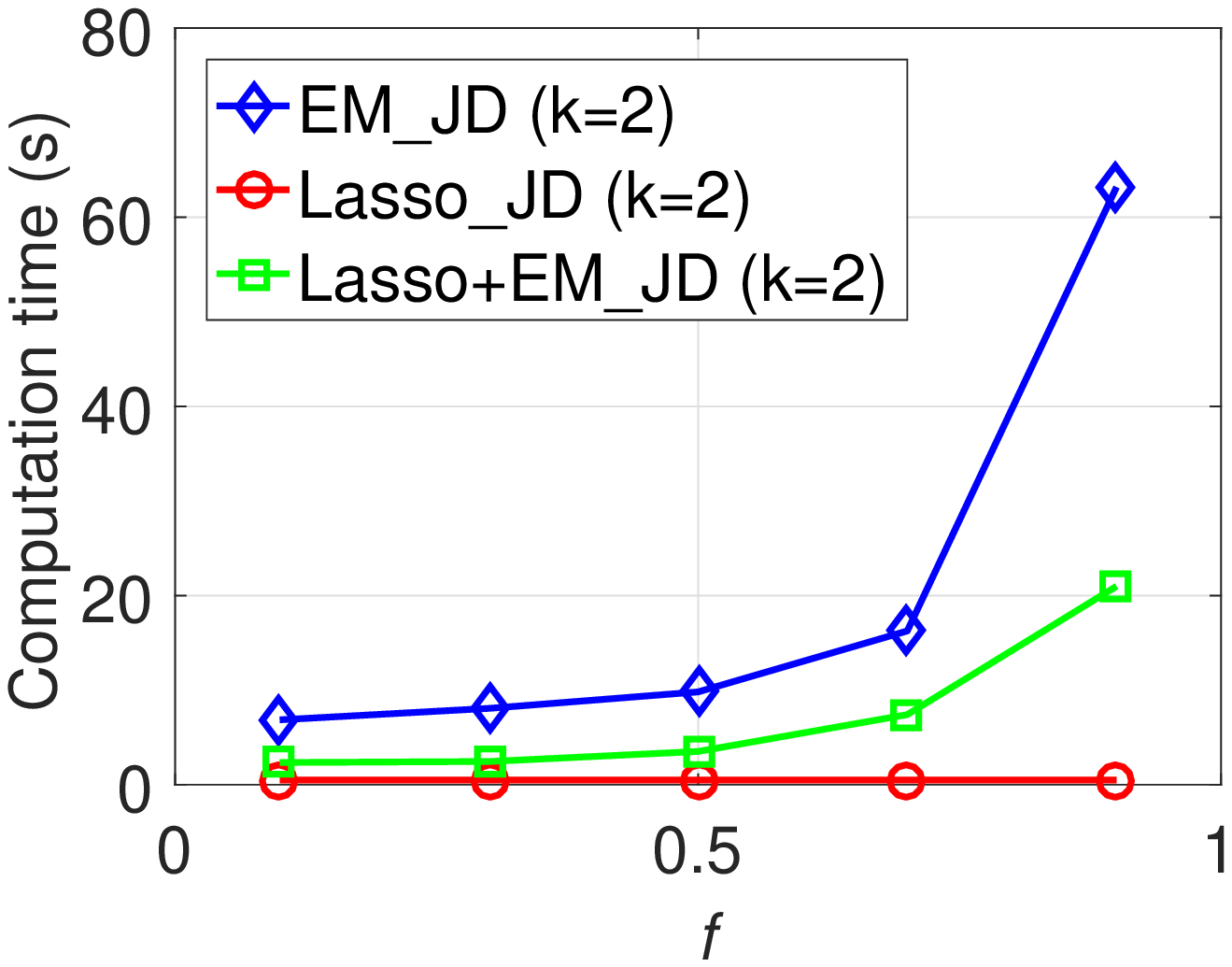, width=125pt}
\caption{Estimation Time (\textsf{Adult}, $k=2$)\label{timea2}}
\end{minipage}
\ \
\begin{minipage}[t]{4.2cm}
\centering \epsfig{file=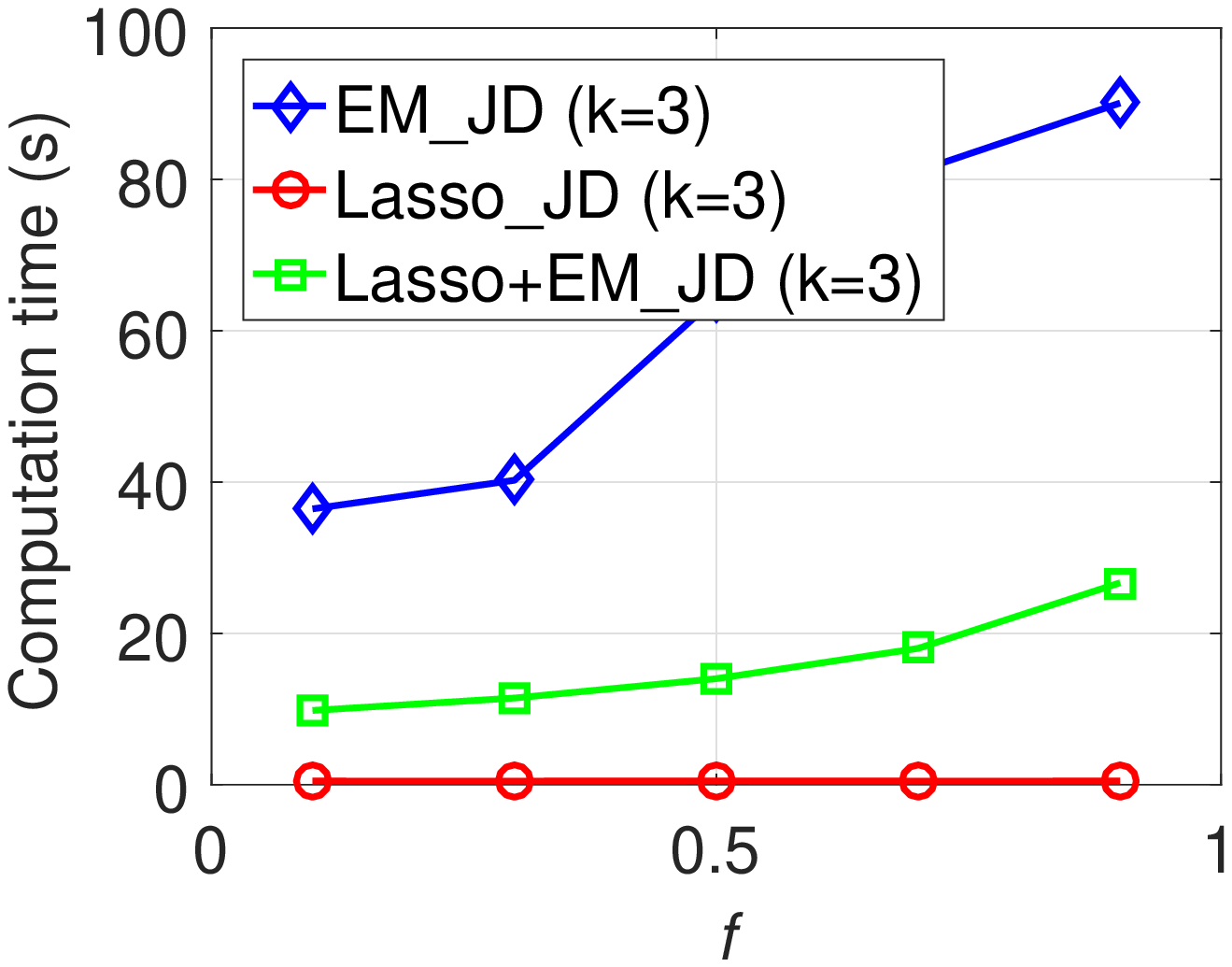, width=125pt}
\centering\caption{Estimation Time (\textsf{Adult}, $k=3$)\label{timea3}}
\end{minipage}
\vspace{-0.3cm}
\end{figure*}

Figures~\ref{timer3} and \ref{timer5} compare the computation time on the binary dataset \textsf{Retail} with both $k=3$ and $k=5$. It can be noticed that, for each dimension $k$, \textsf{Lasso\_JD} is consistently much faster than \textsf{EM\_JD} and \textsf{Lasso+EM\_JD}, especially when $k$ is large. This is because \textsf{EM\_JD} has to repeatedly scan each user's bit string. Particularly, the time consumption of \textsf{EM\_JD} increases with $f$ because there will be more iterations for the fixed convergence gap. In contrast, \textsf{Lasso\_JD} uses the regression to estimate the joint distribution more efficiently. Furthermore, the complexity of \textsf{Lasso+EM\_JD} is much less than \textsf{EM\_JD} as the initial estimation of \textsf{Lasso\_JD} can greatly reduce the candidate attribute space and the number of iterations needed. When $k$ is growing, the computation time of \textsf{Lasso\_JD} increases slowly, unlike \textsf{EM\_JD} that has a dramatic increase. This is because the time complexity of \textsf{Lasso\_JD} is mainly subject to the number of users.

Figures~\ref{timea2}, \ref{timea3}, \ref{timet2}, and \ref{timet3} depict the computation time on non-binary datasets (\textsf{Adult} and \textsf{TPC-E}) when $k=2$ and $k=3$. As we can see, \textsf{EM\_JD} runs with acceptable complexity on low dimension $k=2$. When $k=3$, the time complexity of \textsf{EM\_JD} increases sharply by several times. When $k$ further increases, it does not return any result within an unacceptable time during our experiment. However, \textsf{Lasso\_JD} takes less than a few seconds. This discrepancy is consistent with our complexity analysis, where we envision that the exponential growth of the candidate set will have a significant impact on \textsf{EM\_JD}. So, with the initial estimation of \textsf{Lasso\_JD}, the combined estimation \textsf{Lasso+EM\_JD} can run relatively faster than \textsf{EM\_JD} with limited candidate set. The computation time of \textsf{EM\_JD} and \textsf{Lasso\_JD} on \textsf{TPC-E} dataset with different $k=2$ and $k=3$ exhibits a similar tendency, as shown in Figures~\ref{timet2} and \ref{timet3}. We omitted the detailed report here due to the space constraint. It should be noted that the general time complexity on \textsf{TPC-E} is larger than \textsf{Adult} since the average candidate domain of \textsf{TPC-E} is larger.

\subsubsection{Accuracy}
Next, we compare the estimation accuracy of \textsf{EM\_JD},\textsf{Lasso\_JD}, and \textsf{Lasso+EM\_JD} on real datasets.

\begin{figure*}[htbp]
\begin{minipage}[t]{4.2cm}
\centering \epsfig{file=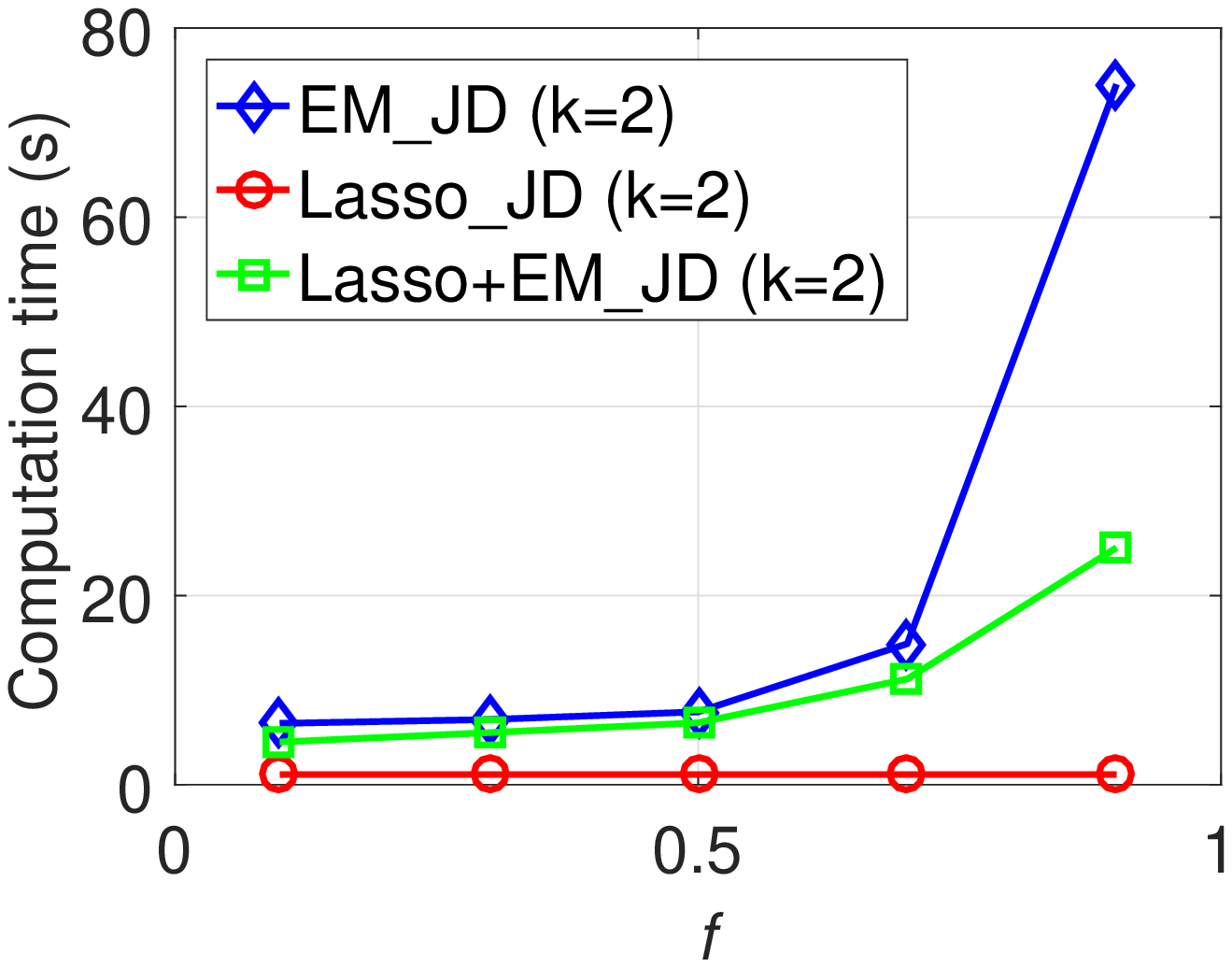, width=125pt}
\caption{Estimation Time (\textsf{TPC-E}, $k=2$)\label{timet2}}
\end{minipage}
\ \
\begin{minipage}[t]{4.2cm}
\centering \epsfig{file=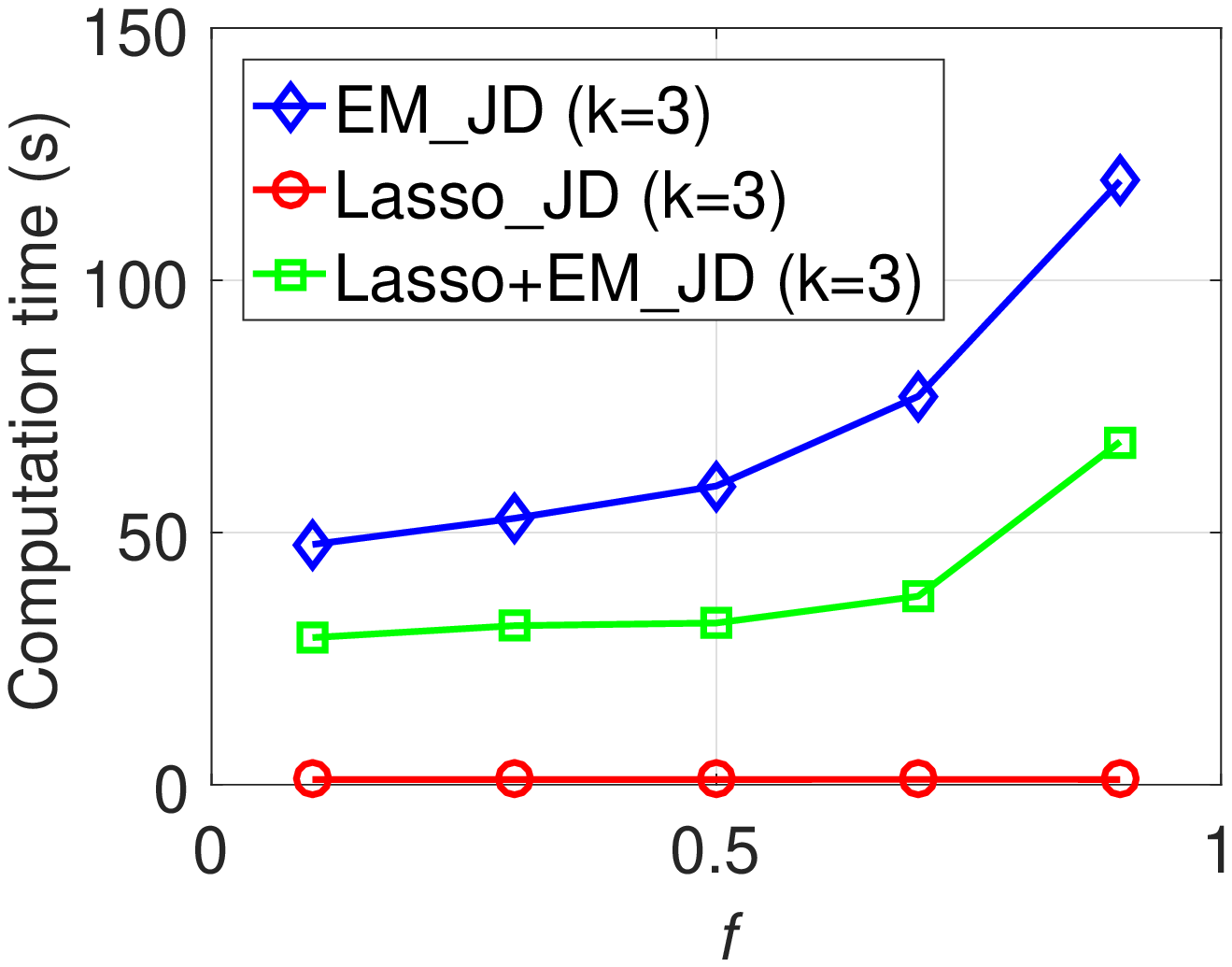, width=125pt}
\centering\caption{Estimation Time (\textsf{TPC-E}, $k=3$)\label{timet3}}
\end{minipage}
\ \
\begin{minipage}[t]{4.2cm}
\centering\epsfig{file=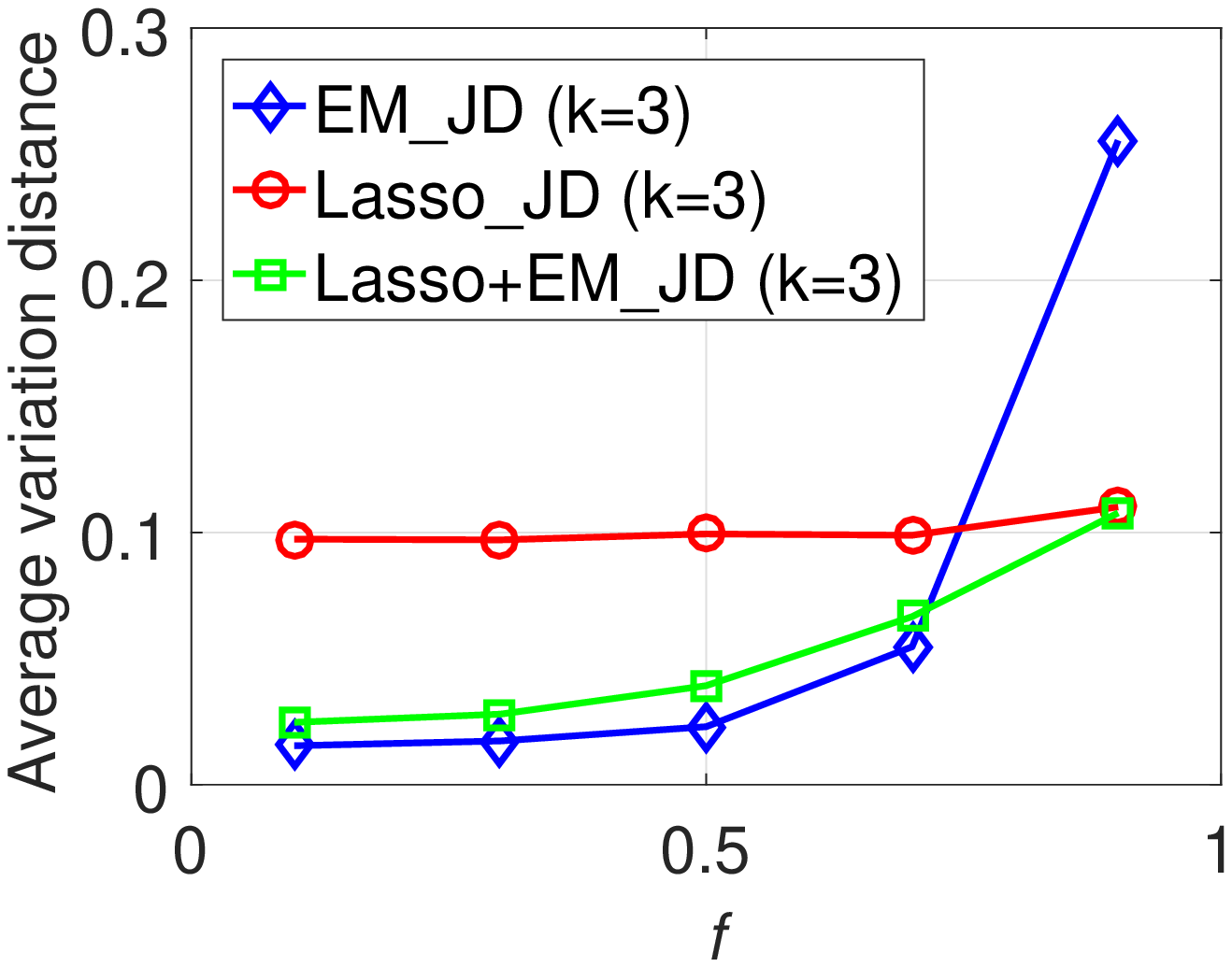, width=125pt}
\caption{Estimation Accuracy (\textsf{Retail}, $k=3$)\label{avdr3}}
\end{minipage}
\ \
\begin{minipage}[t]{4.2cm}
\centering \epsfig{file=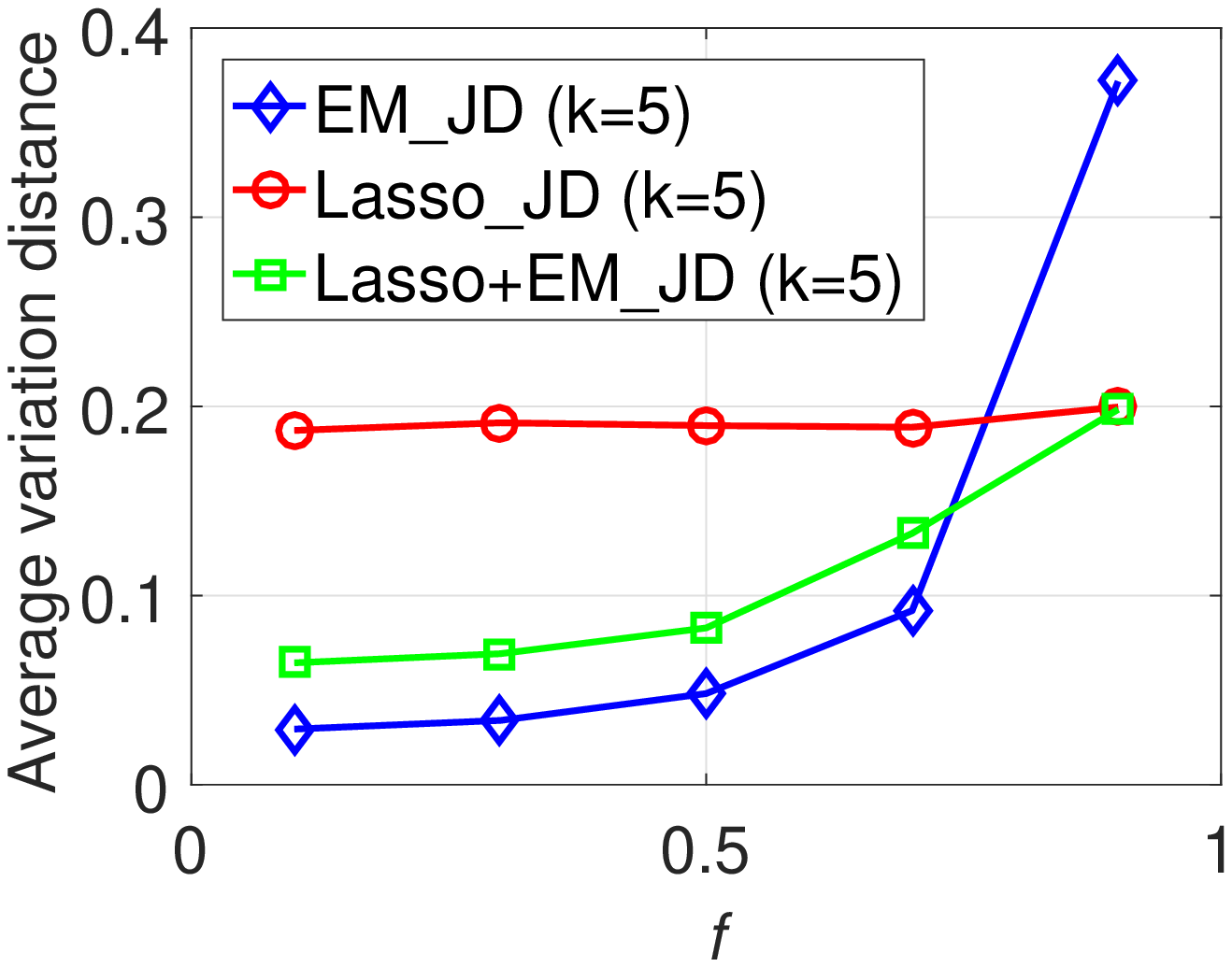, width=125pt}
\centering\caption{Estimation Accuracy (\textsf{Retail}, $k=5$)\label{avdr5}}
\end{minipage}
\vspace{-0.3cm}
\end{figure*}

\begin{figure*}[htbp]
\begin{minipage}[t]{4.2cm}
\centering\epsfig{file=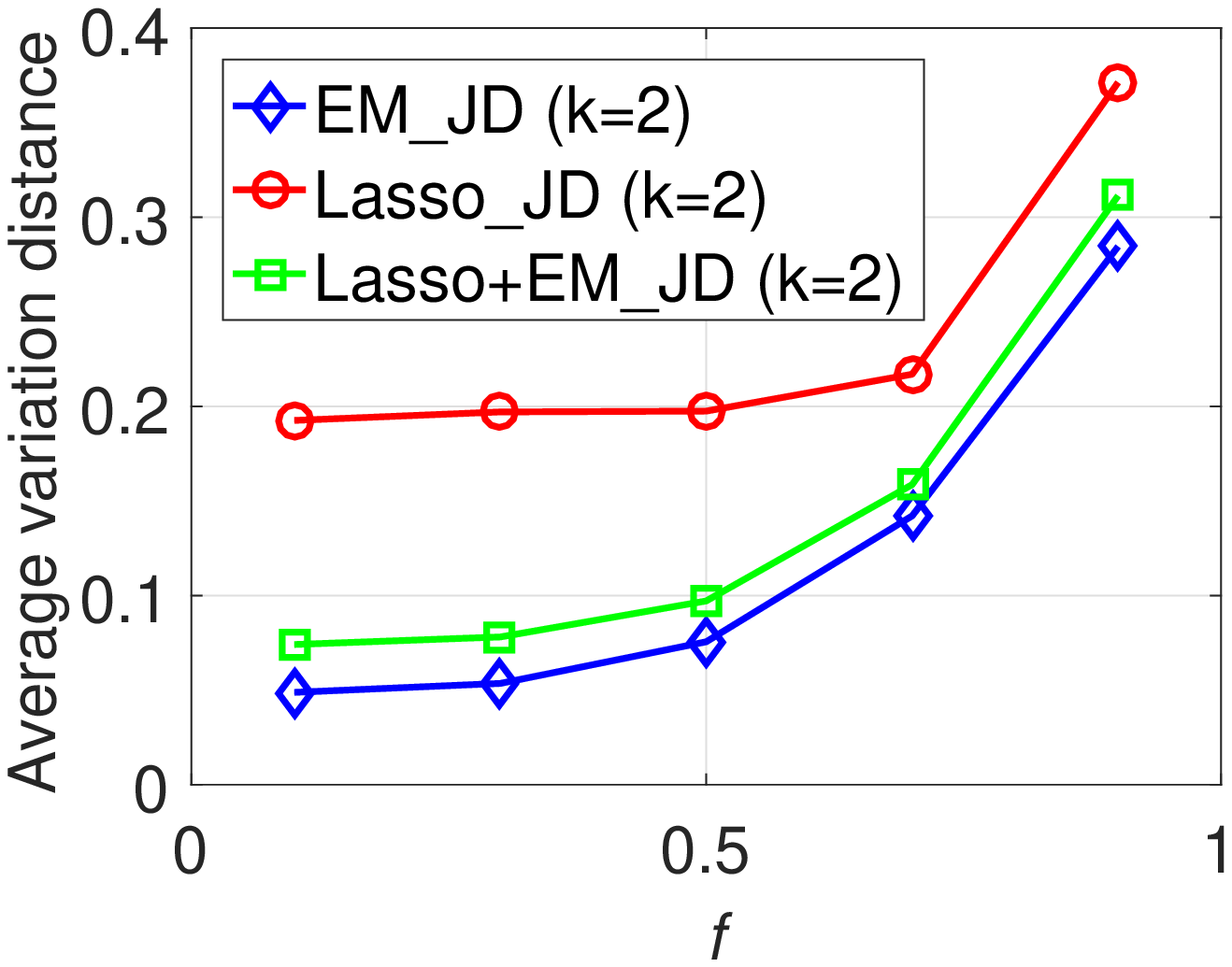, width=125pt}
\caption{Estimation Accuracy (\textsf{Adult}, $k=2$)\label{avda2}}
\end{minipage}
\ \
\begin{minipage}[t]{4.2cm}
\centering \epsfig{file=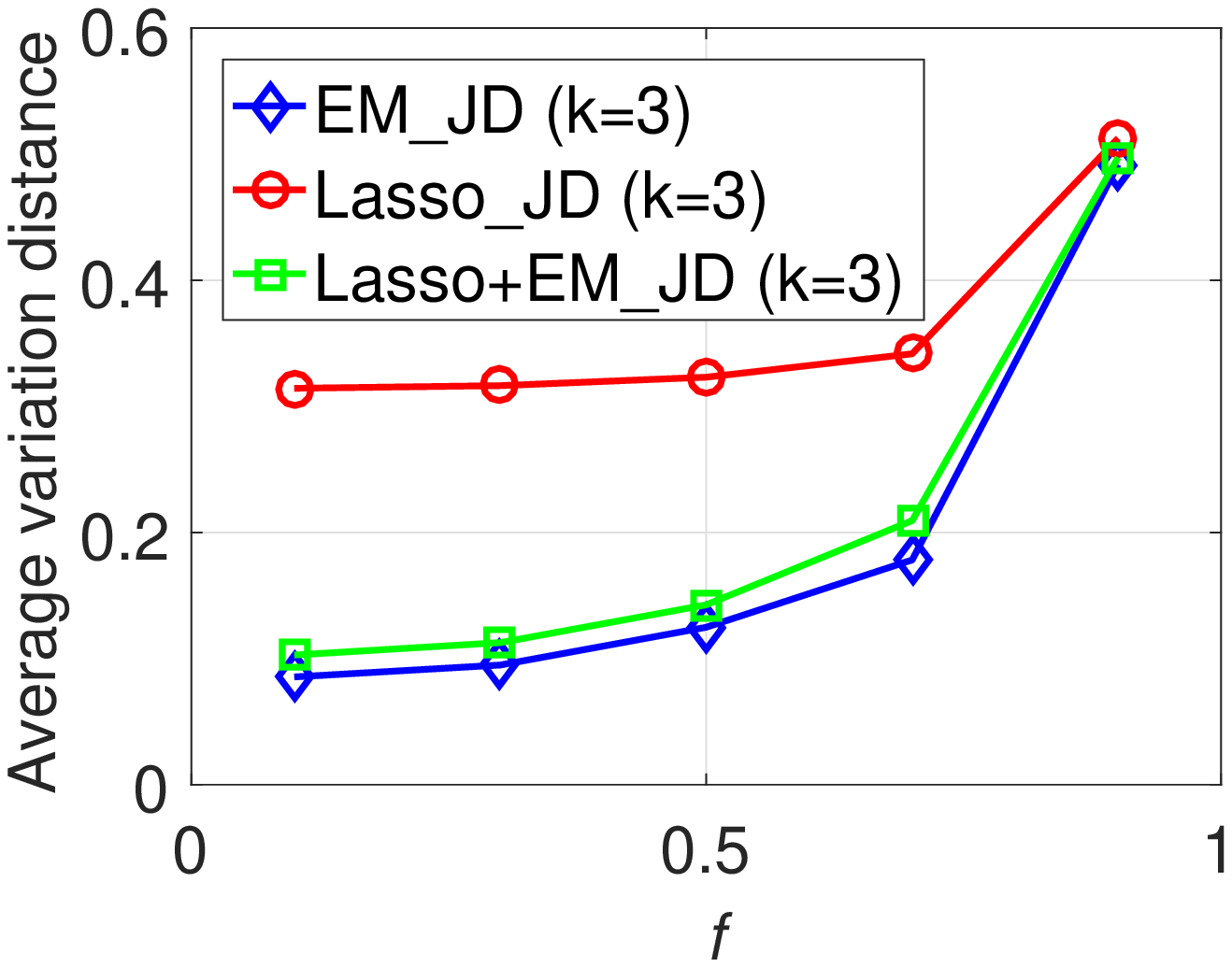, width=125pt}
\centering\caption{Estimation Accuracy (\textsf{Adult}, $k=3$)\label{avda3}}
\end{minipage}
\ \
\begin{minipage}[t]{4.2cm}
\centering\epsfig{file=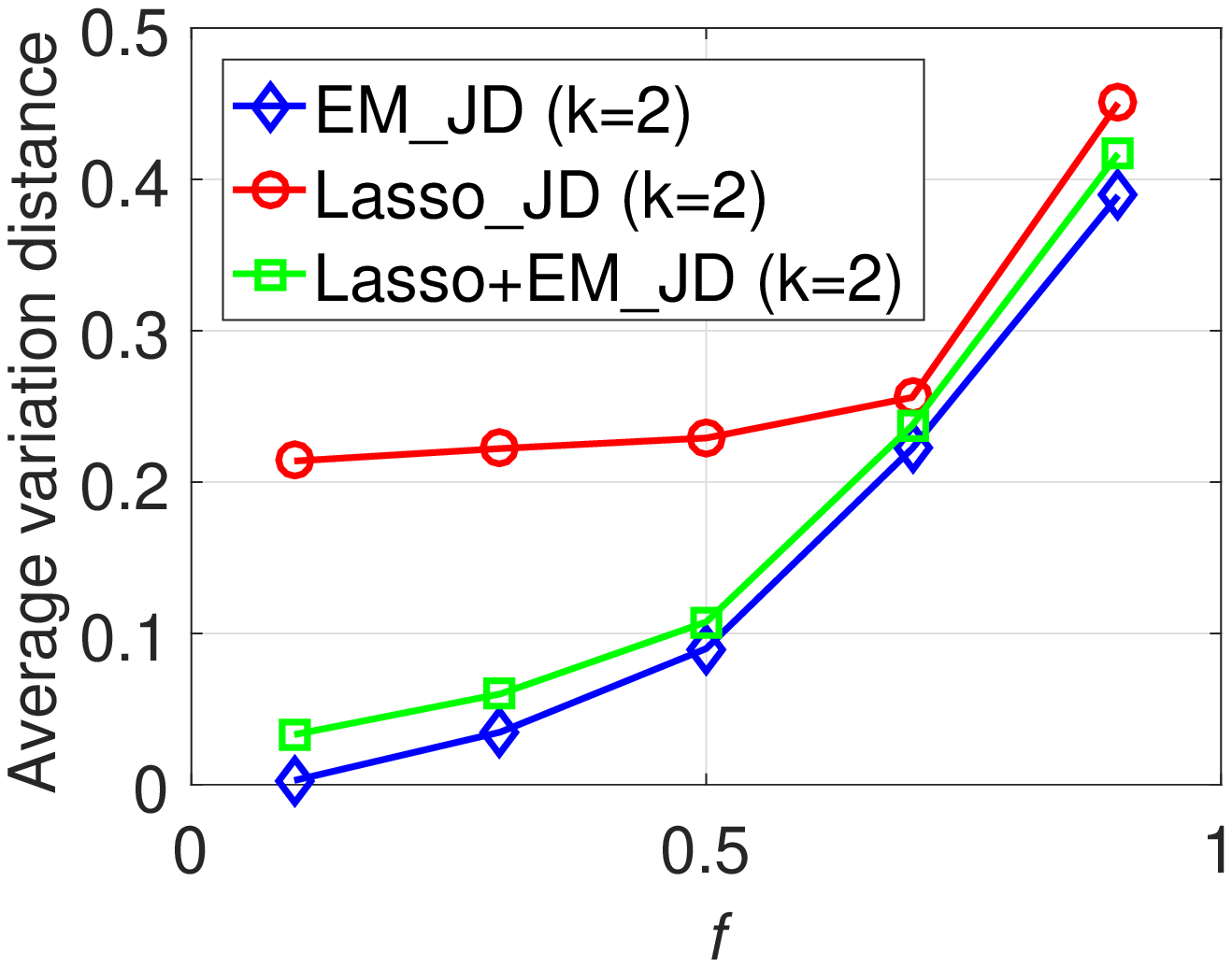, width=125pt}
\caption{Estimation Accuracy (\textsf{TPC-E}, $k=2$)\label{avdt2}}
\end{minipage}
\ \
\begin{minipage}[t]{4.2cm}
\centering \epsfig{file=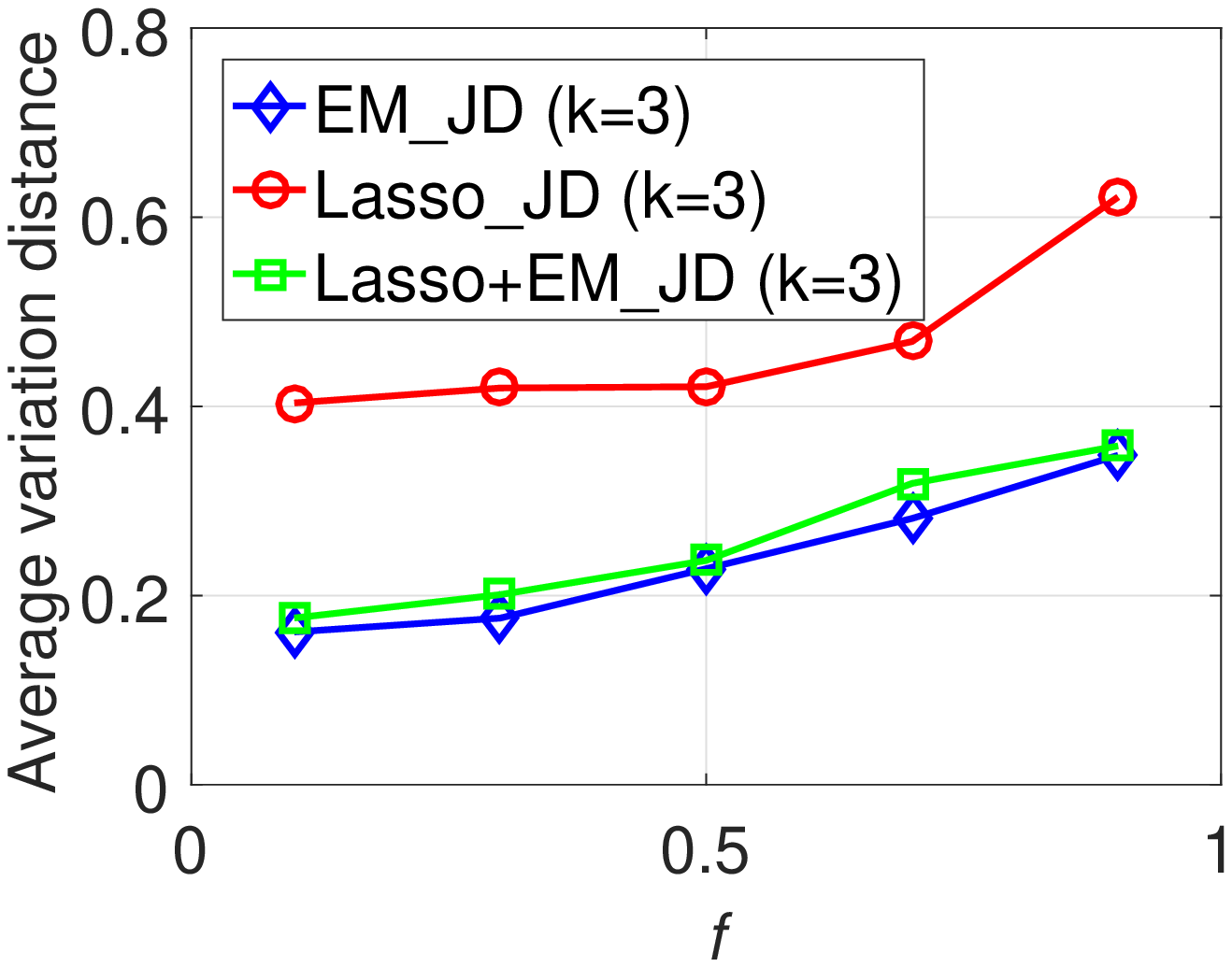, width=125pt}
\centering\caption{Estimation Accuracy (\textsf{TPC-E}, $k=3$)\label{avdt3}}
\end{minipage}
\ \
\centering
\vspace{-0.3cm}
\end{figure*}

Figures~\ref{avdr3} and \ref{avdr5} report the AVD error of \textsf{EM\_JD},\textsf{Lasso\_JD}, and \textsf{Lasso+EM\_JD} on binary dataset \textsf{Retail} with different dimensions $k=3$ and $k=5$. The AVD error of \textsf{EM\_JD} is very small when $f$ is small, but when $f$ grows, it will sharply increase to as high as $0.28$. In contrast, \textsf{Lasso\_JD} retains the error around $0.1$ even when $f=0.9$. However, in practice, when $f$ is small, i.e., $f=0.5$, the differential privacy an individual can achieve is $\epsilon=8.79$ for each dimension, which is insufficient in general. So, when $f$ is large, the AVD error of \textsf{Lasso\_JD} is comparable to or even better than that of \textsf{EM\_JD}. This is because Lasso regression is insensitive to $f$ when estimating the coefficients from the aggregated bit sum vectors. Nonetheless, \textsf{EM\_JD} is sensitive to $f$ and prone to some local optimal value because it scans each record of bit strings. In comparison, \textsf{Lasso+EM\_JD} achieves a better tradeoff between \textsf{Lasso\_JD} and \textsf{EM\_JD}. For example, it has less AVD error than \textsf{Lasso\_JD} when $f$ is small and outperforms \textsf{EM\_JD} when $f$ is large. We can also see that, the AVD error of all estimation algorithms increases with $k$, since the average frequency on $k-$dimensional combined attributes is $N/v^{k}$ 
and its statistical significance decreases with $k$ exponentially.

Figures~\ref{avda2}, \ref{avda3}, \ref{avdt2} and \ref{avdt3} also compare the AVD error of \textsf{EM\_JD},\textsf{Lasso\_JD}, and \textsf{Lasso+EM\_JD} on the non-binary datasets \textsf{Adult} and \textsf{TPC-E} with $k=2$ and $k=3$. As can be seen, when $k=2$, the AVD error of \textsf{Lasso\_JD} does not change with $f$ as the aggregated bit sum vector is insensitive to small $f$. While \textsf{EM\_JD} increases with $f$ gradually due to the scan of each individual bit string. Similar to the conclusion in the binary dataset, when $f$ is large, the trend of \textsf{Lasso\_JD} is very close to \textsf{EM\_JD}. Besides, \textsf{Lasso+EM\_JD} shows very similar performance to \textsf{EM\_JD} and incurs relatively small bias. Therefore, \textsf{Lasso+EM\_JD} achieves a good balance between utility and efficiency as it runs much faster than the baseline \textsf{EM\_JD}. In addition, when $k$ increases $(k=3)$, the estimation error increases as well. However, \textsf{Lasso+EM\_D} can further balance between \textsf{Lasso\_JD} and \textsf{EM\_JD} because the candidate set is much more sparse when $k$ is larger and \textsf{Lasso+EM\_JD} can effectively reduce the redundant of candidate set and iterations. Similar conclusion can be made from the dataset \textsf{TPC-E}. Nonetheless, because of larger candidate domain, the AVD error on \textsf{TPC-E} is generally larger than that on \textsf{Adult}.

\subsection{Correlation Identification}
In this section, we present correlations between the multiple attributes that we can learn from locally privacy-preserved user data. Particularly, we evaluated loss ratio of dependency relationship of attributes in three datasets. The parameters used in the simulation are set as follows. The dependency threshold $0.25$ for \textsf{Retail}, and $0.4$ for \textsf{Adult} and \textsf{TPC-E}. The number of bits and the number of hash functions in the bloom filter are $32$ and $4$ for \textsf{Retail}, and $128$ and $4$ for \textsf{Adult} and \textsf{TPC-E}. The sample rate is $1$ for \textsf{Retail} and $0.1$ for \textsf{Adult} and \textsf{TPC-E}.

\subsubsection{Accuracy}
Figures \ref{Rtf}, \ref{Atf}, and \ref{Ttf} show both the ratio of correct identification (accuracy), added (false positive) and lost (true negative) correlated pairs after estimation, respectively. From these figures, we can see all these estimation algorithms can have a relatively accurate identification among the attributes, especially \textsf{EM\_JD} and \textsf{Lasso+EM\_JD} algorithms. Nevertheless, generally, the accurate rate decreases with $f$ (i.e., privacy level). In Figure \ref{Rtf}, the general accuracy identified rate is about $85\%$ when the privacy is small ($f$ is less than $0.9$). While in Figures \ref{Atf} and \ref{Ttf}, the accuracy rate is as high as $95\%$ because the dependency threshold is relatively loose as $0.4$. High accurate identification guarantees the basic correlations among attributes.

However, the incorrect identification is considered separately with false positive rate and true negative, which reflect the efficiency and effectiveness of dimension reduction. Since false positive identification just adds the correlations that were not exists, this kind of misidentification only incurs no errors but redundant correlations and extra distribution learning. Instead, true negative identification implies the loss of some correlations among attributes, thus causing information loss in our dimension reduction. For false positive identification, we can see that \textsf{EM\_JD} algorithm and \textsf{Lasso+EM\_JD} are less than Lasso. That is because Lasso estimation will choose the sparse probabilities and the mutual information estimated is generally high due to the concentrated probability distribution. Especially in non-binary datasets \textsf{Adult} and \textsf{TPC-E}, the sparsity is much higher, so the estimated probability distribution is more concentrated and the false positive identification rate is high.

The true negative identification in both \textsf{Adult} and \textsf{TPC-E} is small because the true correlations are not very high itself because all attributes have a large domain. Instead, the true correlations in \textsf{Retail} are high and almost any two attributes are dependent. Therefore, the true negative identification is comparatively higher.

\subsubsection{Effectiveness of Pruning Scheme}
We also validated the pruning scheme proposed in Section~\ref{sec:pruning} with simulations on the three datasets. We first defined the dependency loss ratio as the ratio between the dependency loss after pruning with the original number of dependencies in the adjacent matrix $\mathbf{G}_{d \times d}$ of dependency graph. The complexity reduction ratio is defined as the ratio of reduced pairwise comparisons.

\begin{table}\caption{Dependency Loss Ratio and Complexity Reduction Ratio (\textsf{Adult})}\label{pruning:Adult}\centering
\begin{tabular}{|c|c|c|c|c|c|}
  \hline
   $\phi$ & $0.1$ & $0.2$ & $0.3$ & $0.4$ & $0.5$ \\
  \hline
  $\#$. Dep (Pruning) &$88$ &$38$ &$22$ &$12$ &$6$ \\
  $\#$. Dep &$102$ &$42$ &$24$ &$14$ &$8$ \\
  Loss Ratio &$0.137$ &$0.095$ &$0.083$ &$0.143$ &$0.250$  \\
  \hline
  $\#$. Pairs (Pruning) &$91$ &$66$ &$55$ &$36$ &$28$ \\
  $\#$. Pairs  &$105$ &$105$ &$105$ &$105$ &$105$ \\
  Reduction Ratio &$0.133$ & $0.371$ &$0.476$ &$0.657$ &$0.733$ \\
  \hline
\end{tabular}
\end{table}

\begin{table}\caption{Dependency Loss Ratio and Complexity Reduction Ratio (TPC-E)}\label{pruning:TPC-E}\centering
\begin{tabular}{|c|c|c|c|c|c|}
  \hline
   $\phi$ & $0.1$ & $0.2$ & $0.3$ & $0.4$ & $0.5$ \\
  \hline
  $\#$. Dep (Pruning) &$44 $ &$16$ &$16$ &$8$ &$8$ \\
  $\#$. Dep &$46$ &$24$ &$20$ &$10$ &$10$ \\
  Loss Ratio &$0.043$ &$0.333$ &$0.200$ &$0.200$ &$0.200$  \\
  \hline
  $\#$. Pairs (Pruning) &$231$ &$171$ &$136$ &$66$ &$45$ \\
  $\#$. Pairs  &$276$ &$276$ &$276$ &$276$ &$276$ \\
  Reduction Ratio &$0.163$ & $0.380$ &$0.507$ &$0.761$ &$0.837$ \\
  \hline
\end{tabular}
\end{table}

\begin{table}\caption{Dependency Loss Ratio and Complexity Reduction Ratio (Retail)}\label{pruning:Retail}\centering
\scriptsize
\begin{tabular}{|c|c|c|c|c|c|}
  \hline
   $\phi$ & $0.1$ & $0.15$ & $0.2$ & $0.25$ & $0.3$ \\
  \hline
  $\#$. Dep (Pruning) &$256 $ &$256$ &$256$ &$250$ &$244$ \\
  $\#$. Dep &$240$ &$240$ &$238$ &$220$ &$200$ \\
  Loss Ratio &$-0.067$ &$-0.067$ &$-0.076$ &$-0.136$ &$-0.220$  \\
  \hline
  $\#$. Pairs (Pruning) &$91$ &$91$ &$78$ &$66$ &$55$ \\
  $\#$. Pairs  &$120$ &$120$ &$120$ &$120$ &$120$ \\
  Reduction Ratio &$0.242$ & $0.242$ &$0.350$ &$0.450$ &$0.512$ \\
  \hline
\end{tabular}
\end{table}

Tables~\ref{pruning:Adult}, \ref{pruning:TPC-E}, and \ref{pruning:Retail} illustrate the effectiveness of our proposed heuristic pruning scheme. Particularly, as shown in Tables~\ref{pruning:Adult} and \ref{pruning:TPC-E}, with the increase of $\phi$, which shows the strength of correlations, the number of original dependencies in dataset \textsf{Adult} decreases dramatically. Also, the dependencies after the heuristic pruning decrease accordingly and their number is quite close to the original. However, when $\phi$ increases, the number of pairwise comparison becomes less compared to the full pairwise comparison. So, it shows that the heuristic pruning scheme can effectively reduce the complexity with fairly small sacrifice of dependency accuracy. Similar conclusion can be found in Table~\ref{pruning:TPC-E} on non-binary dataset \textsf{TPC-E}. On the binary dataset \textsf{Retail}, due to the prior knowledge that binary datasets normally have strong mutual dependency, we changed the pruning scheme a little. Particularly, we assume all the attributes are dependent with each other and our pruning scheme aims at finding the non-dependency from that attributes $A_j$ with less entropy $H(A_j)$. According to Table~\ref{pruning:Retail}, the number of dependencies after pruning decreases slowly and the minus symbol in the dependency loss ratio means that there is no loss of dependencies but there are redundant dependencies that should not exist in original datasets. It should be noted that redundant dependencies cover all the original dependencies. Therefore, the redundancy will not cause the degrade of data utility since more correlations are kept. However, the efficiency of dimension reduction, which should cut off as many unnecessary correlations as possible, is hindered. So, according to Table~\ref{pruning:Retail}, we can also say that the heuristic pruning scheme can achieve up to $50\%$ complexity reduction without loss of dependencies.

\subsection{SVM and Random Forest Classifications}

To show the overall performance of LoPub, we evaluated both the SVM and random forest classification error rate in the new datasets synthesized by different versions of LoPub. We first sampled from the three original datasets \textsf{Retail}, \textsf{Adult}, and \textsf{TPC-E} to get both the training sets and test sets. Then, we generated the privacy-preserving synthetic datasets from the training data. Next, we trained three different SVM classifiers and three random forest classifiers on the synthetic datasets. Lastly, we evaluated the classification rate on the original sampled test sets. Particularly, the average random forest classification rate is computed on all the original attributes and the average SVM classification rate is computed on all the original binary-state attributes in each dataset, for example, all attributes in binary dataset \textsf{Retail}, the $10$th (gender) and $15$th (marital) attribute in \textsf{Adult}, and the $2$nd, $10$th, $23$rd, and $24$th attribute in \textsf{TPC-E}. For comparison, we also trained the corresponding SVM and random forest classifiers on each sampled training set and measured their classification rate each time.

\begin{figure*}[htbp]
\begin{minipage}[t]{4.2cm}
\centering\epsfig{file=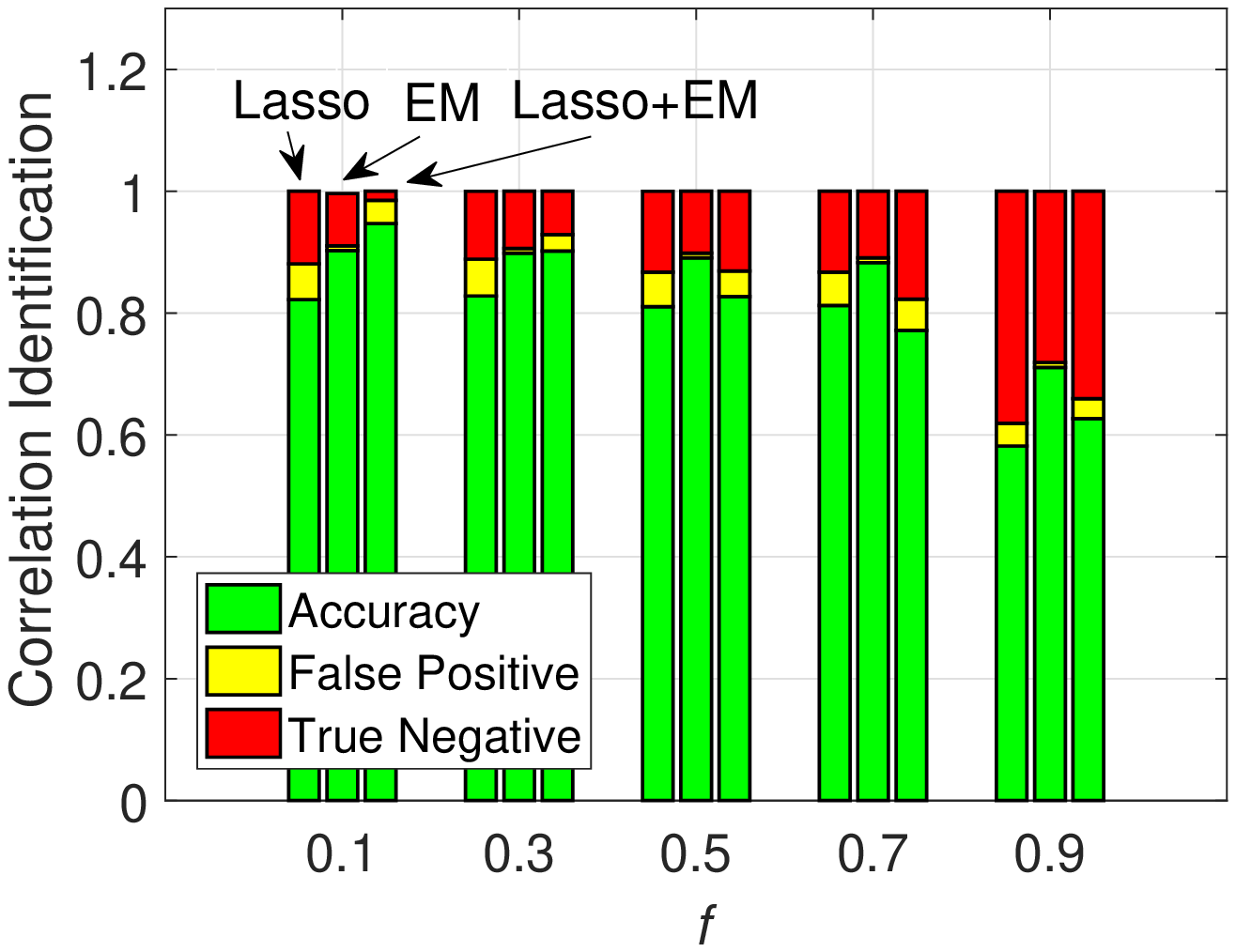, width=140pt}
\caption{Correlation Identification Rate (\textsf{Retail})\label{Rtf}}
\end{minipage}
\ \
\begin{minipage}[t]{4.2cm}
\centering\epsfig{file=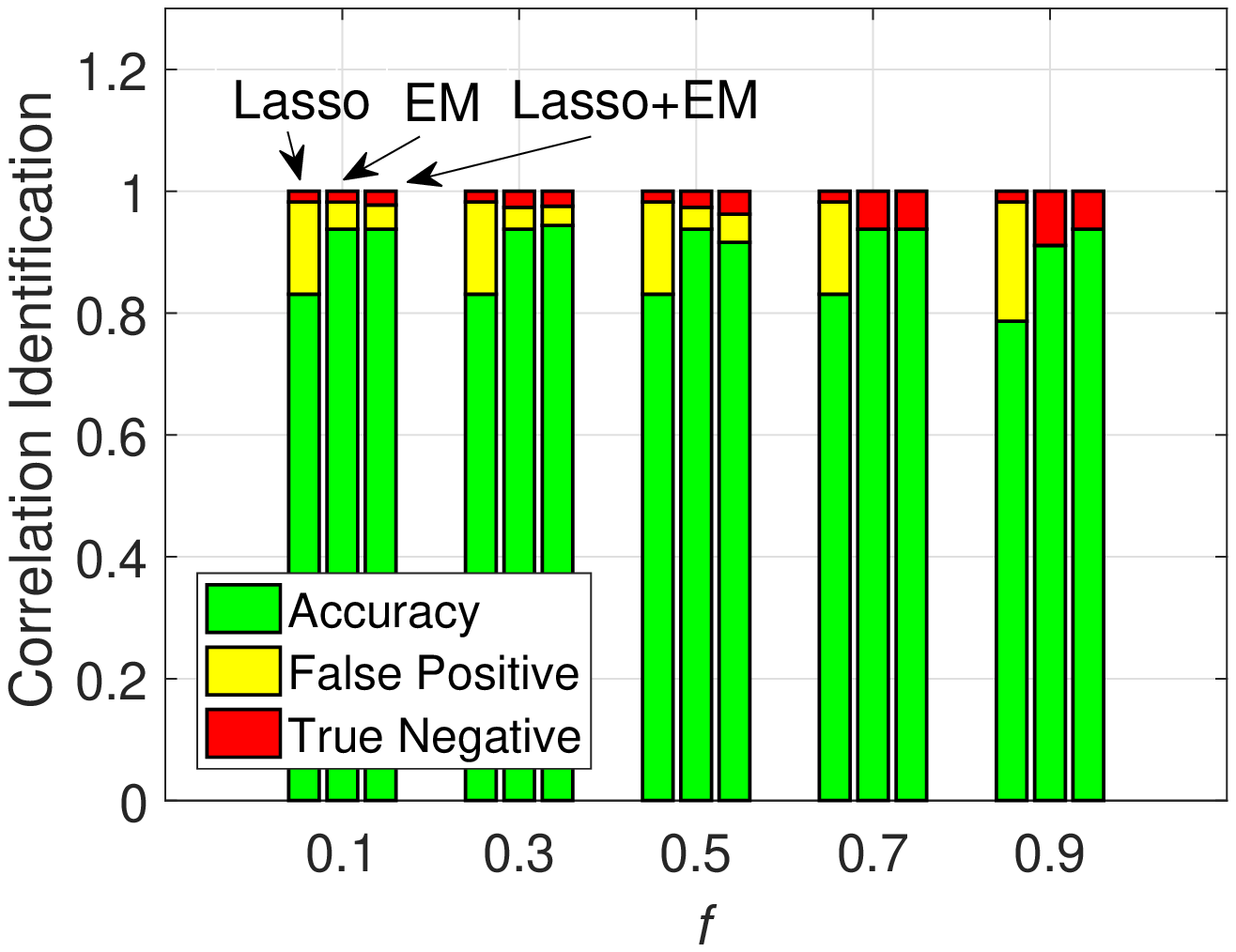, width=140pt}
\caption{Correlation Identification Rate(\textsf{Adult})\label{Atf}}
\end{minipage}
\ \
\begin{minipage}[t]{4.2cm}
\centering\epsfig{file=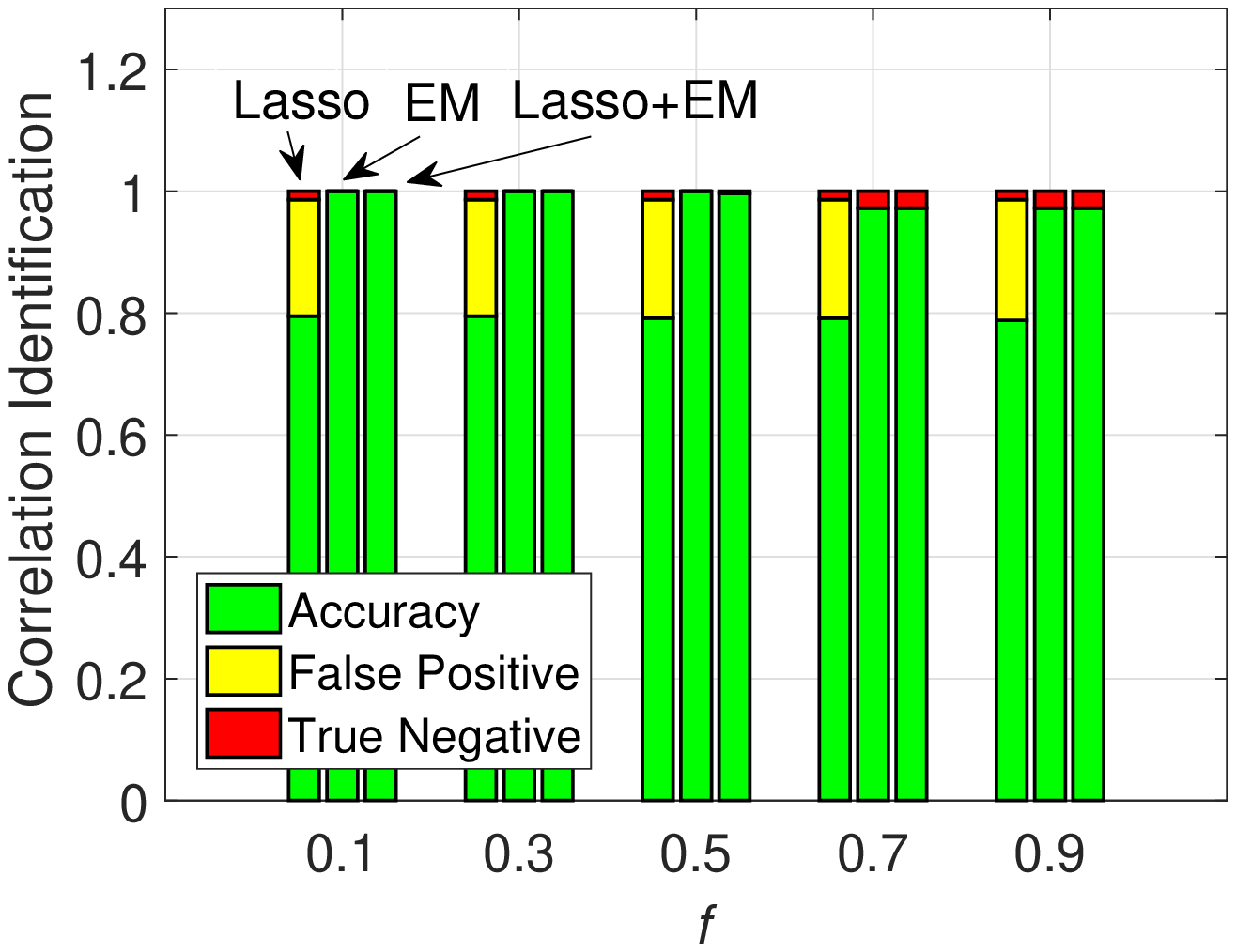, width=140pt}
\caption{Correlation Identification Rate (\textsf{TPC-E})\label{Ttf}}
\end{minipage}
\ \
\begin{minipage}[t]{4.2cm}
\centering\epsfig{file=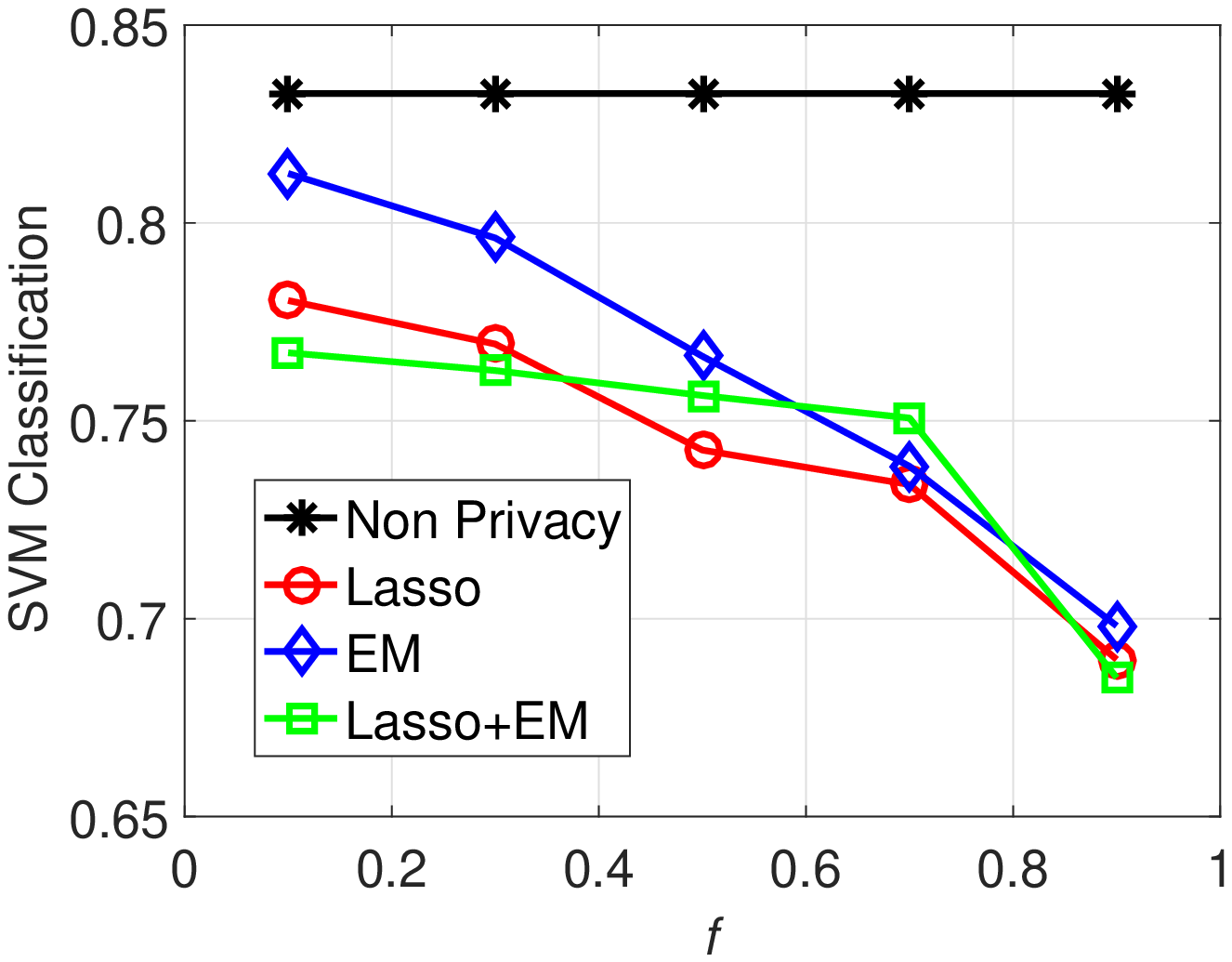, width=125pt}
\caption{SVM Classification Rate (\textsf{Retail})\label{svm4}}
\end{minipage}
\ \
\centering
\vspace{-0.3cm}
\end{figure*}

\begin{figure*}[htbp]
\begin{minipage}[t]{4.2cm}
\centering\epsfig{file=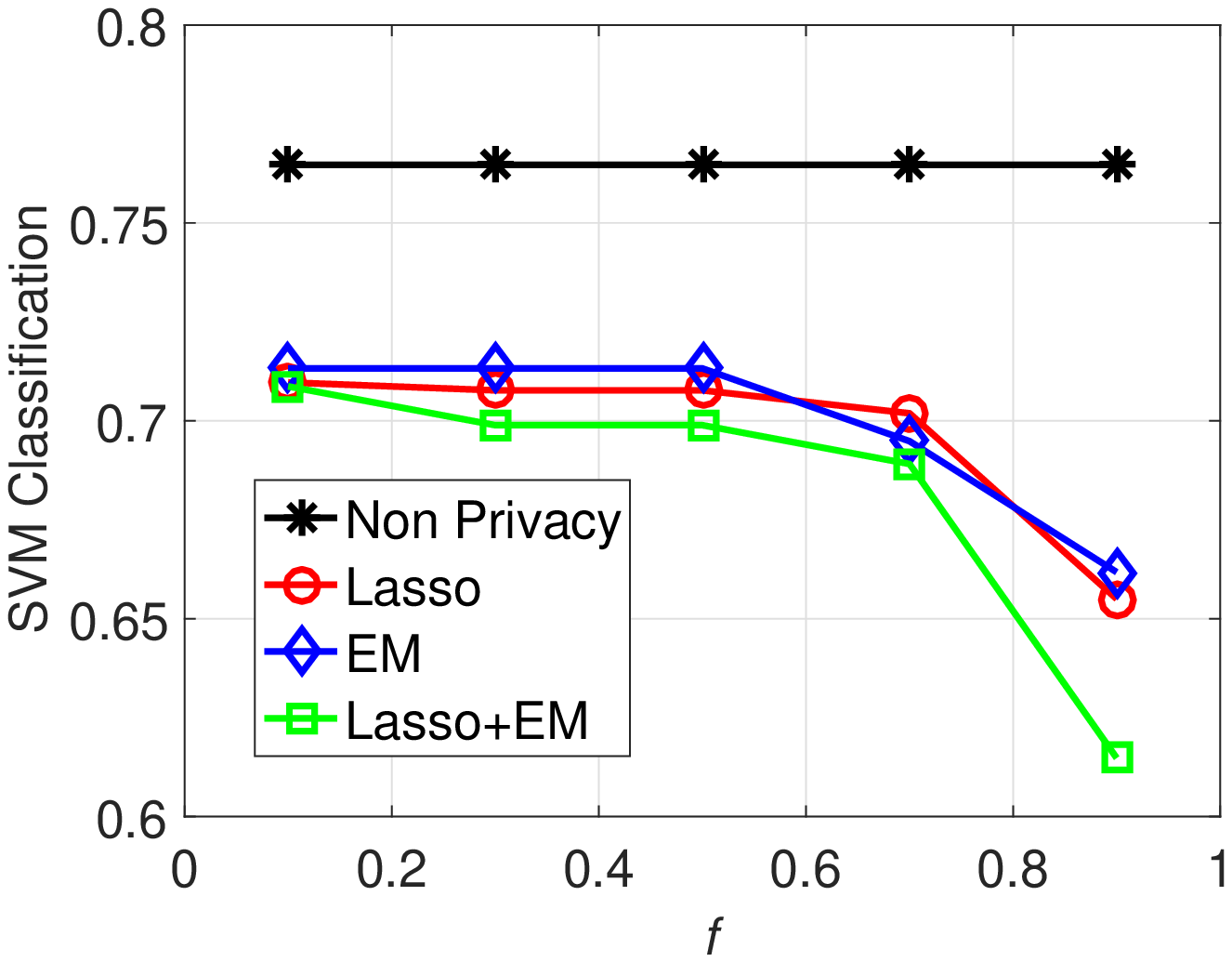, width=125pt}
\caption{SVM Classification Rate (\textsf{Adult})\label{svm2}}
\end{minipage}
\ \
\begin{minipage}[t]{4.2cm}
\centering\epsfig{file=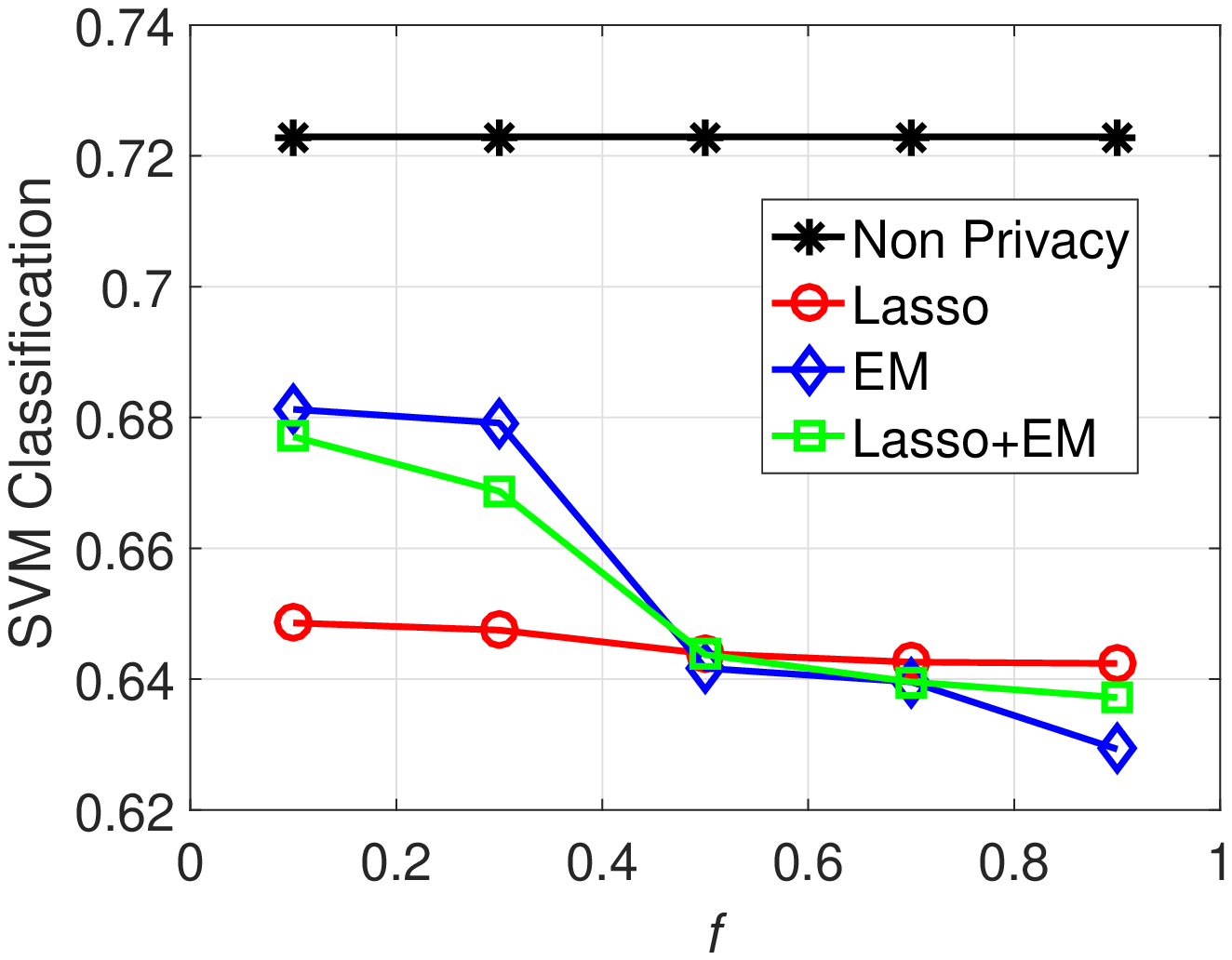, width=140pt}
\caption{SVM Classification Rate (\textsf{TPC-E})\label{svm3}}
\end{minipage}
\ \
\begin{minipage}[t]{4.2cm}
\centering\epsfig{file=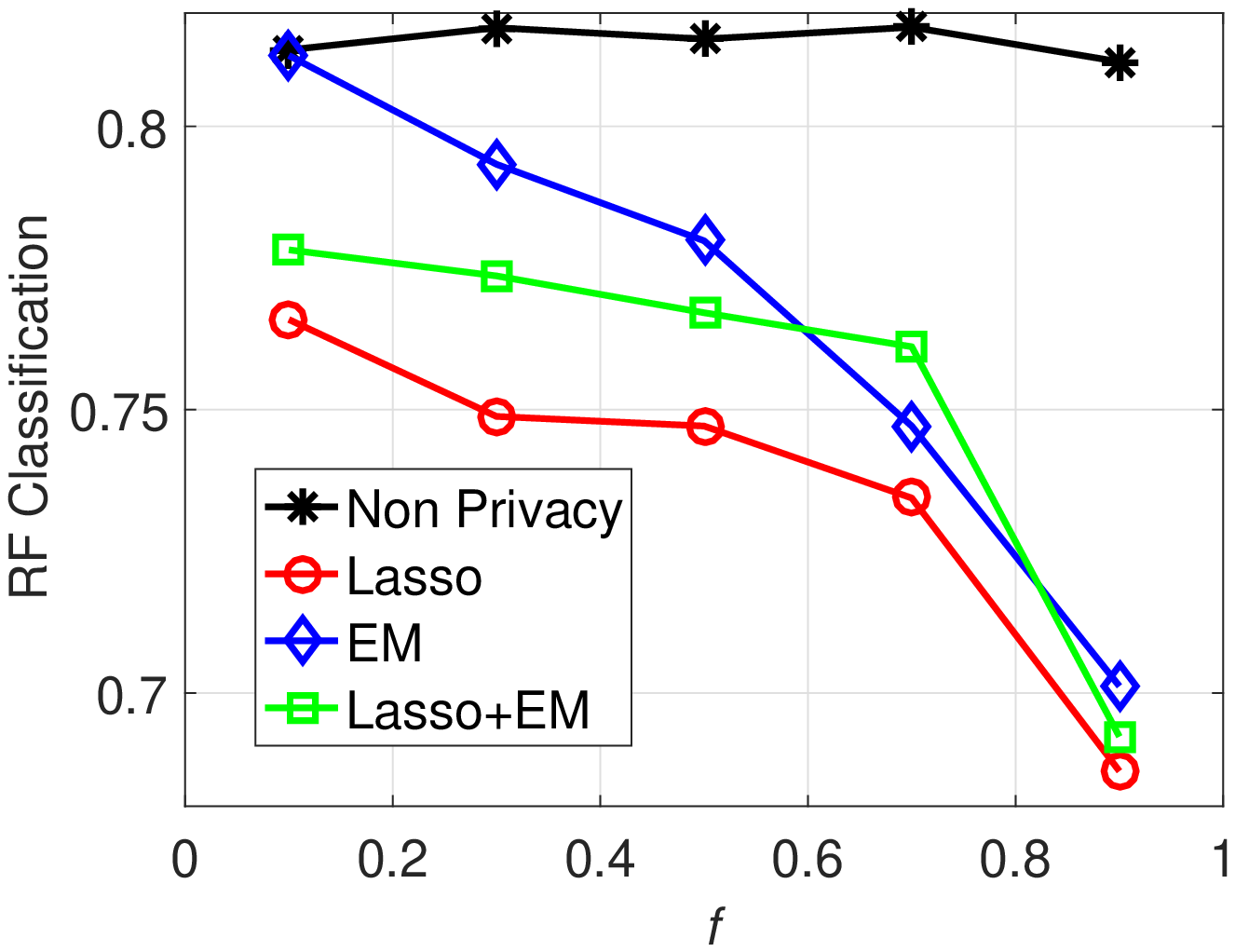, width=125pt}
\caption{Random Forest Classification Rate (\textsf{Retail})\label{rf4}}
\end{minipage}
\ \
\begin{minipage}[t]{4.2cm}
\centering\epsfig{file=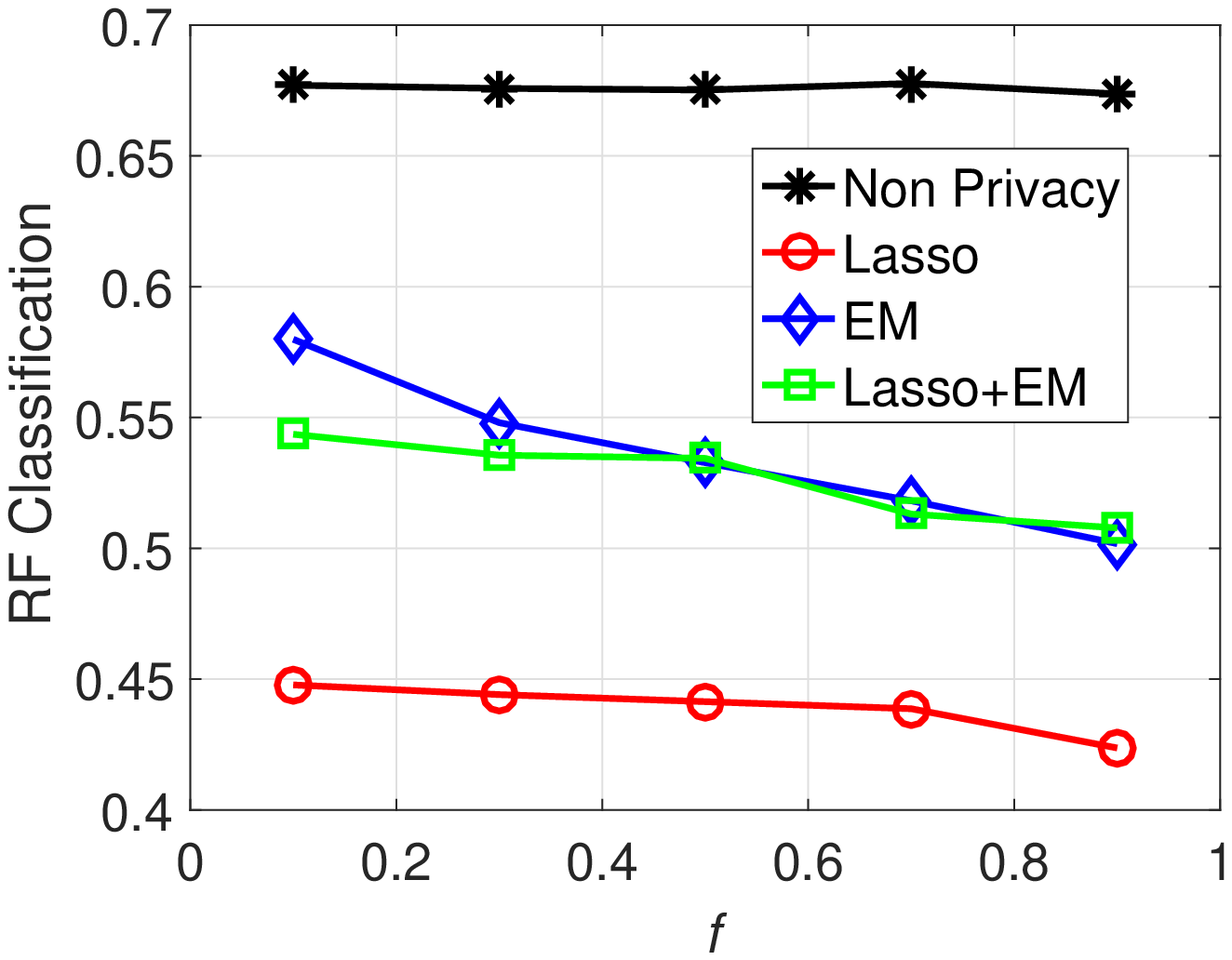, width=140pt}
\caption{Random Forest Classification Rate (\textsf{Adult})\label{rf2}}
\end{minipage}
\ \
\centering
\vspace{-0.3cm}
\end{figure*}

Figures~\ref{svm4}, \ref{svm2}, and \ref{svm3} show the average accurate SVM classification rate on three datasets \textsf{Retail}, \textsf{Adult} and \textsf{TPC-E}. In all figures, the average SVM classification rate decreases with $f$, which reflects the privacy level. Generally, when $f$ is small $f<0.9$, the classification rate drops slowly. Nevertheless, when $f=0.9$, there will be a large gap. This is because the differential privacy level changes as shown in Equation~(\ref{eq:dp}). For SVM, the classification rate is relatively close to the that of non-privacy case. This is because SVM classification only considers binary-state attributes and the distribution estimation on binary-state attributes can be more accurate than non-binary attributes, which have sparser distribution. In all figures, we can see that Lasso based estimation has generally smaller classification rate because its biased estimation. EM-based estimation generally outperforms others but still showed performance degradation when $f$ is large, while \textsf{Lasso+EM\_JD} could find a better balance between alternative methods.

\begin{figure*}[htbp]
\begin{minipage}[t]{4.2cm}
\centering\epsfig{file=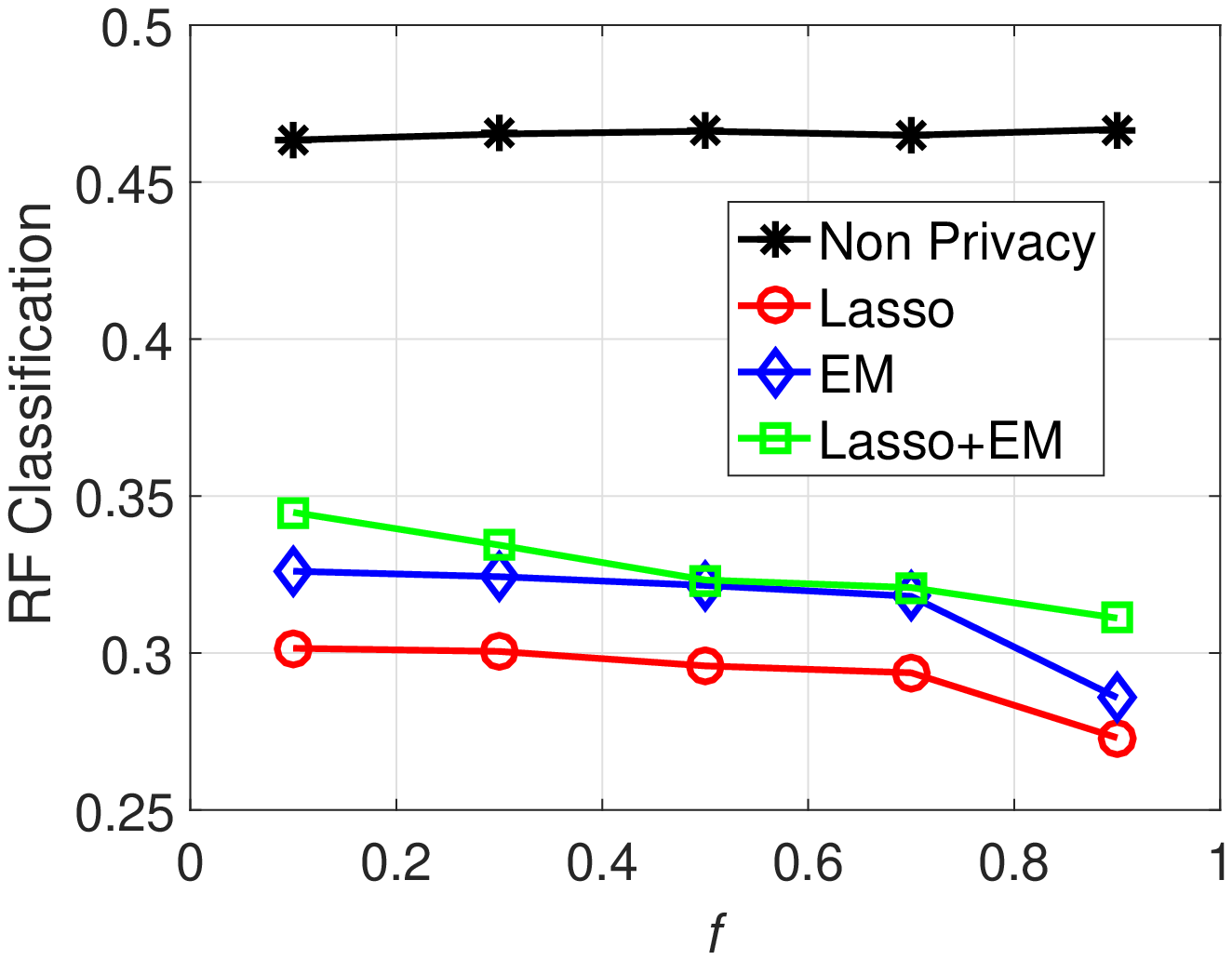, width=140pt}
\caption{Random Forest Classification Rate (\textsf{TPC-E})\label{rf3}}
\end{minipage}
\ \
\begin{minipage}[t]{4.2cm}
\centering\epsfig{file=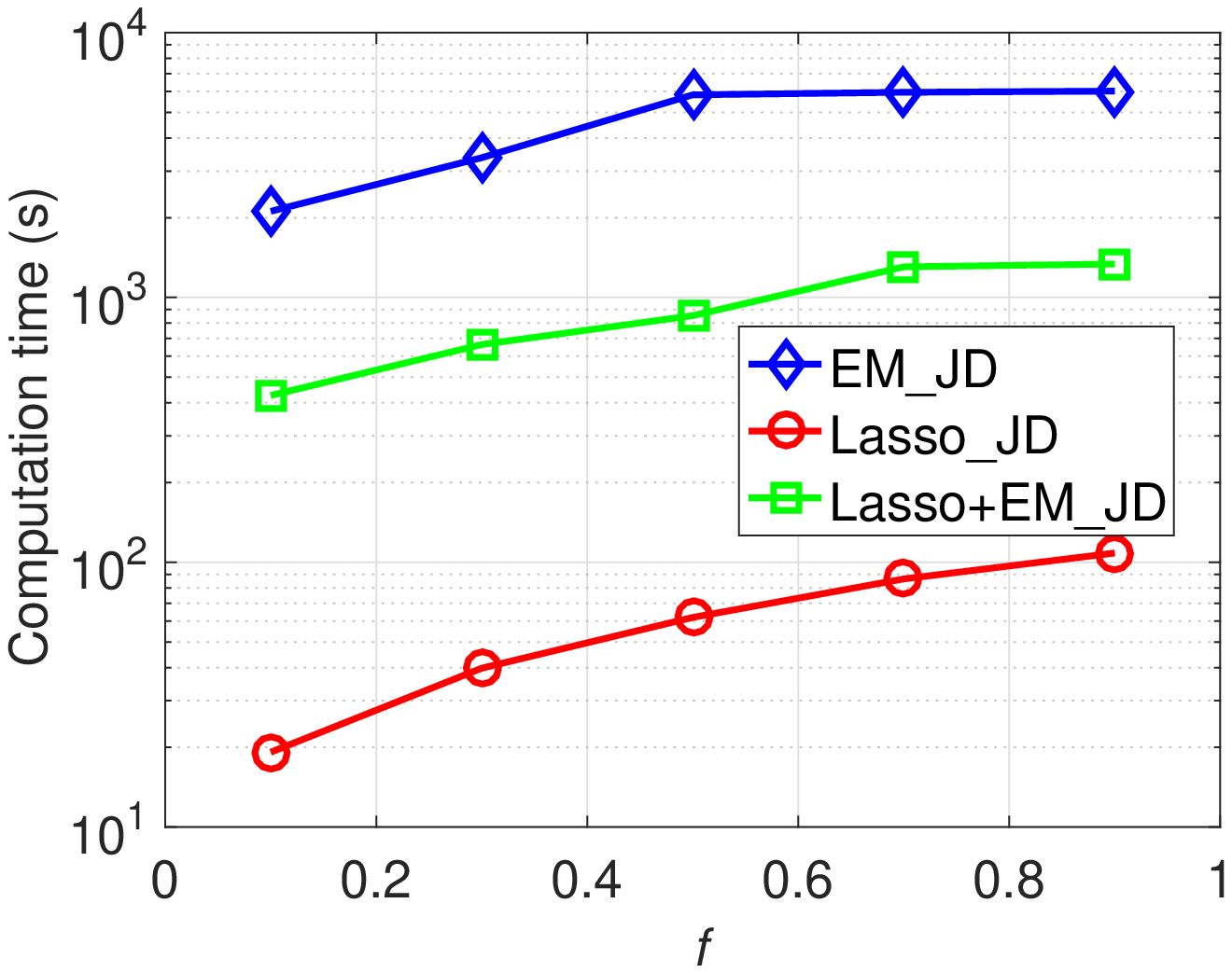, width=125pt}
\caption{Overall Time of LoPub (\textsf{Retail})\label{tttime4}}
\end{minipage}
\ \
\begin{minipage}[t]{4.2cm}
\centering\epsfig{file=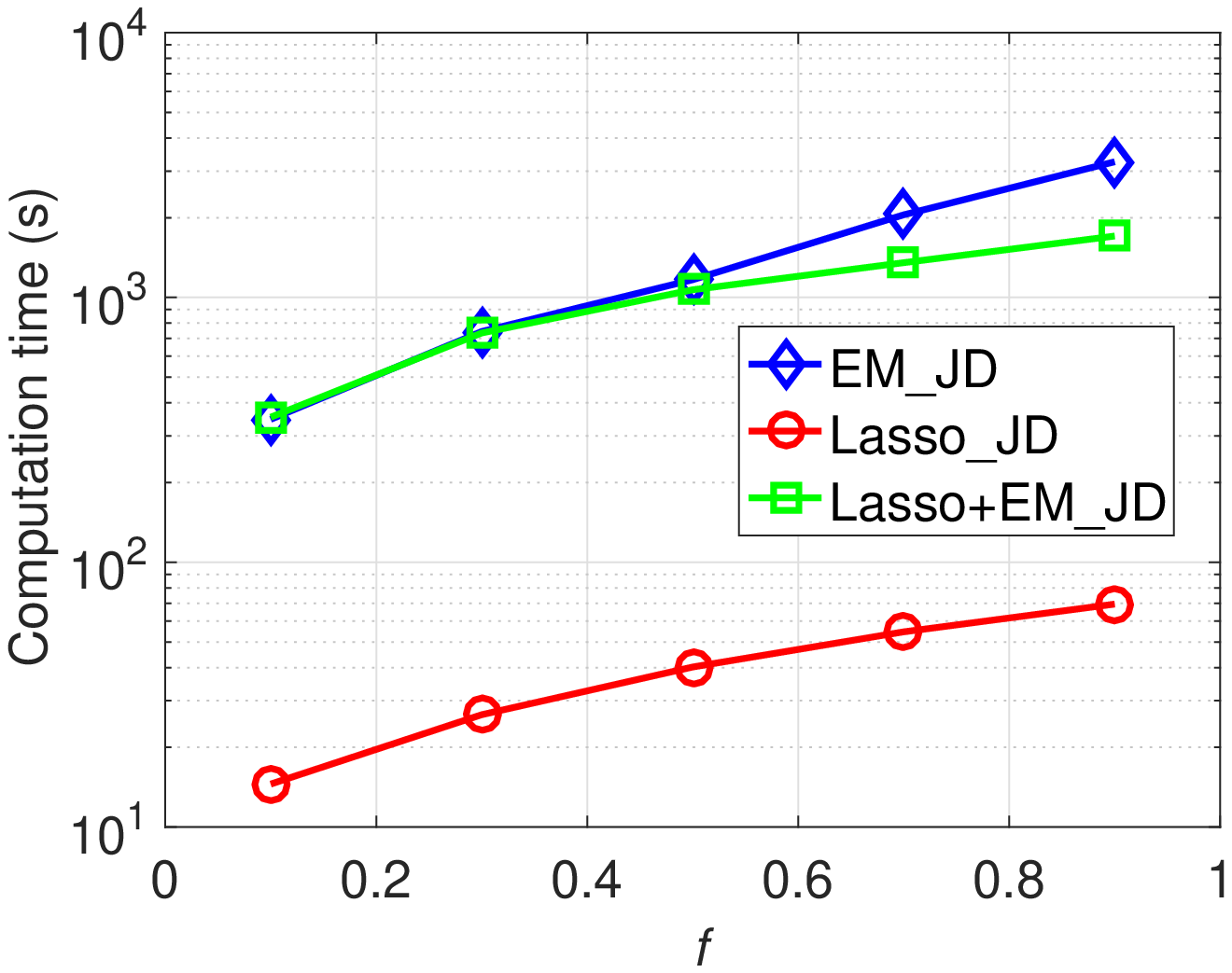, width=125pt}
\caption{Overall Time of LoPub (\textsf{Adult})\label{tttime2}}
\end{minipage}
\ \
\begin{minipage}[t]{4.2cm}
\centering\epsfig{file=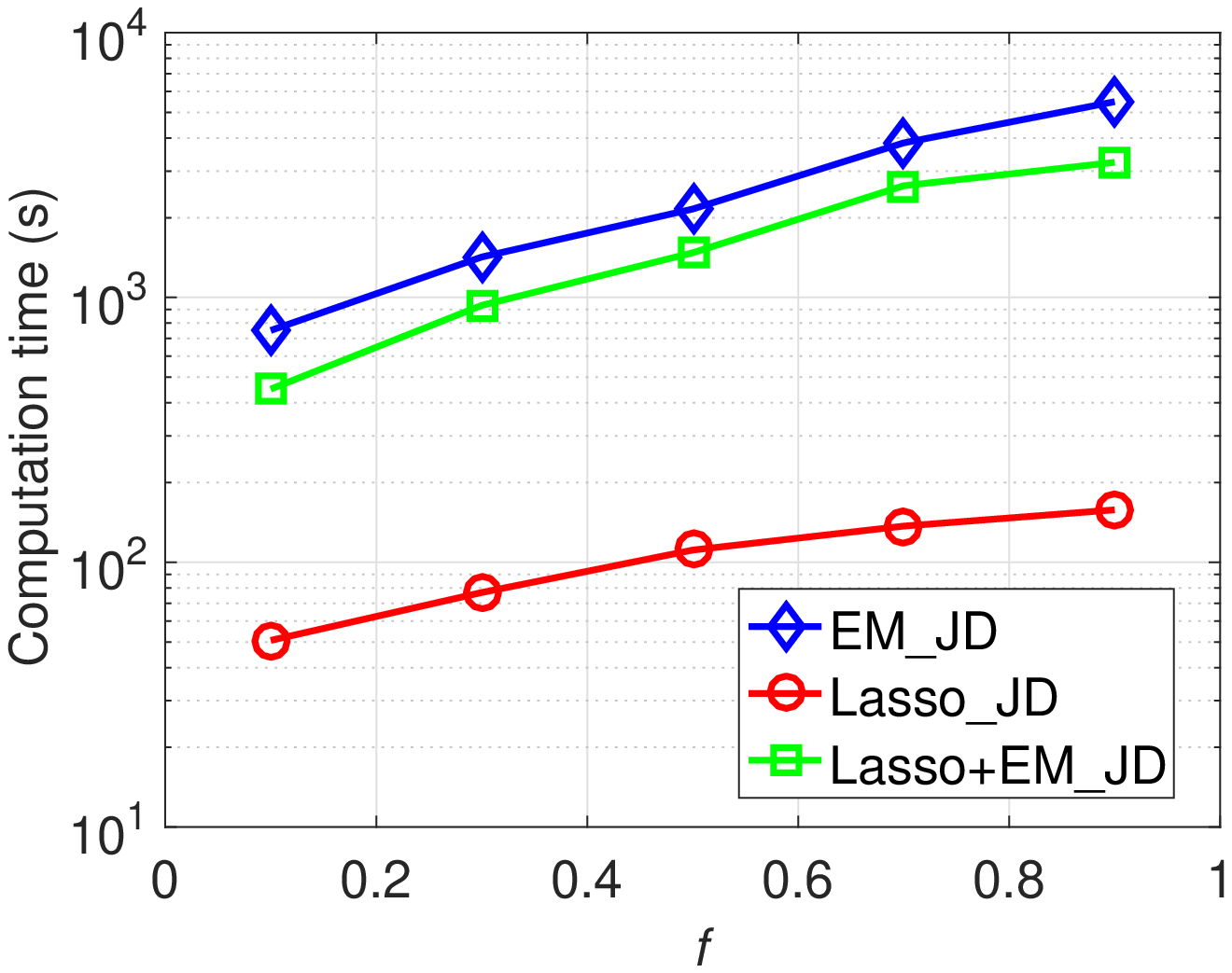, width=125pt}
\caption{Overall Time of LoPub (\textsf{TPC-E})\label{tttime3}}
\end{minipage}
\ \
\centering
\vspace{-0.3cm}
\end{figure*}

However, in Figures~\ref{rf4}, \ref{rf2}, and \ref{rf3}, due to the high sparsity in the distribution of non-binary attributes, the joint distribution estimation on non-binary attributes may be biased and that is why the random forest classification on our synthetic datasets is not as good as SVM classification. Nonetheless, the synthetic data still keeps sufficient information of original crowdsourced datasets. For example, the worst random forest classification rate in the three datasets is $67\%$, $42\%$, and $26\%$, which are much larger than the average random guess rate of $50\%$, $15\%$, and $13\%$, respectively. In detail, EM-based estimation worked relatively well to generate the synthetic datasets and Lasso estimation caused larger bias in the random forest classification. However, with the initial estimation of Lasso estimation, \textsf{Lasso+EM\_JD} works also well and degrades slowly with $f$.

For reference, the overall computational time for synthesizing new datasets are also presented in Figures~\ref{tttime4}, \ref{tttime2}, and \ref{tttime3}. Despite the worst utility, Lasso-based algorithm is the most efficient solution, which achieves approximately three orders of magnitudes faster than the EM-based method. As mentioned before, that is because it can estimate the joint distribution regardless of the number of bit strings. With the initial estimation of \textsf{Lasso\_JD}, the \textsf{EM\_JD} can then be effectively simplified from two aspects: the sparse candidates can be limited and the initial value is well set. Instead, the baseline \textsf{EM\_JD} not only needs to build prior probability distribution for all candidates but also begins the convergence with a randomness value.

\section{Conclusion}\label{sec:conclusion}
In this paper, we propose a novel solution, LoPub, to achieve the high-dimensional data release with local privacy in crowdsourced systems. Specifically, LoPub learns from the distributed data records to build the correlations and joint distribution of attributes, synthesizing an approximate dataset for privacy protection. To realise the effective multi-variate distribution estimation, we proposed EM-based and Lasso-based joint distribution estimation algorithms. The experiment results on the real datasets show that LoPub is an efficient and effective mechanism to release a high-dimensional dataset while providing sufficient local privacy guarantee for crowdsourced users.

\bibliographystyle{abbrv}
\bibliography{refs}

\end{document}